\documentclass[12pt,a4paper]{article}
\usepackage[utf8]{inputenc}
\usepackage[T1]{fontenc}
\usepackage[english]{babel}
\usepackage[a4paper,top=2cm,bottom=2cm,left=2cm,right=2cm,marginparwidth=1.75cm]{geometry}
\usepackage{authblk}
\usepackage{balance}
\usepackage{bm}
\usepackage{siunitx}
\usepackage{graphicx}
\usepackage{array}
\usepackage[usenames,dvipsnames]{xcolor}
\usepackage{booktabs}
\usepackage{multirow}
\usepackage{epstopdf}
\usepackage{caption}
\usepackage{subcaption}
\usepackage[version=3]{mhchem}
\usepackage{soul}
\usepackage{url}
\usepackage{hyperref}
\usepackage{doi}
\usepackage{amsmath, amssymb}
\usepackage[export]{adjustbox}
\usepackage[super,sort&compress,comma]{natbib}
\usepackage[prefix]{nomencl}
\usepackage{changes}
\usepackage{tikz}
\usetikzlibrary{tikzmark}
\usepackage{pdflscape}
\usepackage[nottoc]{tocbibind}
\usepackage[symbol]{footmisc}
\DeclareSIUnit\bar{bar}
\DeclareSIUnit\pixel{px}

\makenomenclature

\renewcommand{\nomgroup}[1]{%
  \ifthenelse{\equal{#1}{A}}{\item[\textbf{}]}{%
  \ifthenelse{\equal{#1}{B}}{\item[\textbf{Greek Symbols}]}{%
  \ifthenelse{\equal{#1}{C}}{\item[\textbf{Subscripts}]}{}}}}
\hypersetup{colorlinks = true,linkcolor = blue,anchorcolor =red,citecolor = blue,filecolor = red,urlcolor = red, pdfauthor=author}

\title{Boosting electrode performance and bubble management via Direct Laser Interference Patterning\footnotemark[2]}

\author[1,2]{Hannes Rox$^{\ast,\ddag,}$}
\author[3]{Fabian Ränke$^{\ddag,}$}
\author[2]{Jonathan Mädler}
\author[4]{Mateusz M. Marzec}
\author[4]{Krystian Sokołowski}
\author[3]{Robert Baumann}
\author[2]{Homa Hamedi}
\author[1]{Xuegeng Yang}
\author[1]{Gerd Mutschke}
\author[2]{Leon Urbas}
\author[3]{Andrés Fabián Lasagni}
\author[1,2,5]{Kerstin Eckert} 

\affil[1]{Institute of Fluid Dynamics, Helmholtz-Zentrum Dresden-Rossendorf, 01328 Dresden, Germany. E-mail: h.rox@hzdr.de; k.eckert@hzdr.de}
\affil[2]{Institute of Process Engineering and Environmental Technology, Technische Universität Dresden, 01062 Dresden, Germany.}
\affil[3]{Institute of Manufacturing Science and Engineering, TU Dresden, 01062 Dresden, Germany.}
\affil[4]{Academic Centre for Materials and Nanotechnology, AGH University of Krakow, 30-059 Krakow, Poland.}
\affil[5]{Hydrogen Lab, School of Engineering, Technische Universität Dresden, 01062 Dresden, Germany.}

\begin{document}
\date{}	
\maketitle
\footnotetext[1]{Corresponding author: h.rox@hzdr.de}
\footnotetext[2]{Electronic Supplementary Information (ESI) available.}
\footnotetext[3]{These authors contributed equally to this work.} 
\begin{abstract}
	Laser-structuring techniques like Direct Laser Interference Patterning show great potential for optimizing electrodes for water electrolysis. Therefore, a systematic experimental study based on statistical design of experiments is performed to analyze the influence of the spatial period and the aspect ratio between spatial period and structure depth on the electrode performance for pure Ni electrodes. The electrochemically active surface area could be increased by a factor of 12 compared to a non-structured electrode. For oxygen evolution reaction, a significantly lower onset potential and overpotential ($\approx$ \SI{-164}{\milli\volt} at \SI{100}{\milli\ampere\per\centi\metre\squared}) is found. This is explained by a lower number of active nucleation sites and, simultaneously, larger detached bubbles, resulting in reduced electrode blocking and thus, lower ohmic resistance. It is found that the spatial distance between the laser-structures is the decisive processing parameter for the improvement of the electrode performance.
\end{abstract}
\clearpage

\section*{Broader Context} 
Central to the efficiency of water electrolysis are gas-evolving electrodes, where the oxygen and hydrogen evolution reaction takes place. However, the  gas bubbles evolving cause considerable losses by blocking the electrode surface and increasing the electrolyte resistance and thus, increasing the ohmic losses. Therefore, optimizing the electrode material and surface to manage bubble growth and detachment is a promising approach for enhancing the overall efficiency and cost-effectiveness of water electrolysis. However, it is important that the approach chosen is industrially applicable. Consequently, the focus must be on the utilization of readily available materials, and straightforward and scalable manufacturing techniques. For this reason, nickel was chosen as material as it is widely used in alkaline electrolyzers. By using laser-structuring, a process suitable for industry, the electrode surface can be optimized for water electrolysis to increase the overall efficiency.

\section{Introduction} 
\label{sec:intro}
Green hydrogen produced by renewable energies using water electrolysis has become a central technology for the transition towards carbon-neutral industry.\cite{Spek2022, Staffell2019} Therefore, fossil energies have to be replaced by low or zero-carbon energy sources like solar- or wind-derived electricity to produce hydrogen and replace fossil fuels.\cite{Smolinka2021, Hermesmann2021} This is particularly necessary in end uses that are difficult to electrify, like heavy transport,\cite{Valente2021} maritime applications\cite{Hoecke2021} or high-temperature processes like steel\cite{Wang2021} or glass industry.\cite{FurszyferDelRio2022}

Besides Proton Exchange Membrane (PEM) and Solid Oxide Electrolyzers (SOE), the most mature technology is still alkaline water electrolysis (AWE).\cite{ElEmam2019} However, large-scale production of green hydrogen still lacks in terms of efficiency, and hence economic competiness.  Considerable losses are caused by the evolving hydrogen and oxygen bubbles by increasing ohmic resistances and blocking electrochemically active sites.\cite{He2023, Swiegers2021, Angulo2020, Zeng2010} In addition, the mass transfer and the actual current density is influenced by the electrode coverage.\cite{Vogt2017, Vogt2015, Vogt2011,Balzer2003, Eigeldinger2000} These effects can be reduced by applying external forces,\cite{Rocha2022, Li2021b,  Darband2019, Baczyzmalski2017, Koza2011, Iida2007, Balzer2003} enhancing bubble coalescence,\cite{Bashkatov2024, Lv2021} optimizing the electrode's electrocatalysts,\cite{Jiao2021} morphology\cite{Rocha2022, Yang2022a, Kou2020,  Fujimura2021, Darband2019} or surface,\cite{Andaveh2022, Darband2019} to achieve an optimized bubble nucleation, growth and detachment.

Knowledge of the forces acting on electrogenerated bubble is crucial here. In general, buoyancy,\cite{Zhang2012} hydrodynamical\cite{Baczyzmalski2017, Zhang2012} and Marangoni\cite{Bashkatov2022, Massing2019, Yang2018} interfacial tension,\cite{Meulenbroek2021,Darband2019, Zhang2012,Duhar2006} contact pressure\cite{Duhar2006} and electrical forces\cite{Haverkort2024,Bashkatov2022, Brandon1985a, Brandon1985b} act on an individual bubble. In this study, we focus on the interfacial tension force $F_\text{S}$, given by
\begin{align}
	F_\text{S} = -2 \pi r_\text{c}\gamma sin\theta, \label{eq:Fs}
\end{align}
which acts as an retarding force at the three-phase contact line.\cite{Bashkatov2022, Meulenbroek2021} Here, $\gamma$ denotes the surface tension. As shown in Eq. \ref{eq:Fs}, besides the contact radius $r_\text{c}$, the contact angle $\theta$ has to be taken into account. By changing the wetting behaviour of the surface and thus, increasing or decreasing $\theta$ as well as $r_\text{c}$, $F_\text{S}$ can be reduced. This results in earlier bubble detachment, thus releasing nucleation sites more quickly and finally increasing efficiency.\cite{Shi2021}

Electroless deposition,\cite{Kim2021a} electrodeposition,\cite{Krause2023, Skibinska2023, Ren2023} lithography,\cite{Akay2022, Brinkert2018} UV lithography\cite{Fujimura2021} and laser techniques\cite{Raenke2024, Raenke2022, Baumann2020, Koj2019, Rauscher2017, Neale2014}  have recently been reported to optimize the electrode surface. Laser texturing of surfaces offers several advantages, primarily due to its precise control over the size, shape, and distribution of surface features. Laser-based techniques also enable a reproducible process on an industrial scale,\cite{Lasagni2017} free from chemical reagents and with minimal waste.\cite{Wahab2016}

Previous studies have demonstrated the use of laser-based methods to create complex patterns on metallic surfaces, featuring repetitive or periodic arrangements. These structures have been employed to significantly increase electrode surface area, thereby enhancing electrochemical performance of gas evolution reactions.\cite{Raenke2024, Raenke2022, Baumann2020} Moreover, when ultrashort laser pulses are applied to metallic surfaces, self-organizing nano- and microstructures are formed, which further increase the electrode surface area and the number of possible nucleation sites. These features, known as Laser-Induced Periodic Surface Structures (LIPSS), arise when materials are irradiated with energy densities near the ablation threshold.\cite{Roemer2014, Sipe1983, Bonse2017}

For instance, Direct Laser Writing (DLW) has been used to structure nickel electrodes for applications in electrocatalysis and energy storage.\cite{Hoffmann2022, Cai2023} In this context, \citeauthor{Rauscher2017} employed a femtosecond pulsed laser to fabricate self-organized conical microstructures on nickel electrodes, achieving a \SI{45}{\%} improvement in Hydrogen Evolution Reaction (HER) efficiency.\cite{Rauscher2017}  However, when features with lateral sizes below a few microns are required, DLW faces limitations in throughput and resolution due to the diffraction limit. A promising alternative capable of overcoming these challenges is Direct Laser Interference Patterning (DLIP). This method, when combined with a high-power picosecond laser source, has been used to create periodic line-like patterns with spatial periods of 11 and \SI{25}{\micro\metre}, improving the efficiency of nickel electrodes for HER up to \SI{22}{\%}.\cite{Raenke2024} Additionally, \citeauthor{Raenke2024b} employed the DLIP technique in conjunction with a femtosecond pulsed laser to generate highly periodic pillar-like patterns with spatial periods of \SI{3}{\micro\metre}, increasing the Electrochemically Active Surface Area ($ECSA$) of nickel-based electrodes by almost 10 times and hence achieving a reduction of overpotential of HER by 49\%.\cite{Raenke2024b}

As stated above, most reported studies are dealing with HER. However, the exact reaction mechanism and ideal electrodes for the Oxygen Evolution Reaction (OER) are still being researched.\cite{Fabbri2018} In general, it can be expected that OER requires an higher overpotential to overcome of the kinetic barrier, since it is a four electron-proton coupled reaction compared to the two electron-transfer reaction of HER.\cite{Suen2017} 

Therefore, this study systematically investigates the influence of spatial period $\Lambda$ and structure depth of laser-structured surfaces produced by DLIP on the overall electrode performance during OER. For this purpose, these structures are applied to high-purity Ni as a standard material in alkaline electrolyzers. In addition to surface characterization, the electrochemical performance of the electrodes and the bubble dynamics in terms of detached bubble sizes and number of nucleation sites are analyzed. Using a statistical design of experiment (DoE) approach, models are developed to further optimize the laser structures.

\section{Material and methods}    
\label{sec:methods}
\subsection*{Design of Experiments}
In order to study the influence of the laser-structuring on the electrode performance, DoE was applied to cover the widest possible range of process parameters and determine the significant structuring parameters.\cite{lee_statistical_2019} The design considered three factors: the spatial period $\Lambda$, the aspect ratio $AR$ between $\Lambda$ and structure depth (as shown in Fig. \ref{fgr:methods_laser_structuring}), and the current density $j$. Applying galvanostatic measurement, the influence of these factors on several responses were studied: the quasi-steady state electrode potential $E_\text{SS}$, the bubble size (mode ($d_\text{m}$) and median ($d_{50}$) value of the bubble size distribution), and the mean number of active nucleation centers $\bar{n}_\text{nucl}$. Hereby, $E_\text{SS}$ was defined as the time-averaged potential $\bar{E}$ of the last \SI{20}{\second} of each measurement. Since, nonlinear effects were expected, a full-factorial design on three levels was chosen to study the relationships. The levels are summarized in Tab. \ref{tbl:methods_parameters}. A detailed measurement plan is shown in Tab. S1.$^\dag$  In order to estimate the measurement noise, the center point was measured independently three times. For better comparison a non-structured reference electrode was studied as well. Thus, $N = 3^3 + 2 + 3 = 32$ experiments were scheduled.

\begin{table}[ht]
	\centering
	\small
	\caption{\ Experimental parameters with highlighted center point}
	\label{tbl:methods_parameters}
	\begin{tabular*}{0.48\textwidth}{@{\extracolsep{\fill}}ll}
		\hline
		Parameter & Range \\
		\hline
		Spatial period $\Lambda$ (\si{\micro\metre}) & 6, \textbf{15}, 30\\
		Aspect ratio $AR$ (-) & 0.33, \textbf{0.67}, 1\\
		Current density $j$ (\si{\milli\ampere\per\centi\metre\squared}) & 10, \textbf{31.62}, 100 \\
		\hline
	\end{tabular*}
\end{table}

\subsection*{Electrode fabrication}
Ni-foils with a thickness of \SI{0.12}{\milli\metre} (GoodFellow, purity \SI{99.95}{\%}) were used as substrate for all electrodes. For better comparison, a non-structured sample (NSE) of the same substrate was cut into the same dimensions of \SI{10}{\milli\metre} $\times$ \SI{50}{\milli\metre}.

\begin{figure*}[h]
	\centering
	\includegraphics[width=\textwidth]{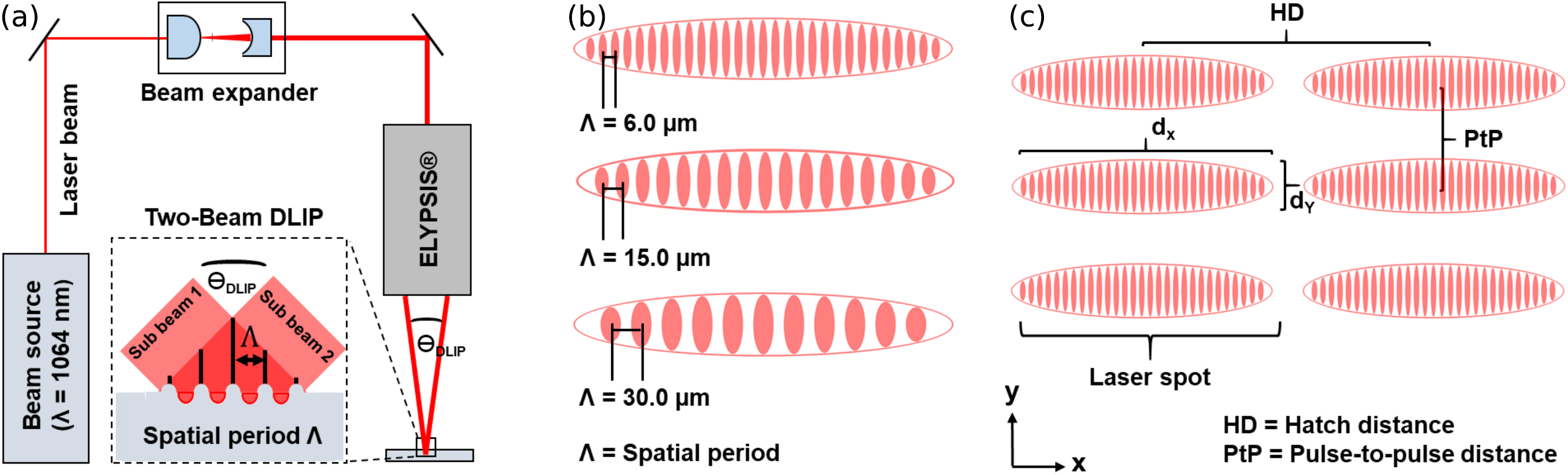}
	\caption{(a) Schematic drawing of the laser texturing showing the two-beam DLIP optical configuration combined with ELIPSYS\textregistered \,head. The inset denotes the two overlapping sub-beams producing a line like interference pattern. (b) Resulting interference profiles for spatial periods $\Lambda = 6$, 15 and \SI{30}{\micro\metre}. (c) Process strategy for structuring surfaces with elongated laser spots and corresponding structure parameter (spot dimensions $d_\text{x} = \SI{0.85}{\milli\metre}$, $d_\text{y} = \SI{0.08}{\milli\metre}$), pulse-to-pulse distance $PtP$ and hatch distance $HD$.}
	\label{fgr:methods_laser_structuring}
\end{figure*}

The laser texturing was performed by employing an optical configuration with two-beam interference optics. The experimental setup consists of a picosecond solid-state laser (Innoslab PX, EdgeWave, Germany) delivering laser pulses with a pulse duration $\tau$ of \SI{12}{\pico\second} and a maximal average laser power of \SI{60}{\watt}. The infrared beam ($\lambda = \SI{1064}{\nano\metre}$) emitted from the laser source is expanded using a two-lens telescope system and guided into the recently developed optical DLIP head (ELIPSYS\textregistered, SurFuntion GmbH, Germany)\cite{Lasagni2021} that utilizes a Diffractive Optical Element (DOEl) to split the incoming main beam into two sub-beams, which later are shaped to elongated lines (see Fig. \ref{fgr:methods_laser_structuring}a). The introduced optical head enables an impressive depth of focus of \mbox{$\approx$ \SI{10}{\milli\metre}} and generates an elliptically shaped laser spot with dimensions ($d_\text{y} \times d_\text{x}$) of \SI{0.08}{\milli\metre} $\times$ \SI{0.85}{\milli\metre} in the focal plane. Using this optical setup as shown in Fig. \ref{fgr:methods_laser_structuring}, considering interference angle $\theta_\text{DLIP}$ as well as the applied laser wavelength $\lambda$, the spatial period $\Lambda$ of the interference pattern can be calculated by 
\begin{align}
	\Lambda = \frac{\lambda}{2 \cdot \sin \frac{\theta_\text{DLIP}}{2}} \label{eq:spatial_period}
\end{align}

For this study the spatial period $\Lambda$ was changed by swapping the DOEl within the optical configuration. The beam splitting mechanism relies on the optical diffraction grating principle, resulting from a periodic structure atop the DOEl surface. Depending on the geometric characteristics of the optical grating, the interference angle $\theta_\text{DLIP}$ of the overlapping beams is modified. For the experimental process three distinct DOEls, each corresponding to spatial periods of 6, 15, and \SI{30}{\micro\metre}, were employed. The movement of the metallic substrates in  two orthogonal directions was realized with mechanical stages (Aerotech PRO155-05, USA). The texturing was consistently executed at a fixed repetition rate $f_\text{rep}$ of \SI{10}{\kilo\hertz}, using pulse-to-pulse distance $PtP$ of \SI{5}{\micro\metre} (see \mbox{Fig. \ref{fgr:methods_laser_structuring} (c)}). For the treatment of large areas, the hatch distance $HD$, which is the lateral distance between pulses (see Fig. \ref{fgr:methods_laser_structuring} (c)), was set to \SI{300}{\micro\metre} except for $\Lambda = \SI{6}{\micro\metre}$, where it was adjusted to \SI{360}{\micro\metre}. This increase in $HD$ for the smaller $\Lambda$ was necessary to prevent partial remelting of the sub-micro textures due to higher localized thermal loads.

Based on a preliminary study (see Sec. S2)$^\dag$ on the influence of the number of consecutive passes $N$ on the resulting structure morphology and Aspect Ratio $AR$, the fabrication of line shaped DLIP features with specific $AR$ of 0.33, 0.67 and 1.0 were conducted by adjusting the number of scans from 1 to 27, utilizing distinct single pulse fluences $\Phi_\text{sp}$ ranging from 0.27 to \SI{0.71}{\joule\per\centi\metre\squared}. $AR$ is defined to:
\begin{align}
	AR = \frac{Structure \, \, depth}{\Lambda}
\end{align}

\subsection*{Experimental methods}
\begin{figure*}[ht]
	\centering
	\includegraphics[width=\textwidth]{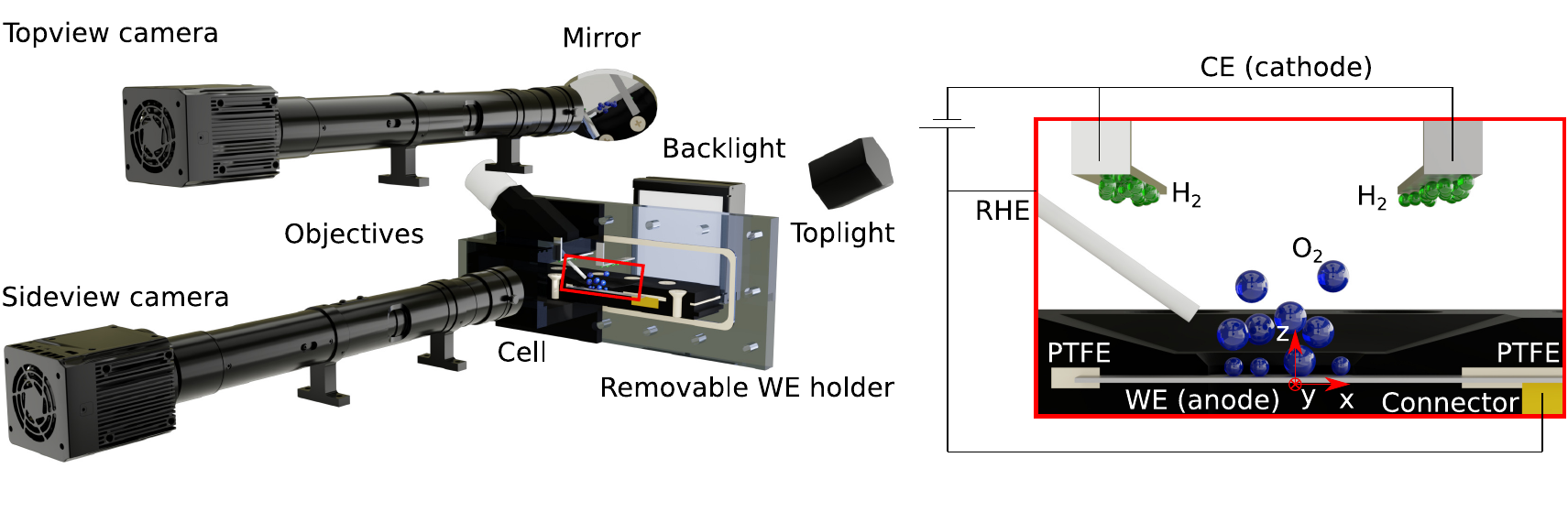}
	\caption{Schematic drawing of the membraneless cell and the optical measurement system used to perform electrochemical measurements and simultaneously study the O$_2$ bubble evolution.}
	\label{fgr:methods_experimental}
\end{figure*}

A membraneless cell out of PVC (total electrolyte volume \mbox{$V \approx \SI{60}{\milli\litre}$)} was used to perform all electrochemical experiments, as shown in Fig. \ref{fgr:methods_experimental}. The working electrode (WE) was mounted on an removable holder and pressed between two sheets of compressible PTFE (PTFE.EXS.100, High-tech-flon, Germany) by 10 M4 screws to ensure proper sealing of the active WE area. The open area of the holder was an elongated hole with a diameter of \SI{2}{\milli\metre} and a length of \SI{10}{\milli\metre} (see Fig. S2)$^\dag$, which corresponds to an accessible area of the WE of $\approx$ \SI{0.23}{\centi\metre\squared}. The counter electrode (CE) consists out of two pieces of Pt-foil (GoodFellow, purity 99.95 \%) with a total area of $A_\text{el} \approx \SI{3}{\centi\metre\squared}$ and was placed horizontally at the top of the cell. A reversible hydrogen reference electrode (Mini RHE, Gaskatel, Germany) was placed inside the cell with tip pointing towards the WE, as it is shown in the highlighted section in Fig. \ref{fgr:methods_experimental}. The design of the cell and the WE holder was an adapted version of the cell setup used in \citeauthor{Rox2023}.\cite{Rox2023} This cell was used for cyclic voltammetry (CV), linear sweep voltammetry (LSV) and galvanostatic measurements.

All experiments were carried out in a \SI{1}{M} KOH (Titripur, Merck, Germany) solution under ambient conditions $\left(T = \SI{293}{\kelvin}, p = \SI{1}{\bar}\right)$. Prior to the experiments, the electrolyte was purged with N$_2$ for \SI{20}{\minute} and pumped into the cell under a N$_2$ atmosphere by means of a OB1 MK3+ (Elveflow, France). All WEs were cleaned in a ultrasonic bath with Isopropanol for \SI{5}{\minute}, rinsed with deionized (DI) water and stored in DI water for at least \SI{48}{\hour} to ensure a superhydrophilic surface.\cite{Heinrich2024} This was proven using a contact angle measurement system (OCA 200, DataPhysics Instruments GmbH, Germany) by applying a water droplet with a volume of \SI{5}{\micro\liter} on the surface, which spreaded directly over the entire structured area.$^\dag$ For this purpose, the samples were dried with compressed air before the measurements. Before each measurement performed in the electrochemical cell, the WE was cleaned with ethanol and subsequently rinsed with DI water to remove any remaining contamination before mounting them onto the electrode holder.

\subsubsection*{Characterization of electrode surfaces.}
For evaluating the surface topography of the laser structured samples, White Light Interferometric (WLI) images (Sensofar S-Neox, Spain) were recorded by employing 50x magnification objective. The surface profiles and average structure depth values were obtained using the SensoMAP Advanced Analysis Software (Sensofar, Spain). In addition, high resolution images of the treated substrates were taken using Scanning Electron Microscopy (SEM) operating at an acceleration voltage of \SI{12}{\kilo\volt} (Quattro ESEM, Thermo Fischer Scientific, Germany).

X-ray photoelectron spectroscopy (XPS) measurements were performed to characterize the surface composition. Therefore, monochromatic Al K$\alpha$ (\SI{1486.6}{eV}) X-rays were focused to a \SI{100}{\micro\metre} spot using a PHI VersaProbeII Scanning XPS system (ULVAC-PHI). The photoelectron takeoff angle was \SI{45}{\degree} and the pass energy in the analyzer was set to \SI{117.50}{eV} (\SI{0.5}{eV} step) for survey scans and \SI{46.95}{eV} (\SI{0.1}{eV} step) to obtain high energy resolution spectra for the C 1s, O 1s, N 1s, P 2p, S 2p and Ni 2p regions. A dual beam charge compensation with \SI{7}{eV} Ar+ ions and \SI{1}{eV} electrons was used to maintain a constant sample surface potential regardless of the sample conductivity. All XPS spectra were charge referenced to the unfunctionalized, saturated carbon (C-C) C 1s peak at \SI{285.0}{eV}. The operating pressure in the analytical chamber was less than \SI{3e-9}{\milli\bar}. Deconvolution of spectra was carried out using PHI MultiPak software (v.9.9.3).$^\dag$ Spectrum background was subtracted using the Shirley method. Details about the deconvolution and fitting of the XPS spectra can be found in Sec. S8.$^\dag$ In addition, for individual electrodes detailed XPS measurements were performed to differentiate between bright and dark areas caused by $HD$ of the laser-structuring. Therefore, three bright (b1, b2, b3) and three dark (d1, d2, d3) areas where analyzed on each sample.

\subsubsection*{Characterization of the electrode performance.}
Prior to all electrochemical measurements, each electrode was activated by ensuring a constant open circuit potential (OCP) over \SI{5}{\minute} and afterwards running 200 cycle voltammograms (CV) at a scan rate of $\nu = \SI{500}{\milli\volt\per\second}$ from \SI{0.2}{\volt} to \SI{1}{\volt}  vs. RHE. These CVs were also used to specify the potential region where non-Faradaic currents occur to run series of CVs in a range of $\pm \SI{50}{\milli\volt}$ around a starting potential within this region. In total 14 different $\nu$ (0.02, 0.04, 0.06, 0.08, 0.1, 0.2, 0.3, 0.4, 0.5, 0.6, 0.7, 0.8, 0.9 and \SI{1}{\volt\per\second}) were applied and for each $\nu$ five cycles were performed with a $\SI{1}{\minute}$ break after each set. For the calculation of the double-layer capacitance $C_\text{dl}$ the last three measurements were taken. Linear sweep voltammetry (LSV) was used to calculate the onset potential $E_\text{on}$ of the OER. A potential range from \SI{0}{\volt} to \SI{2.5}{\volt} vs. RHE at a scan rate of \SI{100}{\milli\volt\per\second} was selected for this purpose. Finally, galvanostatic measurements at fixed current densities (10, 31.62 and \SI{100}{\milli\ampere\per\centi\metre\squared}) according to Tab. S1$^\dag$ were performed over a time of $t=\SI{1}{\minute}$. Therefore, the applied current was calculated for all electrodes by using the open area of the electrode holder of \SI{0.23142}{\centi\metre\squared}. For all electrochemical measurements a Modulab Xm with a Pstat 1MS/s module (Solartron analytical, Ametek, USA) was used as electrochemical workstation.

\subsubsection*{Characterization of the bubble evolution.}
During the galvanostatic measurements the bubble evolution and detachment were optically investigated from two perspectives (see Fig. \ref{fgr:methods_experimental}). Therefore, two high-speed cameras (1920 $\times$ \SI{760}{\pixel}, IDT OS-7 S3, USA) were used, each equipped with a precision micro-imaging lens with a magnification of 2 and a 12.5:1 zoom-module (Optem® FUSION, USA). This resulted in a resolution of \SI{867.38}{\pixel\per\milli\metre} and \SI{543.124}{\pixel\per\milli\metre} for sideview and topview, respectively.  The depth of field was determined to be \SI{335}{\micro\metre} using a calibration plate. A LED-panel (CCS TH2, Japan) as back illumination completed the shadowgraphy measurement system, while the top view was lit at an angle of $\approx \SI{60}{\degree}$ by a M-LED 3000 plus (ILO electronic, Germany). The field of view (FOV) for the sideview was adjusted right above the electrode cover, to prevent detached bubbles from dissolving, at a sample rate of \SI{250}{\hertz}. To capture the bubble growth the sample rate for topview images was set to \SI{1000}{\hertz} and the FOV was centered in the middle of the $xy$-plane of the WE surface. 

The grayscale images, taken with a 12-bit depth, were analyzed using Python 3.9. The image processing followed the  segmentation method introduced in \citeauthor{Rox2023}.\cite{Rox2023} Thus, a machine-learning (ML) based approach was chosen to segment the bubbles in both sideview and topview. Therefore, for each camera perspective a stardist model (v.0.8.5)\cite{Schmidt2018, Weigert2020} was trained with a randomly chosen set of 240 and 100 manual labelled images for sideview and topview, respectively. After segmentation of the detached bubbles in the sideview images, all objects were linked using trackpy\cite{Allan2023} and finally, blurred bubbles were eliminated by calculating the size-normalized variance of the image Laplacian (Var($\Delta$)$\cdot d_\text{B}$). Therefore, all segmented bubbles below the 50 \% quantile of this metric were excluded from further analysis, as they correspond to the most blurred bubbles.\cite{Rox2023, Krause2023} For the topview evaluation, the number of detected bubbles were taken as measure of the active nucleation centers. The stardist model was trained to differentiate between bubbles sitting on the electrode and already detached bubbles by means of the shadow of the inclined illumination. Examples for the image processing steps are shown in Fig. S3 and \mbox{Fig. S4.$^\dag$} All processed data including all relevant metadata is available at \mbox{\href{https://doi.org/10.14278/rodare.3064}{10.14278/rodare.3064}.$^\dag$}

\subsubsection*{Multiple Regression Analysis.}
For the multiple regression analysis, the factors ($\Lambda$, $AR$ and $j$) were transformed into the space between -1 and 1. Therefore, the transformation rules shown in Tab. \ref{tbl:factor_transformation_rules} were applied to convert the factors from the real experiment to values between -1 and 1. Afterwards, a response surface model with
\begin{align}
	y = &\beta_0 + \beta_1 \, \bar{\Lambda} + \beta_2 \, \bar{AR} + \beta_3 \, \bar{j} + \beta_{12} \, \bar{\Lambda} \, \bar{AR} + \beta_{13} \, \bar{\Lambda} \, \bar{j} \label{eq:respnse_surface_model} \\ 
	&+ \beta_{23} \, \bar{AR} \, \bar{j} + \beta_{11} \, \bar{\Lambda}^2 + \beta_{22} \, \bar{AR}^2 + \beta_{33} \, \bar{j}^2 \notag
\end{align}
was fitted for each response $y$ (see Eq. \ref{eq:respnse_surface_model}), where $y$ refers to $E_\text{SS}$, $d_\text{m}$, $d_{50}$ and $\bar{n}_\text{nucl}$. 
Then, the response surface model was reduced applying backward elimination discarding the least relevant factors (threshold: p-Value > 0.05). Thus, at the end a reduced model for each response was derived. 

\begin{table}[b]
	\small
	\centering
	\caption{\ Transformation rules for the multiple regression analysis to scale the factors from the real experiment to values between -1 and 1)}
	\begin{tabular*}{0.48\textwidth}{@{\extracolsep{\fill}}ll}
		\hline
		Parameter & Rule \\
		\hline \\
		Spatial period $\Lambda$ & $\bar{\Lambda} = \frac{1}{0.5108} \ln \left( \frac{\frac{\Lambda}{\mathrm{\mu m}} + 7.5}{22.5} \right)$ \\ \\
		Aspect ratio $AR$ & $\bar{AR} = \frac{2 \left(AR - \frac{1}{3} \right)}{1 - \frac{1}{3}} - 1$ \\ \\
		Current density$j$ & $\bar{j} = \frac{1}{1.151} \cdot \ln \left( \frac{\frac{j}{\mathrm{mA / cm^2}}}{31.62} \right)$\\ \\
		\hline
	\end{tabular*}
	\label{tbl:factor_transformation_rules}
\end{table}

Due to high number of scans needed for the electrode with $\Lambda = \SI{30}{\micro\metre}$ and $AR = 1.0$ and thus, excessive fluence $\Phi$ for the given thickness of the substrate, this structure could not be produced reproducibly.$^\dag$ Therefore, all measurement points with \mbox{$\Lambda$ = \SI{30}{\micro\metre}} and $AR = 1.0$ were neglected in this study and $N$ was reduced by $3$. Thus, all developed models can only provide a first approximation since this corner point of the full-factorial design is missing.

\begin{figure*}[ht]
	\centering
	\includegraphics[width=\textwidth]{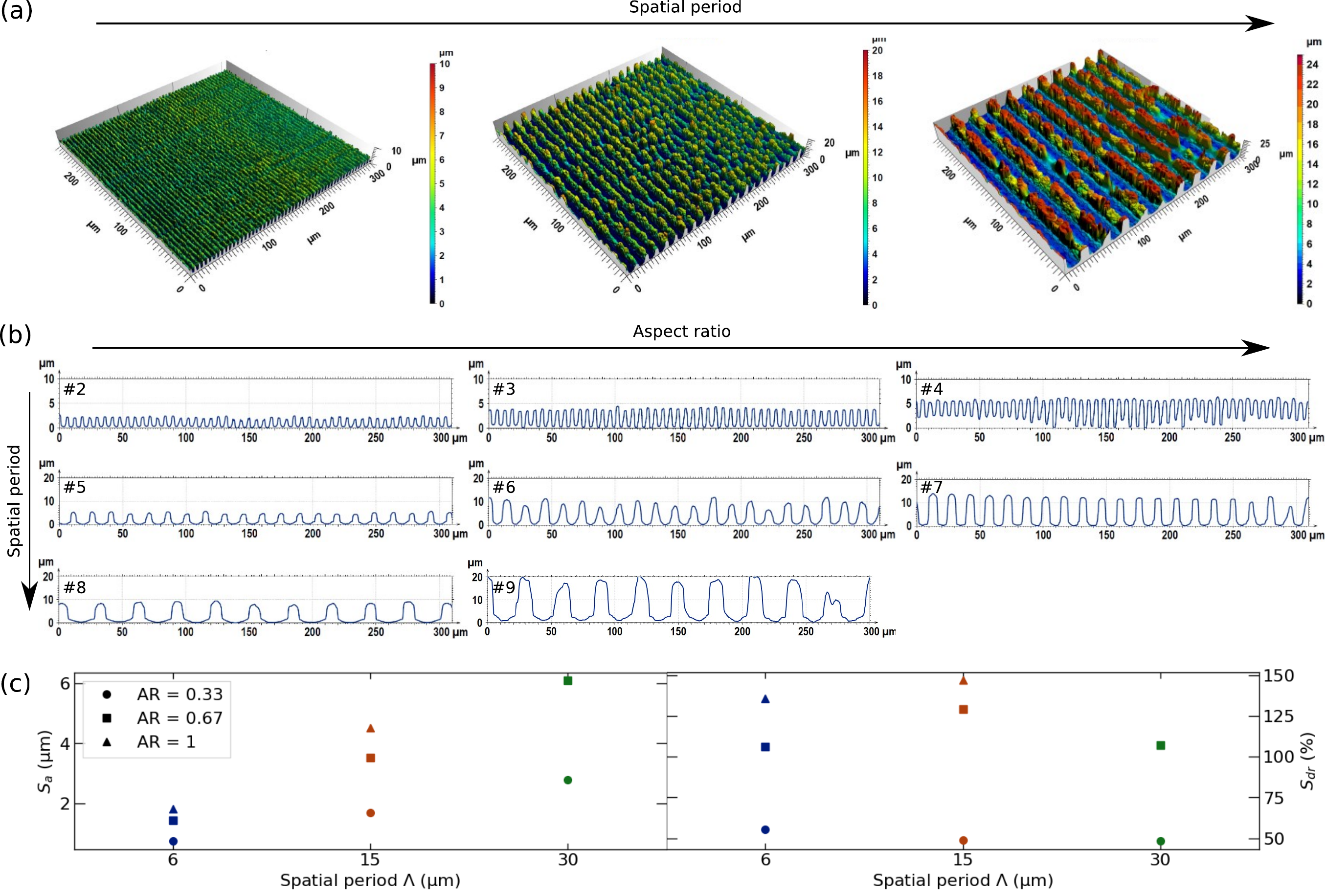}
	\caption{(a) 3D confocal images of DLIP line-like structures with $\Lambda = 6, 15$ and \SI{30}{\micro\metre} at $AR = 0.67$. (b) Height profiles of structured Ni-surfaces for all studied electrodes ($\Lambda = 6, 15$ and \SI{30}{\micro\metre} and $AR = 0.33, 0.67$ and $1.00$). The electrode ID (see Tab. \ref{tbl:electrode_metrics}) is provided in the upper left corner.  (c) Average surface roughness $S_\text{a}$ and developed interfacial area ratio $S_\text{dr}$ as a function of $\Lambda$ and $AR$.}
	\label{fgr:results_laser_structure}
\end{figure*}

\section{Results and discussion}
\label{sec:results}

\subsection*{Characterization of the electrode structure}
As reported in \citeauthor{Heinrich2024}, by storing the DLIP-structures in DI water the adsorption of organic compounds is limited and thus, the wettability is increased.\cite{Heinrich2024} The hydrophilicity of the DLIP-structures and their surface chemistry\cite{Zhu2019} is supported by the capillary effects of the channels. In comparison to the the non-structured electrode (NSE), with a measured static water contact angle (WCA) of $\theta_\text{NSE} = \SI{38.5}{\degree} \pm \SI{2.6}{\degree}$, WCA measurements were not possible for the superhydrophilic DLIP-structures, since the applied droplets spread directly across the entire electrode surface, as it is shown in Fig. S7.$^\dag$ Thus, for all DLIP-structures it can be stated that $\theta \ll \theta_\text{NSE}$.

Exemplary confocal microscopic images of the DLIP line-like features with an $AR = 0.67$ are shown in Fig. \ref{fgr:results_laser_structure} (a) revealing that with increasing spatial period $\Lambda$ the structure regularity is decreasing. In this context, the number of applied scans $N$ is the decisive factor, which, as indicated in Fig. \ref{fgr:results_laser_structure} (a), was significantly higher for larger structure periods. Generally, the structure formation process was characterized not only by the ablation of the nickel substrate, but also by the redeposition of removed material from the ablation plume.\cite{Schaefer2024} The amount of redeposited material grows continuously with the increasing number of over scans and therefore had a stronger influence on the regularity of the line-like pattern with $\Lambda$ = 15 and \SI{30}{\micro\metre}. Furthermore, the redeposition occurred predominantly in the areas of interference minima and did not happen uniformly over the whole surface, leading to the formation of a more irregular DLIP texture. This could also be concluded from the plotted height profiles in Fig. \ref{fgr:results_laser_structure} (b), in which, e.g., electrode \#9 shows a cut-off peak in the structure.

However, as expected the average surface roughness $S_\text{a}$ followed a linear trend with $\Lambda$ (see Fig. \ref{fgr:results_laser_structure} (c)). In contrast, the developed interfacial area ratio $S_\text{dr}$, as a measure of the additional surface area created by the texture compared to the ideal flat substrate, showed a maximum at $\Lambda = \SI{15}{\micro\metre}$ with an enlargement of $\approx$ 150 \%. The complete confocal images of all DLIP-structures can be found in Fig. S5.$^\dag$

\begin{figure*}[ht]
	\centering
	\includegraphics[width=\textwidth]{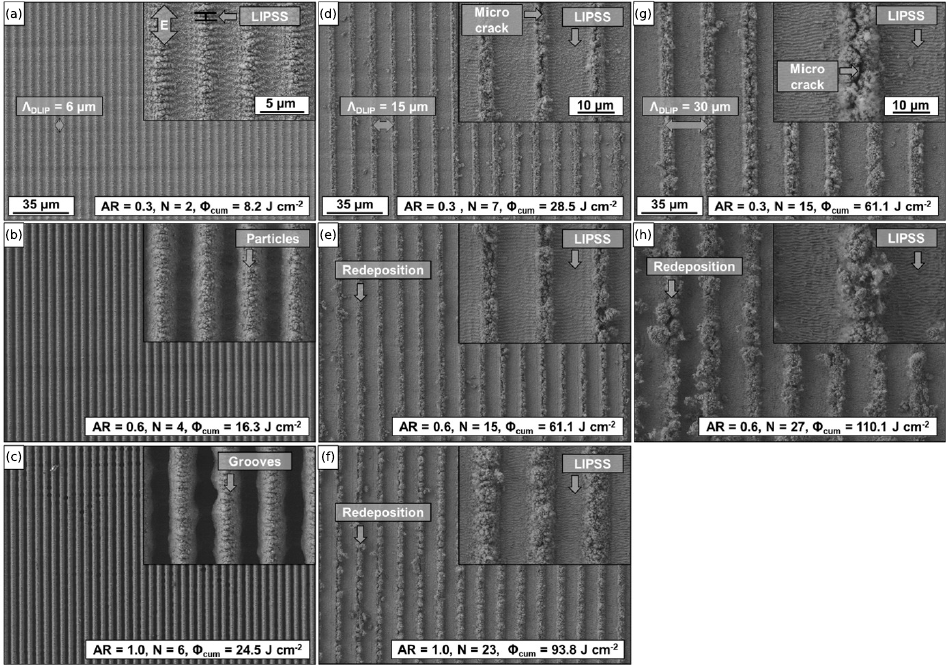}
	\caption{SEM images of DLIP line-like with a spatial period $\Lambda$ of \SI{6}{\micro\metre} (a,b,c), \SI{15}{\micro\metre} (d,e,f) and \SI{30}{\micro\metre} (g,h) fabricated on nickel foils with a single pulse fluence ($\Phi_\text{sp}$) of \SI{0.25}{\joule\per\centi\metre\squared} and a pulse-to-pulse distance ($PtP$) of \SI{5}{\micro\metre}. The generated aspect ratio $AR$, the total number of applied passes $N$ and cumulated fluence $\Phi_\text{cum}$ are displayed in the corresponding labels. The scale bars in the first row are representative of all columns.}
	\label{fgr:results_sem}
\end{figure*}

For a more detailed analysis of the surface topographies generated, SEM images were recorded. The line-like DLIP patterns for different $AR$ are shown in Fig. \ref{fgr:results_sem}. The patterns displayed in \mbox{Fig. \ref{fgr:results_sem} (a-c)} depict DLIP structures fabricated with $\Lambda = \SI{6}{\micro\metre}$, whereas those in (d-f) exhibit $\Lambda = \SI{15}{\micro\metre}$ and (g-h) present \mbox{$\Lambda = \SI{30}{\micro\metre}$.} The resulting cumulated fluence $\Phi_\text{cum}$, the number of scans $N$, as well as $AR$ are given in the labels for each display.

For the samples exhibiting an aspect ratio $AR = 0.33$ the appearance of a homogeneous line-like DLIP pattern decorated with a sub-structure could be observed for all spatial periods. Upon inspecting the samples structured with $\Lambda = \SI{6}{\micro\metre}$ (Fig. \ref{fgr:results_sem} (a)), \SI{15}{\micro\metre}  (Fig. \ref{fgr:results_sem} (d)) and \SI{30}{\micro\metre} (Fig. \ref{fgr:results_sem} (g)) at a higher magnification (see insets), the presence of a wavy texture could be observed. For the line texture with $\Lambda = \SI{6}{\micro\metre}$, all areas of the microstructure shown in Fig. \ref{fgr:results_sem} (a) were completely covered with the wavy sub-textures. In contrast, for the larger structure periods, ripple textures only occurred in the regions corresponding to the interference maxima positions. However, these ripples were oriented perpendicular to the polarization of the applied laser radiation (double arrow E in Fig. \ref{fgr:results_sem} (a)) and cross the generated DLIP lines pattern at an angle of \SI{90}{\degree}. The measured periodicity  of the ripples $\Lambda_\text{LIPSS}$ ranged from \SI{740}{\nano\metre} to \SI{930}{\nano\metre}, which corresponds to \SI{70}{\%} - \SI{87}{\%} of the used laser wavelength ($\lambda = \SI{1064}{\nano\metre}$). These characteristics suggested that the sub-structure can be regarded as LIPSS and further classified as Low-Spatial Frequency LIPSS (LSFL), according to previous studies.\cite{Bonse2012, Bonse2020, Aguilar2018} Furthermore, the linear textures with $\Lambda$  equal to 15 and \SI{30}{\micro\metre}, displayed the redeposition of ablated material in the region of minima interference. Apart from this, the formation of microcracks along the structural peaks of DLIP structure became visible for both cases (see Fig.  \ref{fgr:results_sem} (d) and (g)).

An increase in $AR$ to 0.67 led to the homogenization of the DLIP patterns for the \SI{6}{\micro\metre} period, causing the previously visible LSFL substructures to disappear. The redeposition of material in form of nano- and micro particles from the ablation plume was also observed on the structural peaks (Fig.  \ref{fgr:results_sem} (b)).

The resulting DLIP line patterns with an $AR$ of 0.67, which are shown in Fig. \ref{fgr:results_sem} (e) and (h), were characterized by the formation of a partially irregular DLIP texture, with increased redeposition of material in the areas of the interference minima. The degree of redeposition was directly dependent on the number of consecutive scans $N$. For instance, processing with a cumulative fluence $\Phi_\text{cum}$ of \SI{110.1}{\joule\per\centi\metre\squared}, resulting from 27 consecutive passes in Fig. \ref{fgr:results_sem} (h), led to the continuous growth of redeposition clusters, which started to partially shield the areas of the interference maxima. In general, the homogeneity of the surface structures with an $AR$ of 0.67 was observed to steadily decrease with increasing spatial period. Upon closer inspection of the magnified SEM sections for the \SI{15}{\micro\metre} and \SI{30}{\micro\metre} periods, it was apparent that LIPSS textures were still present in the areas of maxima interference.

The deepest line structures were fabricated for $AR = 1$. The SEM images of the resulting line pattern with $\Lambda = \SI{6}{\micro\metre}$ continued to show a high degree of homogeneity. Though it became evident that 6 successive passes had influenced the geometric shape of the individual DLIP features. As a result, the borders between interference maxima and minima developed a wavy characteristic and partially even hole-like structures were formed (Fig. \ref{fgr:results_sem} (c) and S6 (b)).\cite{Liu2019}$^{,\dag}$ For $\Lambda = \SI{15}{\micro\metre}$ in Fig. \ref{fgr:results_sem} (f), 23 over scans resulted in a higher amount of redeposited material, leading to an enlargement of line-like DLIP features.

The XPS results in Fig. S10 and Tab. S2$^\dag$ show the relative, chemical composition of the electrode samples. The chemical states and surface concentrations of C, O, and Ni were deconvoluted by fitting the XPS spectra. No clear tendency of the influence of $\Lambda$ and $AR$ was found for the surface composition. With the exception of the group of aliphatic carbon \ce{C-C} at a binding energy of \SI{285}{eV} only little differences were found and thus, the further discussion is focused on electrochemical properties.

\subsection*{Electrochemical characterization of the electrodes}
\begin{figure}[t]
	\centering
	\includegraphics[width=0.48\textwidth]{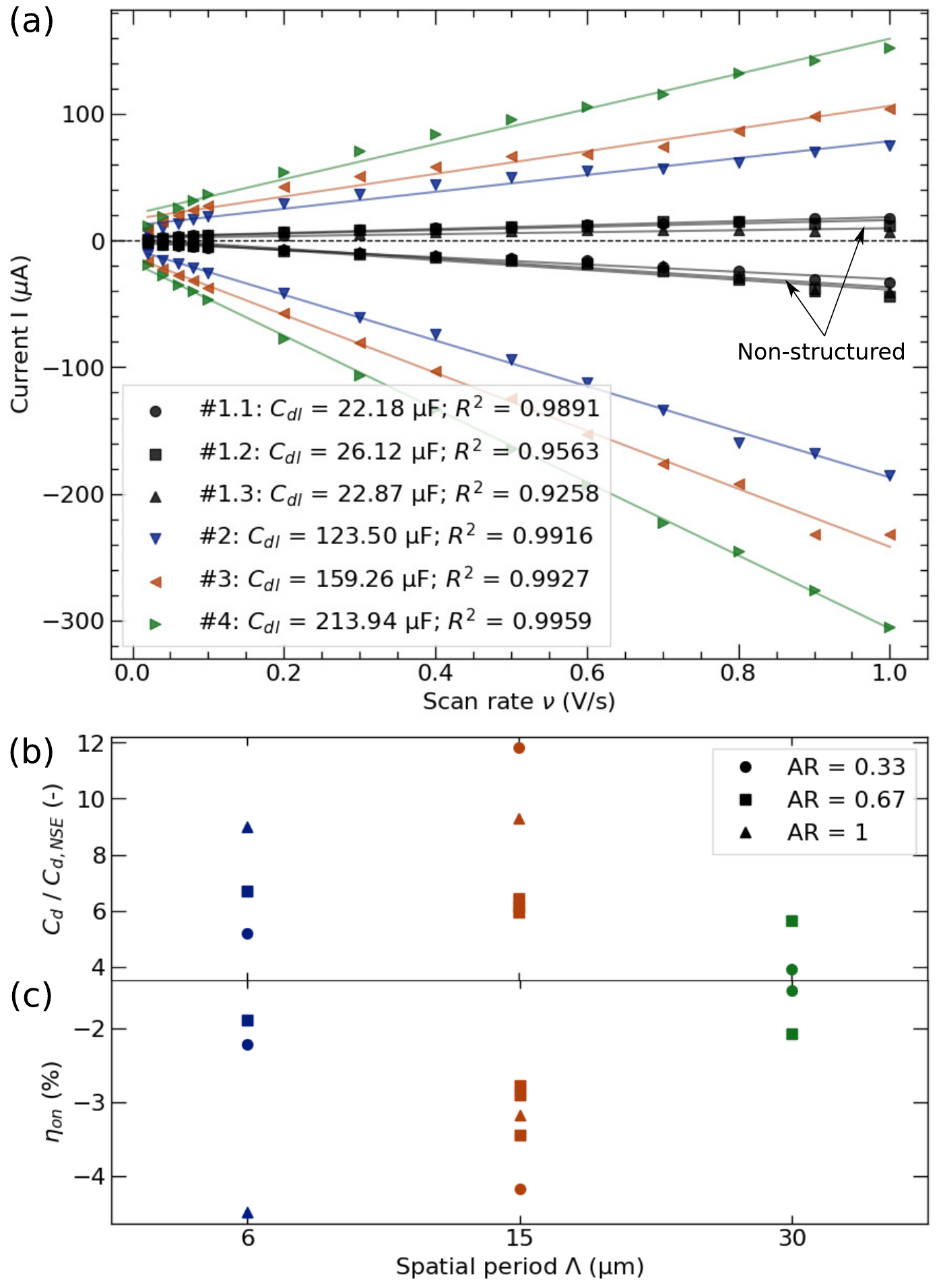}
	\caption{(a) Linear fit of the average anodic and cathodic currents $I$ over $\nu$ measured during CV and calculated $C_\text{dl}$ equal to the slope of the fit for a period length of \SI{6}{\micro\metre} at different aspect ratios. (b) Normalized $C_\text{dl}$ and (c) onset potential $\eta_\text{on}$ (see Eq. \ref{eq:efficiency_onset}) for all studied electrodes.}
	\label{fgr:results_ec_characterization}
\end{figure}

For further characterization of the DLIP-structures, $C_\text{dl}$ as a measure of $ECSA$ was analyzed. $C_\text{dl}$ was calculated as the slope of the linear fits, shown in Fig. \ref{fgr:results_ec_characterization} (a) using
\begin{align}
	C_\text{dl} = \frac{\bar{I}_\text{anodic} + \left|\bar{I}_\text{cathodic}\right|}{2 \cdot \nu}.
\end{align}
By running multiple CVs at different $\nu$ (see Fig. S8)$^\dag$, the average currents $\bar{I}$ of the last three cycles could be plotted against $\nu$, revealing the linear correlation. In addition to the definition of an unique electrode ID for further discussion of the results, the obtained electrode performance metrics can be found in Tab. \ref{tbl:electrode_metrics}.

\begin{table}[ht]
	\centering
	\small
	\caption{\ Average surface roughness $S_\text{a}$ (CM), double-layer capacitance $C_\text{dl}$ (CV) and onset potential $E_\text{on}$ (LSV) of all electrodes \textendash\ Standard deviation was calculated for middle and reference point of DoE}
	\label{tbl:electrode_metrics}
	\begin{tabular*}{0.6\textwidth}{@{\extracolsep{\fill}}llllll}
		\hline
		\multirow{2}{*}{ID} & $\Lambda$ & AR & $S_\text{a}$ & $C_\text{dl}$ & $E_\text{on}$ \\
		& (\si{\micro\metre}) &  (-) & (\si{\micro\metre}) & (\si{\micro\farad}) & (\si{\volt})\\
		\hline
		\#1 & \multicolumn{2}{c}{NSE} & & 23.73 $\pm$ 1.72 & 1.743 $\pm$ 0.021\\
		\#2 & 6 & 0.33 & 0.749 & 123.50 & 1.704\\
		\#3 & 6 & 0.67 & 1.43 & 159.26 & 1.710\\
		\#4 & 6 & 1 & 1.83 & 213.94 & 1.664\\
		\#5 & 15 & 0.33 & 1.70 & 280.48 & 1.670\\
		\#6 & 15 & 0.67 & 3.53 & 146.61 $\pm$ 4.70 & 1.690 $\pm$ 0.005\\
		\#7 & 15 & 1 & 4.53 & 220.74 & 1.687\\
		\#8 & 30 & 0.33 & 2.80 & 92.76 & 1.717\\ 
		\#9 & 30 & 0.67 & 6.10 & 133.64 & 1.707\\
		\hline
	\end{tabular*}
\end{table}

The shown increase of $C_\text{dl}$ can be attributed to the increase of the developed interfacial area ratio $S_\text{dr}$, since the normalized $C_\text{dl}$ in Fig. \ref{fgr:results_ec_characterization} (b) follows a similar trend like $S_\text{dr}$ in Fig. \ref{fgr:results_laser_structure} (c) with a maximum at $\Lambda = \SI{15}{\micro\metre}$. In addition, when electrode \#5 was neglected, the tendency was found that an increasing $AR$ leads to an higher $C_\text{dl}$. This also clearly follows the geometric surface enlargement in terms of $S_\text{a}$ and $S_\text{dr}$. However, at low $AR$ and $\Lambda = \SI{15}{\micro\metre}$ or \SI{30}{\micro\metre}, LSFL were detected in the SEM images, as shown in Fig. \ref{fgr:results_sem}, which contributed to the increased $C_\text{dl}$. In general, it was found that the applied DLIP structuring results in a significant, up to $\approx 12\times$ increase of $C_\text{dl}$, which exceeds the increase of $C_\text{dl}$ achieved by laser-structuring in \citeauthor{Bernaecker2022} and \citeauthor{Baumann2020}. However, \citeauthor{Bernaecker2022} used short pulse laser-structuring with non-regular laser structures\cite{Bernaecker2022} and \citeauthor{Baumann2020} used DLIP structuring at a lower spatial period of $\Lambda = \SI{5.8}{\micro\metre}$.\cite{Baumann2020}

This improvement was also evident in the measured onset potential $E_\text{on}$. For this purpose, the intersection of the non-Faradaic and Faradaic current regions of the recorded LSV curves was defined as $E_\text{on}$, as shown in Fig. S9$^\dag$. For better comparability, the efficiency $\eta_\text{on}$ was defined as
\begin{align}
	\eta_\text{on} &= \frac{E_\text{on} - E_\text{on, NSE}}{E_\text{on, NSE}}. \label{eq:efficiency_onset}
\end{align}
As a result, it was possible to deduce that the DLIP-structures lead to a reduction of $E_\text{on}$ up to $\approx$ 4.5 \%. 

In addition, it was found that the laser-structuring has good reproducibility, as can be seen from the data points at $\Lambda = \SI{15}{\micro\metre}$ and $AR = 0.67$ in Fig. \ref{fgr:results_ec_characterization} (b) and (c) or the calculated standard deviations $\sigma$ for the middle point of the DoE in Tab. \ref{tbl:electrode_metrics} and \mbox{Tab. S3.$^\dag$}

For further characterization of the electrode performance, galvanostatic measurements were run over a time of $t = \SI{60}{\second}$ at three different current densities of $j = 10,\, 31.62$ and \SI{100}{\milli\ampere\per\centi\metre\squared}. According to Faraday's law 
\begin{align}
	I = zFN
\end{align}
and assuming equal electrical contact resistances for all measurements, the molar flux of produced \ce{O2} $\dot{N}_\text{\ce{O2}}$ for all measurements at the same current density is constant:
\begin{align}
	\dot{N}_\text{\ce{O2}} (I) = const.
\end{align}
Since the CE had an area $\approx 12 \times$ larger than the WE, it was ensured that the HER is not limiting the OER at the WE. Thus, a change in the measured potential $E$ of the WE could be directly linked to the anodic overpotential $\eta_\text{anode}$ and the ohmic overpotential losses $\eta_\Omega$. As a result, a lower measured $E$ leads to a decrease of the cell potential $E_\text{cell}$ as an important measure of the overall efficiency of water electrolysis,\cite{Pang2020} which is defined as
\begin{align}
	E_\text{cell} = \left|\Delta E^0\right|+ \eta_\text{anode} + \left|\eta_\text{cathode}\right| + \eta_\Omega + \eta_\text{conc}.
\end{align}

As all measurements were carried out in \SI{1}{M} KOH, the concentration overpotential $\eta_\text{conc}$ is negligible.\cite{Pang2020}

Similar to Eq. \ref{eq:efficiency_onset}, the efficiency $\eta_\text{SS}$ was defined for better comparability of the quasi-steady state potential $E_\text{SS}$:
\begin{align}
	\eta_\text{SS}(j) &= \frac{E_\text{SS}(j) - E_\text{SS, NSE}(j)}{E_\text{SS, NSE}(j)} \label{eq:efficiency_ess}
\end{align}

\begin{figure}[t]
	\centering
	\includegraphics[width=0.48\textwidth]{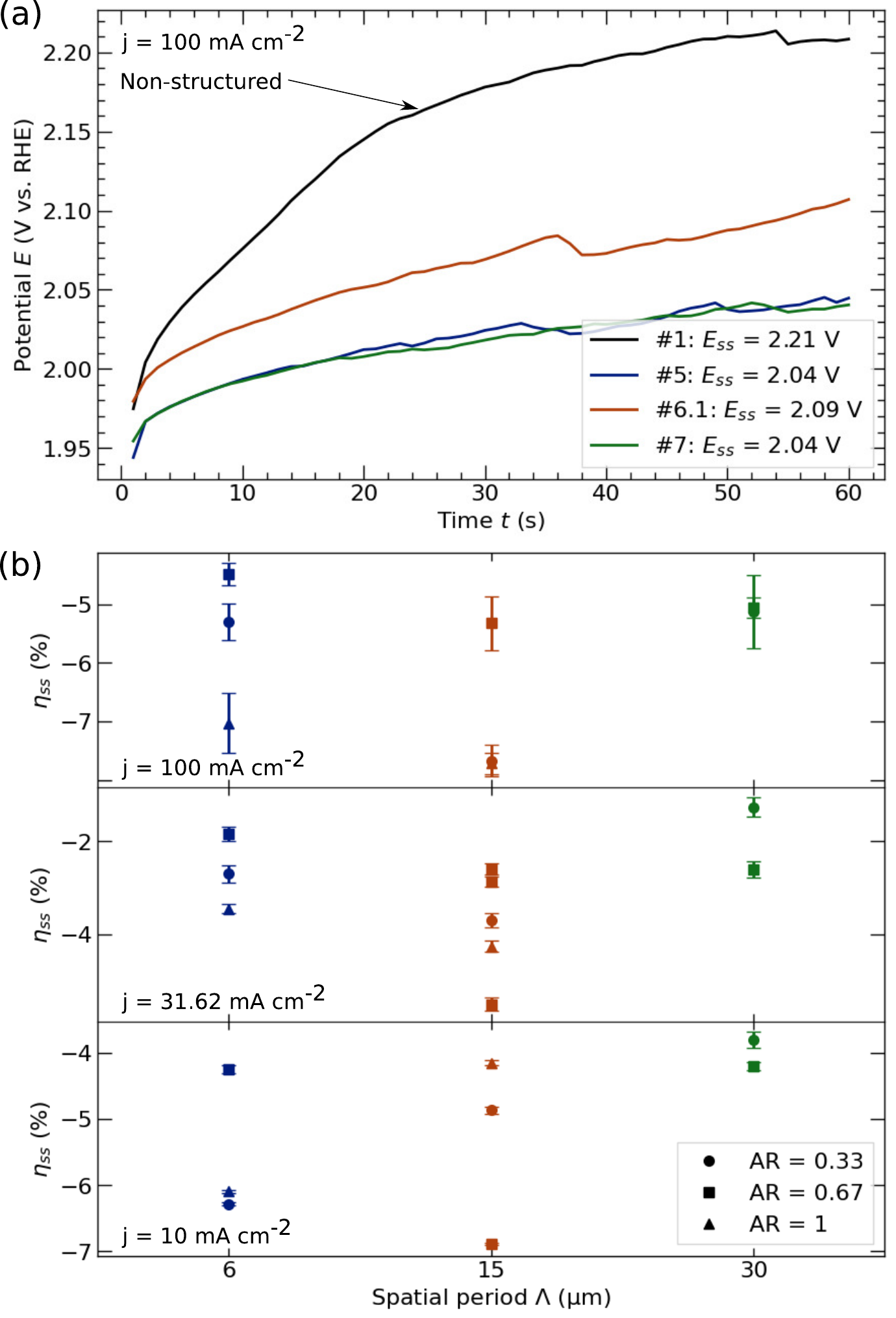}
	\caption{(a) Potential during \SI{60}{\second} galvanostatic measurements for a period length of \SI{6}{\micro\metre} at different aspect ratios. (b) Normalized WE overpotential $\eta_\text{SS}$ (see Eq. \ref{eq:efficiency_ess}) calculated from the quasi-steady state electrode potential $E_\text{SS}$ at $j = 10, 31.62$ and \SI{100}{\milli\ampere\per\centi\metre\squared}.}
	\label{fgr:results_galvanostatic}
\end{figure}

As shown in Fig. \ref{fgr:results_galvanostatic} $E_\text{SS}$ was significantly lower for all studied DLIP-structures. The best-performing electrode \#7 achieved a $\approx$ \SI{164}{\milli\volt} lower $E_\text{SS}$ at $j = \SI{100}{\milli\ampere\per\centi\metre\squared}$ with electrode \#5 performing similarly well, as can be seen in Fig. \ref{fgr:results_galvanostatic} (a). In addition, the calculated measurement noise of $\sigma \approx \SI{0.024}{\volt}$$^\dag$ served to indicate excellent reproducibility of the DLIP-structuring. It should be emphasized that in Fig. \ref{fgr:results_galvanostatic} (b) $\eta_\text{SS}$ follows a similar trend with a maximum at $\Lambda = \SI{15}{\micro\metre}$ like the normalized $C_\text{dl}$ in Fig. \ref{fgr:results_ec_characterization} (b). Since no clear influence on $\eta_\text{SS}$ can be identified for $AR$, the hypothesis can be made that $\Lambda$ is the decisive factor for DLIP-structuring.

This was also confirmed on the basis of the multiple regression analysis.$^\dag$ The obtained model ($R^2 = 0.984$) showed a non-surprising significant influence of $j$. More interesting was the significance of the influence of $\Lambda$ on $E_\text{SS}$ which is confirmed by the significant (thus not eliminated) linear term $\bar{\Lambda}$ and quadratic term for $\bar{\Lambda}^2$ in Eq. \ref{eq:model_potential}. 
\begin{align}
	E_\text{SS} = 1.7927 + 0.0074  \, \bar{\Lambda} + 0.191 \, \bar{j} + 0.0221 \, \bar{\Lambda}^2 + 0.0672 \, \bar{j}^2 \label{eq:model_potential} 
\end{align}
The shown improvement of $\eta_\text{SS}$ of $\approx \SI{-7.7}{\%}$ can be explained by the smaller effective current density $j_\text{eff}$ due to increased $C_\text{dl}$ and following, $ECSA$. At $\dot{N}_\text{\ce{O2}} (I) = const.$ this leads to decreased $\eta_\Omega$, since more electrode surface is available for the electrode-electrolyte interface. 

\subsection*{Detached bubble sizes}

\begin{figure*}[ht]
	\centering
	\includegraphics[width=\textwidth]{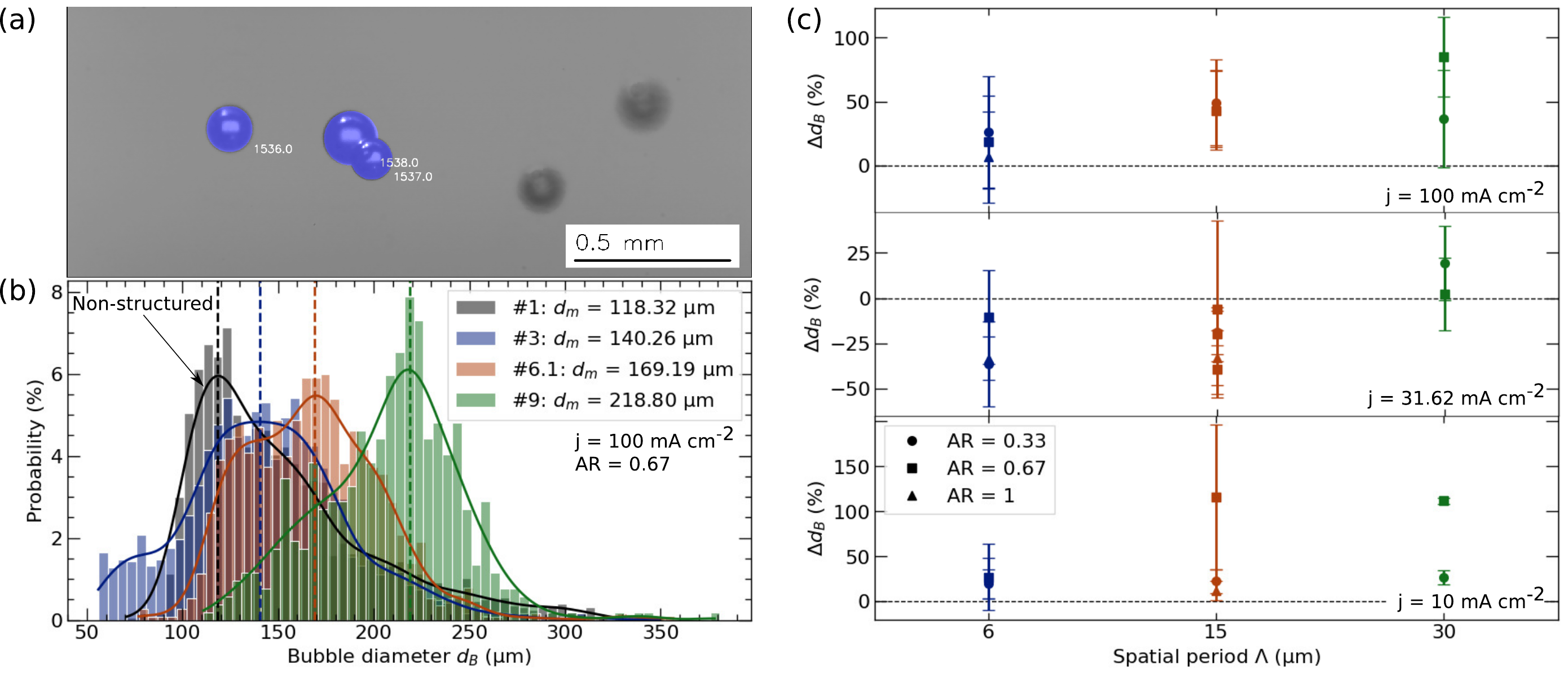}
	\caption{(a) Example sideview image with rising O$_2$-bubbles. Those which are inside the focal plane are highlighted by blue color. The numbers represent the unique bubble IDs. (b) Detached bubble size distribution of all periods ($\Lambda(\text{\#3}) = \SI{6}{\micro\metre}$, $\Lambda(\text{\#6.1}) = \SI{15}{\micro\metre}$ and $\Lambda(\text{\#9}) = \SI{30}{\micro\metre}$) at a constant $AR = 0.67$ in comparison to NSE (\#1) and with mode value $d_\text{m}$ of each bubble size distribution indicated by dash-dotted line. (c) Normalized change of $d_\text{m}$ (see Eq. \ref{eq:delta_bubble}) at $j = 10,\, 31.62$ and \SI{100}{\milli\ampere\per\centi\metre\squared}.}
	\label{fgr:results_bubblesize}
\end{figure*}

For a better understanding of the described improvement of the electrochemical performance of the DLIP-structures, high-speed images of the bubbles were taken. An example of these images with segmented bubbles using the ML-based image analysis is shown in Fig. \ref{fgr:results_bubblesize} (a). It should be pointed out that only sharp bubbles inside the focal plane were included in the further evaluation. As the critical KOH concentration for bubble coalescence of \SI{0.053}{M}\cite{Craig1993} is clearly exceed with $c_\text{KOH} = \SI{1}{M}$ used, bubble coalescence is suppressed. In addition, as the FOV was placed right above the electrode cover, it could be assumed that the bubble diameters $d_\text{B}$ in Fig. \ref{fgr:results_bubblesize} (b) correspond to $d_\text{B}$ at detachment of the electrode surface. With the measured noise of $\sigma = \SI{17.08}{\micro\metre}$ for $d_\text{m}$ and $\sigma = \SI{19.32}{\micro\metre}$ for $d_{50}$ (see Tab. S3)$^\dag$, the DLIP-structuring also showed good reproducibility in terms of bubble development.

A normalized metric was again defined for the discussion of the detached bubble sizes using the mode value of the bubble size distribution $d_\text{m}$ to
\begin{align}
	\Delta d_\text{B}(j) = \frac{d_\text{m}(j) - d_\text{m, NSE}(j)}{ d_\text{m, NSE}(j)}. \label{eq:delta_bubble}
\end{align}
Due to few active nucleation sites at $j = \SI{10}{\milli\ampere\per\centi\metre\squared}$, high spatial-resolution of the camera (\SI{867.38}{\pixel\per\milli\metre}) and therefore few to no detected bubbles in the FOV, the evaluation was focused on higher $j$. In addition, as at $j = \SI{100}{\milli\ampere\per\centi\metre\squared}$ the lowest $\eta_\text{SS}$ was calculated (see \mbox{Fig. \ref{fgr:results_galvanostatic} (b)),} higher $j$ were more relevant for the discussion of the improved electrode performance. 

In Fig. \ref{fgr:results_bubblesize} (c), it can be seen that at $j = \SI{100}{\milli\ampere\per\centi\metre\squared}$ $d_\text{m}$ was larger for all DLIP-structures compared to NSE. Moreover, increasing $\Lambda$ led to increased $d_\text{m}$, with a maximum increase of $\Delta d_\text{B} \approx 80 \%$ at $\Lambda = \SI{30}{\micro\metre}$. However, the influence of $AR$ on $d_\text{B}$ had no clear tendency. At $j = \SI{100}{\milli\ampere\per\centi\metre\squared}$, a clear influence of $AR$ can only be observed for $\Lambda = \SI{30}{\micro\metre}$, assuming larger $d_\text{B}$ with increasing $AR$. However, the exact opposite is seen with a decreased current density of $j = \SI{31.62}{\milli\ampere\per\centi\metre\squared}$. The ambiguity is similarly evident in the obtained models for $d_{m}$ ($R^2 = 0.812$) in Eq. \ref{eq:model_bubble_mode} and $d_{50}$ ($R^2 = 0.777$) in Eq. \ref{eq:model_bubble_median}:
\begin{align}
	d_{m} = &113.17 + 18.745 \bar{\Lambda} - 2.3123 \bar{AR} + 35.806 \, \bar{j} - 17.603 \bar{AR}^2 + 28.066 \, \bar{j}^2 \label{eq:model_bubble_mode} \\
	d_{50} = &106.18 + 18.361 \bar{\Lambda} + 39.635 \, \bar{j} + 27.49 \, \bar{j}^2 \label{eq:model_bubble_median}
\end{align}
Interestingly, a significant influence of $AR$ could be determined for $d_{m}$. Here, $AR$ showed a negative influence on $d_{m}$, whereas $AR$ was eliminated for $d_{50}$. Nevertheless, the factor of $\Lambda$ was again found to be significant and greater for both models. It follows that $\Lambda$ remains the most important parameter of the DLIP-structuring in terms of electrode performance and detached bubble size. Since $\dot{N}_\text{\ce{O2}} = const.$ is still valid for an applied $j$, it could be concluded that the number of bubbles and thus also the number of nucleation centers $n_\text{nucl}$ was lower.

\subsection*{Active nucleation centers}

\begin{figure*}[!ht]
	\centering
	\includegraphics[width=\textwidth]{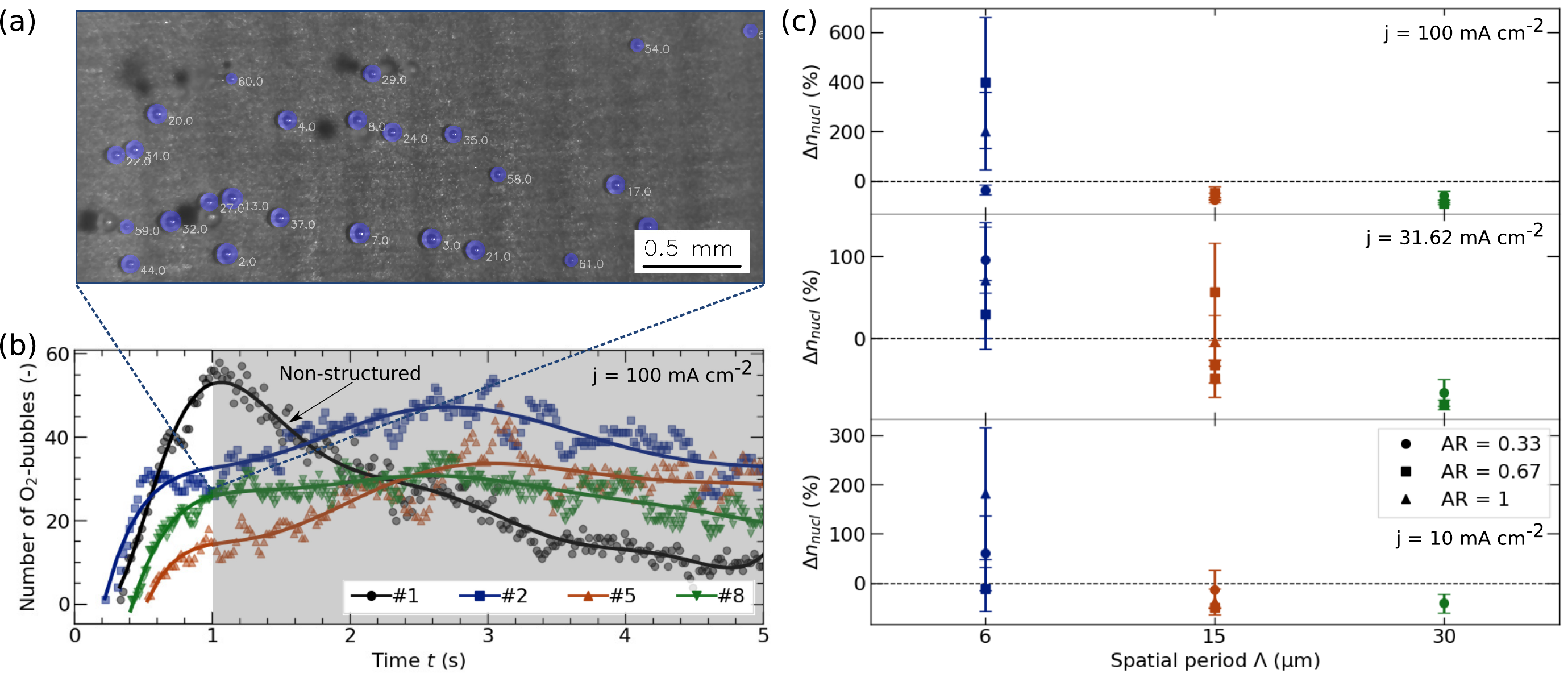}
	\caption{(a) Example topview image with highlighted O$_2$-bubbles growing on electrode \#2 at $j = \SI{100}{\milli\ampere\per\centi\metre\squared}$ and $t = \SI{1}{\second}$.  The numbers represent the unique bubble IDs. (b) Number of O$_2$-bubbles during \SI{5}{\second} galvanostatic measurements as a measure of $n_\text{nucl}$ on all periods ($\Lambda(\text{\#2}) = \SI{6}{\micro\metre}$, $\Lambda(\text{\#5}) = \SI{15}{\micro\metre}$ and $\Lambda(\text{\#8}) = \SI{30}{\micro\metre}$) at a constant $AR = 0.33$ in comparison to NSE (\#1). Only data outside the gray shaded area is included in the further evaluation as rising bubbles are blocking the optical access. (c) Normalized $n_\text{nucl}$ at $j = 10, 31.62$ and \SI{100}{\milli\ampere\per\centi\metre\squared}.}
	\label{fgr:results_coverage}
\end{figure*}

This was confirmed by analysis of the recorded topview images. Therefore, bubbles sitting on the electrode were segmented using a trained stardist model and taken as a measure for active nucleation centers, as shown in Fig. \ref{fgr:results_coverage} (a). Due to rising bubbles, the optical access to the electrode surface was limited after a specific time depending on the applied $j$. As shown in \mbox{Fig. \ref{fgr:results_coverage} (b)} there was a maximum at $t = \SI{1}{\minute}$ for $n_\text{nucl}$ at $j=\SI{100}{\milli\ampere\per\centi\metre\squared}$. Therefore, depending on $j$ only images within the first \SI{4}{\second}, and \SI{1}{\second}, were taken into account to calculate the time-averaged mean $\bar{n}_\text{nucl}$ of the number of nucleation sites $n_\text{nucl}$ at $j = \SI{31.62}{\milli\ampere\per\centi\metre\squared}$ and \mbox{$j = \SI{100}{\milli\ampere\per\centi\metre\squared}$,} respectively. Since few active nucleation sites and small $\dot{N}_\text{\ce{O2}}$, all recorded images were used for $j = \SI{10}{\milli\ampere\per\centi\metre\squared}$. The normalized change in the number of active nucleation sites $\Delta n_\text{nucl}$ was then calculated using the following equation:
\begin{align}
	\Delta n_\text{nucl}(j) = \frac{\bar{n}_\text{nucl}(j) - \bar{n}_\text{nucl, NSE}(j)}{\bar{n}_\text{nucl, NSE}(j)}
\end{align}

With the exception of DLIP-structures with $\Lambda = \SI{6}{\micro\metre}$, all samples showed a clear decrease in $\Delta n_\text{nucl}$ for all applied $j$, as shown in Fig. \ref{fgr:results_coverage} (c). In addition, an increase in $\Lambda$ led to a decrease in $\Delta n_\text{nucl}$. This could be attributed to the decrease in the number of peaks of the DLIP-structure with increasing $\Lambda$, as shown in the height profiles in Fig. \ref{fgr:results_laser_structure} (b), where the bubbles are likely to grow. 

In combination with the developed model for $\bar{n}_\text{nucl}$ in Eq. \ref{eq:model_nucl} and the results of the bubble size analysis, it could be proven that the DLIP-structures strongly influence the bubble dynamic. Hereby, especially $\Lambda$ showed a promising approach to tune $d_\text{B}$ and decrease $E_\text{cell}$, as $\Lambda$ was relevant for all developed models. 
\begin{align}
	\ln{\left( \bar{n}_\text{nucl} \right)} = &1.5035 - 0.74842 \bar{\Lambda} - 0.10964 \, \bar{AR} + 1.4573 \, \bar{j} \label{eq:model_nucl} \\
	&- 0.33154 \, \bar{\Lambda} \, \bar{AR} + 0.36201 \bar{j}^2 \notag
\end{align}
The shown improvement of the electrode performance in Fig. \ref{fgr:results_galvanostatic} (b) could also be explained by the bubble dynamics. As larger bubbles grew in fewer places on the electrode surface, it was ensured that the surface was largely wetted throughout. Thus, O$_2$ could be produced permanently. The dissolved gas was now seemingly collected by the bigger bubbles. From this it could be concluded that there were many active catalytic sites but only a low number of nucleation sites. This results in decreased $\eta_\Omega$ and following lower $E$ were measured.

However, $n_\text{nucl}$ could only be measured during few seconds, whereas $d_\text{B}$ was measured during the full galvanostatic measurements of $t = \SI{60}{\second}$. In addition, as only $\approx$ 35 \% of the entire electrode length of \SI{10}{\milli\metre} was in the FOV in the topview and only \mbox{$\approx$ 22 \%} in the sideview, not all bubbles could be recorded. Especially at low $j$, this could lead to a systematic error if the only active nucleation centers were located outside the FOV. Nevertheless, the measurement noise at the central point of DoE showed only a slight variation ($\sigma = 3.5447$)$^\dag$ for $n_\text{nucl}$. This further indicates that DLIP results in the formation of a homogeneous structure across the entire electrode. Furthermore, the application of a logarithmic transformation to the response prior to fitting resulted in very good results with $R^2 = 0.939$. It should be noted that for the model of $\bar{n}_\text{nucl}$, measurements 21 and 26 were excluded from the regression analysis due to the absence of nucleation sites within FOV.

\begin{figure}[ht]
	\centering
	\includegraphics[width=0.48\textwidth]{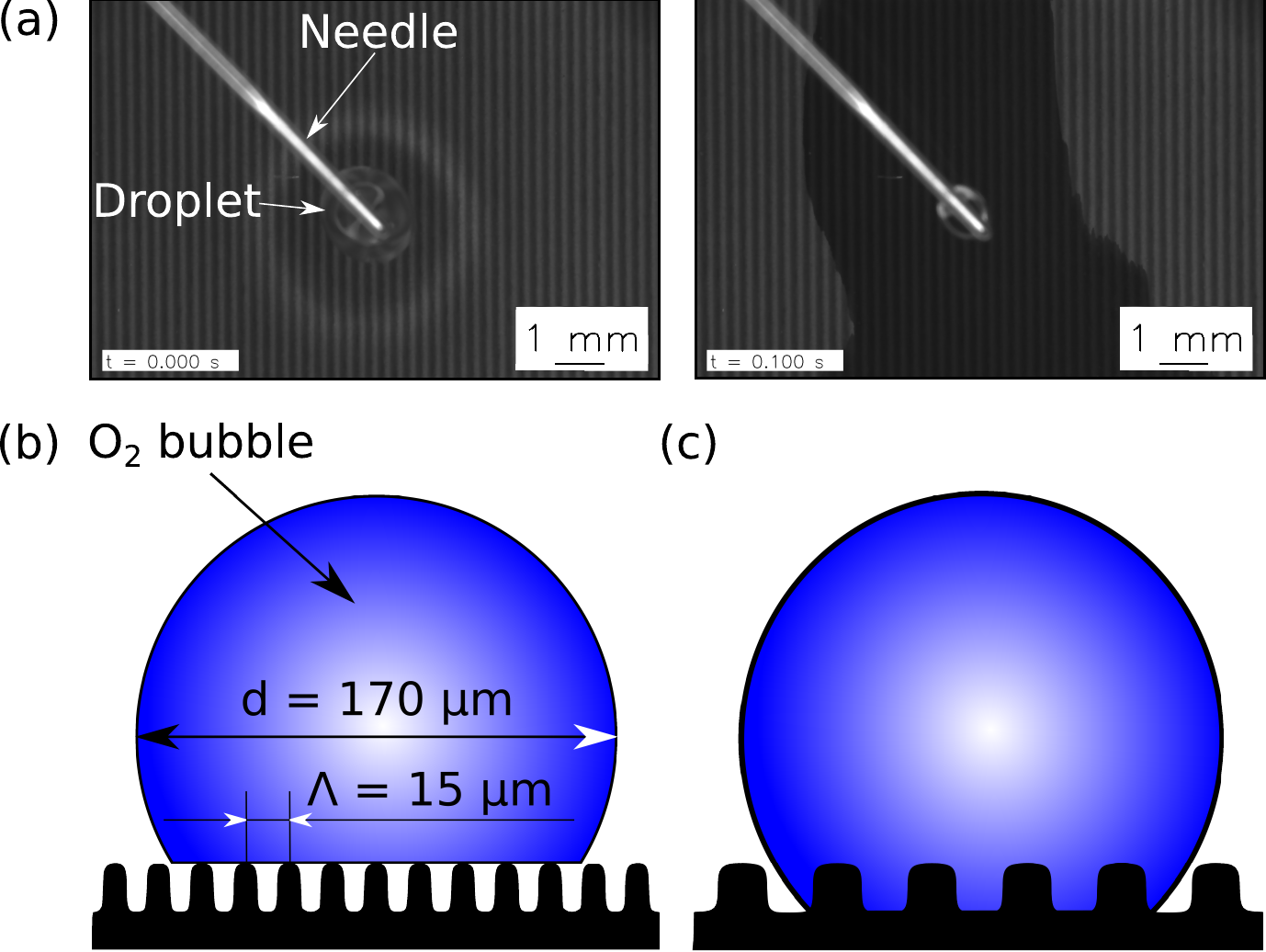}
	\caption{(a) Wetting behaviour of laser-structured surface showing superhydrophilic wetting as applied droplet spreads across entire surface. Full time series can be found in Fig. S7$^\dag$ (b) Possible nucleation of \ce{O2} bubble with fully wetted with an assumed bubble size of $d_\text{m} = \SI{170}{\micro\metre}$ according to Fig. \ref{fgr:results_bubblesize} resulting in total of $\approx$ 11 covered periods for $\Lambda = \SI{15}{\micro\metre}$. (c) Non-wetted electrode surface with increased $\Lambda$. Qualitative sketches throughout.}
	\label{fgr:results_bubble_model}
\end{figure}

Due to limited spatial resolution of the taken topview images, the exact nucleation sites could not be determined. However, as already mentioned above, all laser-structured electrodes showed superhydrophilic wetting behaviour. Exemplary images are shown in Fig. \ref{fgr:results_bubble_model} (a) and Fig. S7 indicating the directional spreading of the DI- water along the DLIP line structures.$^\dag$ In addition, as all samples were cleaned in a ultrasonic bath and stored for at least \SI{48}{\hour} in DI water, it can be assumed that only a negligible amount of gas pockets are present on the electrode surface, where bubbles are most likely to nucleate. It follows that a completely wetted surface is present at the beginning of each measurement. Due to the capillary forces acting in the channels, it can be further assumed that the electrode surface stays in a fully wetted state throughout the measurement. This leads to the assumption that bubble are most likely to be pinned on the tips of the laser-structure as sketched in Fig. \ref{fgr:results_bubble_model} (b). With increasing $\Lambda$ the capillary forces decrease and thus, it can be assumed that the evolving bubbles covers the whole surface structure (see \mbox{Fig. \ref{fgr:results_bubble_model} (c)).} However, these are hypothetical considerations that cannot be substantiated at this time.

\subsection*{Linear patterned nucleation sites}
\begin{figure}[!b]
	\centering
	\includegraphics[width=0.48\textwidth]{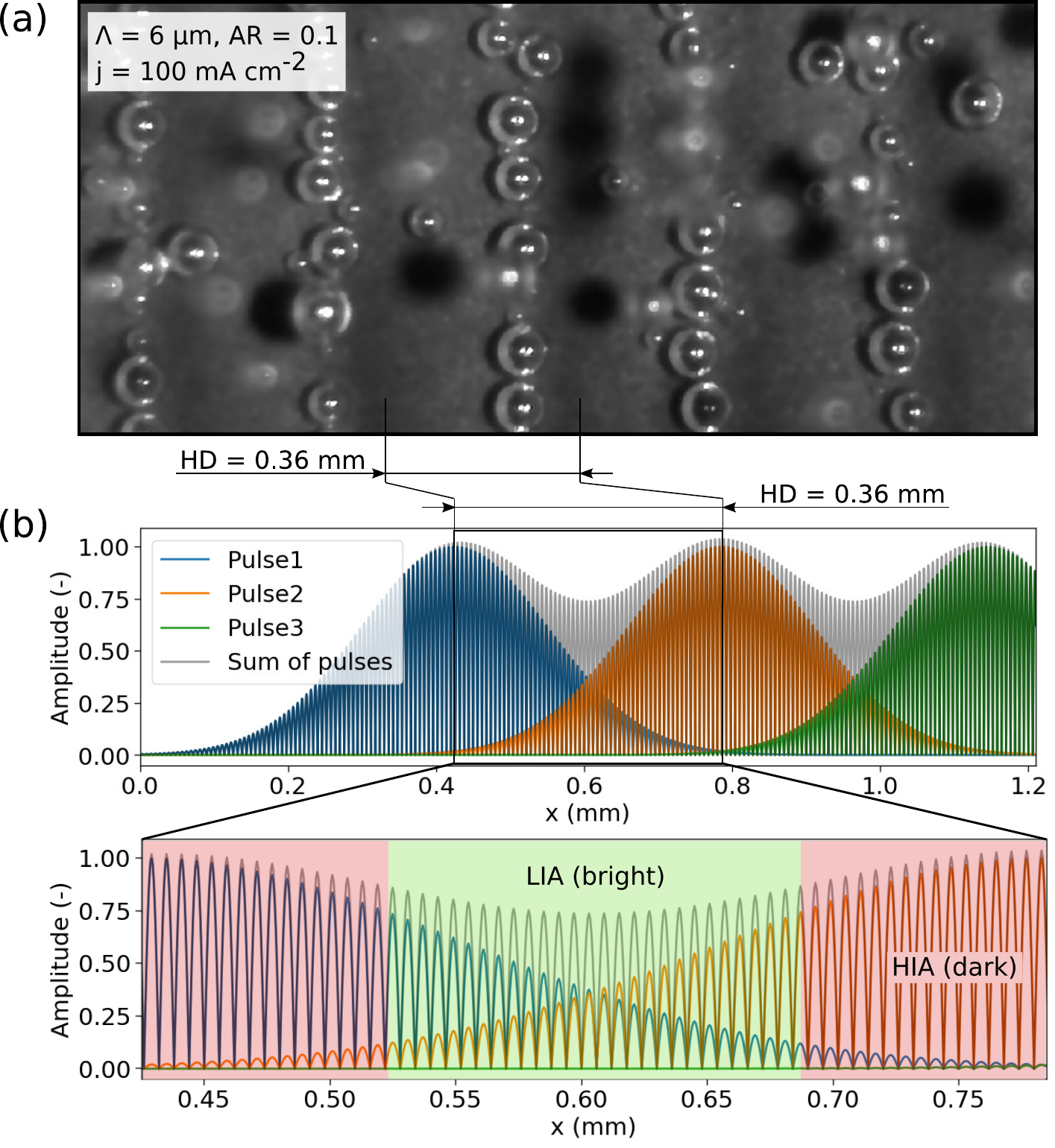}
	\caption{(a) Linear patterned \ce{O2} bubble nucleation. (b) Superposition of several Gaussian interference patterns similar to the DLIP texturing for the period of \SI{6}{\micro\metre} and the sum of all pulses showing a superimposed change in intensity with a period length equal to the hatch-distance. In addition, the high intensity area (HIA) and low intensity area (LIA) are highlighted in red and green, respectively.
	}
	\label{fgr:results_linearpatternednucl}
\end{figure}
A special phenomena could be observed for the electrodes \#3 and \#4 with $\Lambda = \SI{6}{\micro\metre}$ and $AR = 0.67$ and 1, respectively. During OER, the nucleation sites followed a linear pattern, as it is shown in \mbox{Fig. \ref{fgr:results_linearpatternednucl} (a).} The period of the visible dark/bright pattern  of the electrode surface$^\dag$ as well as the distance between the observed lines of active nucleation sites was determined to be equal to $HD$ from the DLIP-process. As each laser pulse overlaps in x-distance with the previous pulse by $HD$, a superimposed intensity profile with a peak-to-peak distance equal to $HD$ is found (see \mbox{Fig. \ref{fgr:results_linearpatternednucl} (b)).} This results in a darker surface in the region of the high intensity area (HIA) and a brighter surface in low intensity areas (LIA).

Initial assumptions of a changed oxide layer at the electrode surface could be invalidated with the help of XPS measurements, as there was no significant difference in the surface composition between dark and bright areas. This is shown in Tab. \ref{tbl:results_xps}, where the chemical states and surface concentrations of \ce{O} and \ce{Ni} were deconvoluted by fitting the XPS spectra (see \mbox{Fig. S10 (c)).} Therefore, the average value of the three measurements performed was calculated for both, HIA and LIA.

\begin{table}[h]
	\centering
	\small
	\caption{\ Averaged surface composition and standard deviation (in atomic \%) of the samples \#3 and \#4 determined by fitting XPS spectra in the high- (bright) and low intensity areas (dark)}
	\label{tbl:results_xps}
	\begin{tabular*}{0.75\textwidth}{@{\extracolsep{\fill}}lccccc}
		\hline
		\textbf{Element} & \multicolumn{3}{c}{\textbf{O}} & \multicolumn{2}{c}{\textbf{Ni}} \\
		BE (eV) & 529.5 & 531.2 & 532.6 & 852.3 & 853.8\\
		& \multirow{3}{*}{O-Ni} & O-Ni, & O-C, & \multirow{3}{*}{Ni$^0$} & \multirow{3}{*}{Ni$^{2+}$}\\
		Groups/ & & O=C, & -OH, && \\
		Ox. state & & O-Si & H$_2$O abs. && \\
		\hline
		\#3 LIA & $10.3\pm0.5$ & $12.7\pm0.6$ & $12.6\pm0.9$ & $1.0\pm0.1$ & $8.6\pm0.4$ \\
		\#3 HIA & $9.8\pm1.1$ & $13.2\pm0.2$ & $12.3\pm0.6$ & $1.1\pm0.1$ & $8.6\pm0.6$ \\
		\#4 LIA & $4.8\pm0.3$ & $22.5\pm0.1$ & $9.1\pm0.1$ & $0.5\pm0.0$ & $5.6\pm0.0$ \\
		\#4 HIA & $5.7\pm0.5$ & $21.1\pm0.3$ & $9.6\pm0.1$ & $0.3\pm0.1$ & $6.2\pm0.4$ \\
		\hline
	\end{tabular*}
\end{table}

Hence, the only significant difference between these areas must lie in the micro- and nanostructure of the electrode surface. This was shown using digital microscopy and SEM images. In lower intensity area (bright) a more shallow profile was detected compared to the higher intensity area (dark). In addition, SEM imaging showed microholes within the maxima region of the interference pattern, which are only present in the higher intensity area (see Fig. \ref{fgr:results_sem} and Fig. S6).$^\dag$ These could serve as cavities for the initial bubble nucleation.

This effect could be interesting with regard to optimized electrode surfaces and morphologies by defining nucleation sites through DLIP-structuring and optimizing the electrolyte flow at these sites. However, more studies are necessary to understand the ongoing mechanisms and long-term measurement have to be performed to proof that this phenomena does not change over time.

\section{Conclusions}
\label{sec:conclusions}
In summary, the influence of the structure parameters of DLIP, spatial period and aspect ratio, on the double-layer capacitance as a measure of the electrochemically active surface area was investigated. Additionally, the onset potential and overpotential during OER were analyzed. The relation between overpotential and bubble dynamics could be studied by determining the bubble size distribution and number of nucleation sites during galvanostatic measurements. 
For that purpose, Ni-foils were structured with line-like surface features with $\Lambda = 6,\, 15,\, \SI{30}{\micro\metre}$ and $AR = 0.33,\, 0.67,\, 1.00$ using a \si{\pico\second} pulsed laser and DLIP. By implementing a statistical design of experiments approach, models were derived from the measurements performed to analyze the significance of the influence of the structure parameters of laser-structuring.

An optimum of the spatial period was found for the double-layer capacitance, and thus the electrochemically active surface area, at $\Lambda = \SI{15}{\micro\metre}$, which led to an increase by a factor of 12. In general, all laser-structured electrodes showed a significant enlargement of the electrochemically active surface area. Furthermore, the onset potential of OER could be decreased by $\approx \, \SI{4.5}{\%}$. However, at higher aspect ratios the homogeneity of the line-patterns was decreasing due to increased ablation and thus, a non-linear trend was observed.

The quasi-steady state potential, as a measure of the electrode overpotential, was decreased by up to $\approx$ \SI{164}{\milli\volt} at \mbox{$j = \SI{100}{\milli\ampere\per\centi\metre\squared}$.} This shows the great potential of DLIP-structures for OER. As the bubble sizes were increased and the number of active nucleation sites was decreased, the ohmic resistances could be decreased as large surface areas were wetted throughout. Furthermore, since no significant changes in the surface composition were found for the DLIP-structures, it could be inferred that the increased electrochemically active surface area provided more active catalytic sites. However, the dissolved gas was then collected by bigger bubbles at fewer nucleation sites on the DLIP-structured electrodes. 

In general, the spatial period had a big impact on the overpotential and bubble dynamics, while the aspect ratio, and thus the depth of the structure, was not relevant for most models developed multiple regression analysis. Therefore, further studies should focus on the structure size instead of the depth.

It was found that the location of bubble nucleation could be tuned by the evolving laser-induced periodic surface structures. In conclusion, DLIP-structuring offers the possibility to enhance the overall efficiency of OER and  thus, an AWE cell. Furthermore, DLIP-structuring might tune the detached bubble sizes to the needs of the periphery. Combined with the eventual possible definition of nucleation centers, this could facilitate the development of novel electrodes and even cell types.

\section*{Author Contributions}
H.R.: Conceptualization, Investigation, Data curation, Formal Analysis, Methodology, Visualization, Writing - original draft, Editing; F.R.: Conceptualization, Investigation, Data curation, Formal Analysis, Methodology, Visualization, Writing - original draft, Editing; J.M.: Conceptualization, Formal Analysis, Methodology, Visualization, Writing - original draft, Editing; M.M.M.: Investigation, Formal Analysis, Writing - review \& editing; K.S.: Investigation, Formal Analysis; R.B.: Investigation, Formal Analysis; H.H.: Conceptualization; X.Y.: Conceptualization, Supervision, Writing - review \& editing; L.U.: Supervision, Writing - review \& editing; A.F.L.: Conceptualization, Supervision, Writing - review \& editing; K.E.: Funding acquisition, Project administration, Supervision, Writing - review \& editing.

\section*{Conflicts of interest}
There are no conflicts to declare.

\section*{Data availability}
Data for this article, including electrochemical measurement data, raw images and relevant metadata of the performed experiments are available at RODARE at \href{https://doi.org/10.14278/rodare.3064}{10.14278/rodare.3064}.

\section*{Acknowledgments}
This project is supported by the Federal State of Saxony in terms of the "European Regional Development Fund" (H2-EPF-HZDR), the Helmholtz Association Innovation pool project "Solar Hydrogen", the Hydrogen Lab of the School of Engineering of TU Dresden, and BMBF (project ALKALIMIT, grant no. 03SF0731A).

\nomenclature[a]{\(d_\text{B}\)}{Bubble diameter, \si{\metre}}
\nomenclature[a]{\(d_\text{m}\)}{Mode of bubble size distribution, \si{\metre}}
\nomenclature[a]{\(d_{50}\)}{Median of bubble size distribution, \si{\metre}}
\nomenclature[a]{\(d_{x,y}\)}{Laser spot size, \si{\metre}}
\nomenclature[a]{\(n_\text{nucl}\)}{Number of nucleation centers, -}
\nomenclature[a]{\(N\)}{Number of experiments/Number of scans, -}
\nomenclature[a]{\(r_c\)}{Contact radius, \si{\metre}}
\nomenclature[a]{\(HD\)}{Hatch distance, \si{\metre}}
\nomenclature[a]{\(PtP\)}{Pulse-to-pulse distance, \si{\metre}}
\nomenclature[a]{\(AR\)}{Aspect ratio between $\Lambda$ and structure depth, -}
\nomenclature[a]{\(ECSA\)}{Electrochemically active surface area, \si{\metre\squared}}
\nomenclature[a]{\(C_\text{dl}\)}{Double-layer capacitance, \si{\farad}}
\nomenclature[a]{\(t\)}{Time, \si{\second}}
\nomenclature[a]{\(E\)}{Electric potential, \si{\volt}}
\nomenclature[a]{\(\Delta E^0\)}{Reversible cell potential, \si{\volt}}
\nomenclature[a]{\(E_\text{cell}\)}{Cell potential, \si{\volt}}
\nomenclature[a]{\(E_\text{SS}\)}{Quasi-steady state potential, \si{\volt}}
\nomenclature[a]{\(E_\text{on}\)}{Onset potential, \si{\volt}}
\nomenclature[a]{\(I\)}{Electric current, \si{\ampere}}
\nomenclature[a]{\(j\)}{Current density, \si{\ampere\per\metre\squared}}
\nomenclature[a]{\(j_\text{eff}\)}{Effective current density, \si{\ampere\per\metre\squared}}
\nomenclature[a]{\(V\)}{Volume, \si{\metre\cubed}}
\nomenclature[a]{\(A_\text{el}\)}{Geometrical electrode area, \si{\metre\squared}}
\nomenclature[a]{\(c\)}{Electrolyte concentration, \si{\mole\per\litre}} 
\nomenclature[a]{\(F_\text{S}\)}{Surface tension force, \si{\newton}}
\nomenclature[a]{\(T\)}{Temperature, \si{\kelvin}}
\nomenclature[a]{\(p\)}{Pressure, \si{\pascal}}
\nomenclature[a]{\(S_\text{a}\)}{Average surface roughness, \si{\metre}}
\nomenclature[a]{\(S_\text{dr}\)}{Interfacial area ratio, -}
\nomenclature[a]{\(F\)}{Faraday constant, \si{\coulomb\per\mole}}
\nomenclature[a]{\(\dot{N}\)}{Molar flux, \si{\mole\per\second}}
\nomenclature[a]{\(z\)}{Charge number, -}
\nomenclature[a]{\(R^2\)}{Coefficient of determination, -}

\nomenclature[b]{\(\beta\)}{Coefficient for multiple regression analysis, -}
\nomenclature[b]{\(\gamma\)}{Surface tension, \si{\newton\per\metre}}
\nomenclature[b]{\(\gamma\)}{Surface tension, \si{\newton\per\metre}}
\nomenclature[b]{\(\theta\)}{Contact angle, \si{\degree}}
\nomenclature[b]{\(\theta_\text{DLIP}\)}{Interference angle, \si{\degree}}
\nomenclature[b]{\(\Lambda\)}{Spatial period length, \si{\metre}}
\nomenclature[b]{\(\lambda\)}{Wavelength, \si{\metre}}
\nomenclature[b]{\(\sigma\)}{Standard deviation, -}
\nomenclature[b]{\(\tau\)}{Pulse duration, \si{\second}}
\nomenclature[b]{\(\nu\)}{Scan rate, \si{\volt\per\second}}
\nomenclature[b]{\(\Phi_\text{sp}\)}{Single pulse fluence, \si{\joule\per\metre\squared}}
\nomenclature[b]{\(\Phi_\text{cum}\)}{Cumulated fluence, \si{\joule\per\metre\squared}}
\nomenclature[b]{\(\eta_\text{on}\)}{Onset potential efficiency, -}
\nomenclature[b]{\(\eta_\text{SS}\)}{Quasi-steady state efficiency, -}
\nomenclature[b]{\(\eta_\text{anode}\)}{Anodic overpotential, \si{\volt}}
\nomenclature[b]{\(\eta_\text{cathode}\)}{Cathodic overpotential, \si{\volt}}
\nomenclature[b]{\(\eta_\Omega\)}{Ohmic overpotential, \si{\volt}}
\nomenclature[b]{\(\eta_\text{conc}\)}{Concentration overpotential, \si{\volt}}

\printnomenclature[2cm]





\bibliography{references} 

\begin{thebibliography}{89}
\providecommand{\natexlab}[1]{#1}
\providecommand{\url}[1]{\texttt{#1}}
\expandafter\ifx\csname urlstyle\endcsname\relax
  \providecommand{\doi}[1]{doi: #1}\else
  \providecommand{\doi}{doi: \begingroup \urlstyle{rm}\Url}\fi

\bibitem[van~der Spek et~al.(2022)van~der Spek, Banet, Bauer, Gabrielli,
  Goldthorpe, Mazzotti, Munkejord, R{\o}kke, Shah, Sunny, Sutter, Trusler, and
  Gazzani]{Spek2022}
Mijndert van~der Spek, Catherine Banet, Christian Bauer, Paolo Gabrielli, Ward
  Goldthorpe, Marco Mazzotti, Svend~T. Munkejord, Nils~A. R{\o}kke, Nilay Shah,
  Nixon Sunny, Daniel Sutter, J.~Martin Trusler, and Matteo Gazzani.
\newblock Perspective on the hydrogen economy as a pathway to reach net-zero
  {CO}2 emissions in europe.
\newblock \emph{Energy Environ. Sci.}, 15\penalty0 (3):\penalty0 1034--1077,
  2022.
\newblock \doi{10.1039/d1ee02118d}.

\bibitem[Staffell et~al.(2019)Staffell, Scamman, Abad, Balcombe, Dodds, Ekins,
  Shah, and Ward]{Staffell2019}
Iain Staffell, Daniel Scamman, Anthony~Velazquez Abad, Paul Balcombe, Paul~E.
  Dodds, Paul Ekins, Nilay Shah, and Kate~R. Ward.
\newblock The role of hydrogen and fuel cells in the global energy system.
\newblock \emph{Energy Environ. Sci.}, 12\penalty0 (2):\penalty0 463--491,
  2019.
\newblock \doi{10.1039/c8ee01157e}.

\bibitem[Smolinka and Garche(2021)]{Smolinka2021}
Tom Smolinka and Jurgen Garche, editors.
\newblock \emph{Electrochemical Power Sources: Fundamentals, Systems, and
  Applications}.
\newblock Elsevier, 2021.
\newblock ISBN 9780128194249.

\bibitem[Hermesmann et~al.(2021)Hermesmann, Grübel, Scherotzki, and
  Müller]{Hermesmann2021}
M.~Hermesmann, K.~Grübel, L.~Scherotzki, and T.~E. Müller.
\newblock Promising pathways: The geographic and energetic potential of
  power-to-x technologies based on regeneratively obtained hydrogen.
\newblock \emph{Renew. Sust. Energ. Rev.}, 138:\penalty0 110644, 2021.
\newblock \doi{10.1016/j.rser.2020.110644}.

\bibitem[Valente et~al.(2021)Valente, Tulus, Gal{\'{a}}n-Mart{\'{\i}}n,
  Huijbregts, and Guill{\'{e}}n-Gos{\'{a}}lbez]{Valente2021}
Antonio Valente, Victor Tulus, {\'{A}}ngel Gal{\'{a}}n-Mart{\'{\i}}n, Mark
  A.~J. Huijbregts, and Gonzalo Guill{\'{e}}n-Gos{\'{a}}lbez.
\newblock The role of hydrogen in heavy transport to operate within planetary
  boundaries.
\newblock \emph{Sustain. Energy Fuels}, pages 4637--4649, 2021.
\newblock \doi{10.1039/d1se00790d}.

\bibitem[Hoecke et~al.(2021)Hoecke, Laffineur, Campe, Perreault, Verbruggen,
  and Lenaerts]{Hoecke2021}
Laurens~Van Hoecke, Ludovic Laffineur, Roy Campe, Patrice Perreault, Sammy~W.
  Verbruggen, and Silvia Lenaerts.
\newblock Challenges in the use of hydrogen for maritime applications.
\newblock \emph{Energy Environ. Sci.}, pages 815--843, 2021.
\newblock \doi{10.1039/d0ee01545h}.

\bibitem[Wang et~al.(2021)Wang, Zhao, Babich, Senk, and Fan]{Wang2021}
R.R. Wang, Y.Q. Zhao, A.~Babich, D.~Senk, and X.Y. Fan.
\newblock Hydrogen direct reduction (h-dr) in steel industry—an overview of
  challenges and opportunities.
\newblock \emph{J. Clean. Prod.}, 329:\penalty0 129797, 2021.
\newblock ISSN 0959-6526.
\newblock \doi{10.1016/j.jclepro.2021.129797}.

\bibitem[Furszyfer Del~Rio et~al.(2022)Furszyfer Del~Rio, Sovacool, Foley,
  Griffiths, Bazilian, Kim, and Rooney]{FurszyferDelRio2022}
Dylan~D. Furszyfer Del~Rio, Benjamin~K. Sovacool, Aoife~M. Foley, Steve
  Griffiths, Morgan Bazilian, Jinsoo Kim, and David Rooney.
\newblock Decarbonizing the glass industry: A critical and systematic review of
  developments, sociotechnical systems and policy options.
\newblock \emph{Renew. Sust. Energ. Rev.}, 155:\penalty0 111885, 2022.
\newblock ISSN 1364-0321.
\newblock \doi{10.1016/j.rser.2021.111885}.

\bibitem[El-Emam and Özcan(2019)]{ElEmam2019}
Rami~S. El-Emam and Hasan Özcan.
\newblock Comprehensive review on the techno-economics of sustainable
  large-scale clean hydrogen production.
\newblock \emph{J. Clean. Prod.}, 220:\penalty0 593--609, 2019.
\newblock ISSN 0959-6526.
\newblock \doi{10.1016/j.jclepro.2019.01.309}.

\bibitem[He et~al.(2023)He, Cui, Zhao, Chen, Shang, and Tan]{He2023}
Yi~He, Yifan Cui, Zhongxi Zhao, Yongtang Chen, Wenxu Shang, and Peng Tan.
\newblock Strategies for bubble removal in electrochemical systems.
\newblock \emph{Energy Rev.}, 2\penalty0 (1):\penalty0 100015, mar 2023.
\newblock \doi{10.1016/j.enrev.2023.100015}.

\bibitem[Swiegers et~al.(2021)Swiegers, Terrett, Tsekouras, Tsuzuki, Pace, and
  Stranger]{Swiegers2021}
Gerhard~F. Swiegers, Richard N.~L. Terrett, George Tsekouras, Takuya Tsuzuki,
  Ronald~J. Pace, and Robert Stranger.
\newblock The prospects of developing a highly energy-efficient water
  electrolyser by eliminating or mitigating bubble effects.
\newblock \emph{Sustain. Energy Fuels}, 5\penalty0 (5):\penalty0 1280--1310,
  2021.
\newblock \doi{10.1039/d0se01886d}.

\bibitem[Angulo et~al.(2020)Angulo, van~der Linde, Gardeniers, Modestino, and
  Rivas]{Angulo2020}
Andrea Angulo, Peter van~der Linde, Han Gardeniers, Miguel Modestino, and
  David~Fern{\'{a}}ndez Rivas.
\newblock Influence of bubbles on the energy conversion efficiency of
  electrochemical reactors.
\newblock \emph{Joule}, 4\penalty0 (3):\penalty0 555--579, mar 2020.
\newblock \doi{10.1016/j.joule.2020.01.005}.

\bibitem[Zeng and Zhang(2010)]{Zeng2010}
Kai Zeng and Dongke Zhang.
\newblock Recent progress in alkaline water electrolysis for hydrogen
  production and applications.
\newblock \emph{Prog. Energy Combust. Sci.}, 36\penalty0 (3):\penalty0
  307--326, 2010.
\newblock \doi{10.1016/j.pecs.2009.11.002}.

\bibitem[Vogt(2017)]{Vogt2017}
H.~Vogt.
\newblock The quantities affecting the bubble coverage of gas-evolving
  electrodes.
\newblock \emph{Electrochim. Acta}, 235:\penalty0 495--499, may 2017.
\newblock \doi{10.1016/j.electacta.2017.03.116}.

\bibitem[Vogt and Stephan(2015)]{Vogt2015}
Helmut Vogt and Karl Stephan.
\newblock Local microprocesses at gas-evolving electrodes and their influence
  on mass transfer.
\newblock \emph{Electrochim. Acta}, 155:\penalty0 348--356, feb 2015.
\newblock \doi{10.1016/j.electacta.2015.01.008}.

\bibitem[Vogt(2011)]{Vogt2011}
H.~Vogt.
\newblock On the gas-evolution efficiency of electrodes {I} - {T}heoretical.
\newblock \emph{Electrochim. Acta}, 56\penalty0 (3):\penalty0 1409--1416, 2011.
\newblock \doi{10.1016/j.electacta.2010.08.101}.

\bibitem[Balzer and Vogt(2003)]{Balzer2003}
R.~J. Balzer and H.~Vogt.
\newblock Effect of electrolyte flow on the bubble coverage of vertical
  gas-evolving electrodes.
\newblock \emph{J. Electrochem. Soc.}, 150\penalty0 (1):\penalty0 E11, 2003.
\newblock ISSN 0013-4651.
\newblock \doi{10.1149/1.1524185}.

\bibitem[Eigeldinger and Vogt(2000)]{Eigeldinger2000}
J.~Eigeldinger and H.~Vogt.
\newblock The bubble coverage of gas-evolving electrodes in a flowing
  electrolyte.
\newblock \emph{Electrochim. Acta}, 45\penalty0 (27):\penalty0 4449--4456,
  2000.
\newblock ISSN 0013-4686.
\newblock \doi{10.1016/S0013-4686(00)00513-2}.

\bibitem[Rocha et~al.(2022)Rocha, Delmelle, Georgiadis, and Proost]{Rocha2022}
Fernando Rocha, Renaud Delmelle, Christos Georgiadis, and Joris Proost.
\newblock Effect of pore size and electrolyte flow rate on the bubble removal
  efficiency of 3d pure ni foam electrodes during alkaline water electrolysis.
\newblock \emph{J. Environ. Chem. Eng.}, 10\penalty0 (3):\penalty0 107648, June
  2022.
\newblock ISSN 2213-3437.
\newblock \doi{10.1016/j.jece.2022.107648}.

\bibitem[Li and Chen(2021)]{Li2021b}
Yan-Hom Li and Yen-Ju Chen.
\newblock The effect of magnetic field on the dynamics of gas bubbles in water
  electrolysis.
\newblock \emph{Sci. Rep.}, 11\penalty0 (1):\penalty0 9346, April 2021.
\newblock ISSN 2045-2322.
\newblock \doi{10.1038/s41598-021-87947-9}.

\bibitem[Darband et~al.(2019)Darband, Aliofkhazraei, and
  Shanmugam]{Darband2019}
Ghasem~Barati Darband, Mahmood Aliofkhazraei, and Sangaraju Shanmugam.
\newblock Recent advances in methods and technologies for enhancing bubble
  detachment during electrochemical water splitting.
\newblock \emph{Renew. Sust. Energ. Rev.}, 114:\penalty0 109300, oct 2019.
\newblock \doi{10.1016/j.rser.2019.109300}.

\bibitem[Baczyzmalski et~al.(2017)Baczyzmalski, Karnbach, Mutschke, Yang,
  Eckert, Uhlemann, and Cierpka]{Baczyzmalski2017}
Dominik Baczyzmalski, Franziska Karnbach, Gerd Mutschke, Xuegeng Yang, Kerstin
  Eckert, Margitta Uhlemann, and Christian Cierpka.
\newblock Growth and detachment of single hydrogen bubbles in a
  magnetohydrodynamic shear flow.
\newblock \emph{Phys. Rev. Fluid}, 2\penalty0 (9):\penalty0 093701, September
  2017.
\newblock ISSN 2469-990X.
\newblock \doi{10.1103/physrevfluids.2.093701}.

\bibitem[Koza et~al.(2011)Koza, Mühlenhoff, Żabiński, Nikrityuk, Eckert,
  Uhlemann, Gebert, Weier, Schultz, and Odenbach]{Koza2011}
Jakub~Adam Koza, Sascha Mühlenhoff, Piotr Żabiński, Petr~A. Nikrityuk,
  Kerstin Eckert, Margitta Uhlemann, Annett Gebert, Tom Weier, Ludwig Schultz,
  and Stefan Odenbach.
\newblock Hydrogen evolution under the influence of a magnetic field.
\newblock \emph{Electrochimica Acta}, 56\penalty0 (6):\penalty0 2665--2675,
  February 2011.
\newblock ISSN 0013-4686.
\newblock \doi{10.1016/j.electacta.2010.12.031}.

\bibitem[Iida et~al.(2007)Iida, Matsushima, and Fukunaka]{Iida2007}
Takami Iida, Hisayoshi Matsushima, and Y.~Fukunaka.
\newblock Water {E}lectrolysis under a {M}agnetic {F}ield.
\newblock \emph{J. Electrochem. Soc.}, 154:\penalty0 E112--E115, 2007.
\newblock \doi{10.1149/1.2742807}.

\bibitem[Bashkatov et~al.(2024)Bashkatov, Park, Demirkır, Wood, Koper, Lohse,
  and Krug]{Bashkatov2024}
Aleksandr Bashkatov, Sunghak Park, {\c{C}}ayan Demirkır, Jeffery~A. Wood, Marc
  T.~M. Koper, Detlef Lohse, and Dominik Krug.
\newblock Performance enhancement of electrocatalytic hydrogen evolution
  through coalescence-induced bubble dynamics.
\newblock \emph{J. Am. Chem. Soc.}, 146\penalty0 (14):\penalty0 10177--10186,
  2024.
\newblock \doi{10.1021/jacs.4c02018}.

\bibitem[Lv et~al.(2021)Lv, Pe{\~{n}}as, The, Eijkel, van~den Berg, Zhang, and
  Lohse]{Lv2021}
Pengyu Lv, Pablo Pe{\~{n}}as, Hai~Le The, Jan Eijkel, Albert van~den Berg,
  Xuehua Zhang, and Detlef Lohse.
\newblock Self-propelled detachment upon coalescence of surface bubbles.
\newblock \emph{Phys. Rev. Lett.}, 127\penalty0 (23):\penalty0 235501, nov
  2021.
\newblock \doi{10.1103/physrevlett.127.235501}.

\bibitem[Jiao et~al.(2021)Jiao, Fu, Wang, and Zhao]{Jiao2021}
Shilong Jiao, Xianwei Fu, Shuangyin Wang, and Yong Zhao.
\newblock Perfecting electrocatalysts via imperfections: towards the
  large-scale deployment of water electrolysis technology.
\newblock \emph{Energy Environ. Sci.}, 14\penalty0 (4):\penalty0 1722--1770,
  2021.
\newblock \doi{10.1039/d0ee03635h}.

\bibitem[Yang et~al.(2022)Yang, Li, Yang, Lan, Liu, Fu, Zhang, Liao, and
  Zhu]{Yang2022a}
Yang Yang, Jun Li, Yingrui Yang, Linghan Lan, Run Liu, Qian Fu, Liang Zhang,
  Qiang Liao, and Xun Zhu.
\newblock Gradient porous electrode-inducing bubble splitting for highly
  efficient hydrogen evolution.
\newblock \emph{Appl. Energy}, 307:\penalty0 118278, feb 2022.
\newblock \doi{10.1016/j.apenergy.2021.118278}.

\bibitem[Kou et~al.(2020)Kou, Wang, Shi, Zhang, Chiovoloni, Lu, Chen, Worsley,
  Wood, Baker, Duoss, Wu, Zhu, and Li]{Kou2020}
Tianyi Kou, Shanwen Wang, Rongpei Shi, Tao Zhang, Samuel Chiovoloni,
  Jennifer~Q. Lu, Wen Chen, Marcus~A. Worsley, Brandon~C. Wood, Sarah~E. Baker,
  Eric~B. Duoss, Rui Wu, Cheng Zhu, and Yat Li.
\newblock Periodic porous 3d electrodes mitigate gas bubble traffic during
  alkaline water electrolysis at high current densities.
\newblock \emph{Adv. Energy Mater.}, 10\penalty0 (46):\penalty0 2002955, oct
  2020.
\newblock \doi{10.1002/aenm.202002955}.

\bibitem[Fujimura et~al.(2021)Fujimura, Kunimoto, Fukunaka, and
  Homma]{Fujimura2021}
Tatsuki Fujimura, Masahiro Kunimoto, Yasuhiro Fukunaka, and Takayuki Homma.
\newblock Analysis of the hydrogen evolution reaction at ni micro-patterned
  electrodes.
\newblock \emph{Electrochim. Acta}, 368:\penalty0 137678, February 2021.
\newblock ISSN 0013-4686.
\newblock \doi{10.1016/j.electacta.2020.137678}.

\bibitem[Andaveh et~al.(2022)Andaveh, Barati~Darband, Maleki, and
  Sabour~Rouhaghdam]{Andaveh2022}
R.~Andaveh, Gh. Barati~Darband, M.~Maleki, and A.~Sabour~Rouhaghdam.
\newblock Superaerophobic/superhydrophilic surfaces as advanced
  electrocatalysts for the hydrogen evolution reaction: a comprehensive review.
\newblock \emph{J. Mater. Chem. A}, 10\penalty0 (10):\penalty0 5147--5173,
  2022.
\newblock ISSN 2050-7496.
\newblock \doi{10.1039/d1ta10519a}.

\bibitem[Zhang and Zeng(2012)]{Zhang2012}
Dongke Zhang and Kai Zeng.
\newblock Evaluating the behavior of electrolytic gas bubbles and their effect
  on the cell voltage in alkaline water electrolysis.
\newblock \emph{Ind. Eng. Chem. Res.}, 51\penalty0 (42):\penalty0 13825--13832,
  2012.
\newblock \doi{10.1021/ie301029e}.

\bibitem[Bashkatov et~al.(2022)Bashkatov, Hossain, Mutschke, Yang, Rox,
  Weidinger, and Eckert]{Bashkatov2022}
Aleksandr Bashkatov, Syed~Sahil Hossain, Gerd Mutschke, Xuegeng Yang, Hannes
  Rox, Inez~M. Weidinger, and Kerstin Eckert.
\newblock On the growth regimes of hydrogen bubbles at microelectrodes.
\newblock \emph{Phys. Chem. Chem. Phys.}, 24\penalty0 (43):\penalty0
  26738--26752, 2022.
\newblock ISSN 1463-9084.
\newblock \doi{10.1039/d2cp02092k}.

\bibitem[Massing et~al.(2019)Massing, Mutschke, Baczyzmalski, Hossain, Yang,
  Eckert, and Cierpka]{Massing2019}
Julian Massing, Gerd Mutschke, Dominik Baczyzmalski, Syed~Sahil Hossain,
  Xuegeng Yang, Kerstin Eckert, and Christian Cierpka.
\newblock Thermocapillary convection during hydrogen evolution at
  microelectrodes.
\newblock \emph{Electrochim. Acta}, 297:\penalty0 929--940, 2019.
\newblock \doi{10.1016/j.electacta.2018.11.187}.

\bibitem[Yang et~al.(2018)Yang, Baczyzmalski, Cierpka, Mutschke, and
  Eckert]{Yang2018}
Xuegeng Yang, Dominik Baczyzmalski, Christian Cierpka, Gerd Mutschke, and
  Kerstin Eckert.
\newblock Marangoni convection at electrogenerated hydrogen bubbles.
\newblock \emph{Phys. Chem. Chem. Phys.}, 20\penalty0 (17):\penalty0
  11542--11548, 2018.
\newblock \doi{10.1039/c8cp01050a}.

\bibitem[Meulenbroek et~al.(2021)Meulenbroek, Vreman, and
  Deen]{Meulenbroek2021}
A.M. Meulenbroek, A.W. Vreman, and N.G. Deen.
\newblock Competing marangoni effects form a stagnant cap on the interface of a
  hydrogen bubble attached to a microelectrode.
\newblock \emph{Electrochim. Acta}, page 138298, 2021.
\newblock \doi{10.1016/j.electacta.2021.138298}.

\bibitem[Duhar and Colin(2006)]{Duhar2006}
Géraldine Duhar and Catherine Colin.
\newblock Dynamics of bubble growth and detachment in a viscous shear flow.
\newblock \emph{Phys. Fluids}, 18\penalty0 (7):\penalty0 077101, jul 2006.
\newblock ISSN 1089-7666.
\newblock \doi{10.1063/1.2213638}.

\bibitem[Haverkort(2024)]{Haverkort2024}
J.W. Haverkort.
\newblock A general mass transfer equation for gas-evolving electrodes.
\newblock \emph{Int. J. Hydrog. Energy}, 74:\penalty0 283--296, July 2024.
\newblock ISSN 0360-3199.
\newblock \doi{10.1016/j.ijhydene.2024.06.010}.

\bibitem[Brandon et~al.(1985)Brandon, Kelsall, Levine, and Smith]{Brandon1985a}
N.~P. Brandon, G.~H. Kelsall, S.~Levine, and A.~L. Smith.
\newblock Interfacial electrical properties of electrogenerated bubbles.
\newblock \emph{J. Appl. Electrochem.}, 15\penalty0 (4):\penalty0 485--493, jul
  1985.
\newblock \doi{10.1007/bf01059289}.

\bibitem[Brandon and Kelsall(1985)]{Brandon1985b}
N.~P. Brandon and G.~H. Kelsall.
\newblock Growth kinetics of bubbles electrogenerated at microelectrodes.
\newblock \emph{J. Appl. Electrochem.}, 15:\penalty0 475--484, 1985.
\newblock \doi{10.1007/BF01059288}.

\bibitem[Shi et~al.(2021)Shi, Shang, and Zhang]{Shi2021}
Run Shi, Lu~Shang, and Tierui Zhang.
\newblock Three phase interface engineering for advanced catalytic
  applications.
\newblock \emph{ACS Appl. Energy Mater.}, 4\penalty0 (2):\penalty0 1045--1052,
  January 2021.
\newblock ISSN 2574-0962.
\newblock \doi{10.1021/acsaem.0c02989}.

\bibitem[Kim et~al.(2021)Kim, Kim, and Kim]{Kim2021a}
Byung~Keun Kim, Myung~Jun Kim, and Jae~Jeong Kim.
\newblock Impact of surface hydrophilicity on electrochemical water splitting.
\newblock \emph{ACS Appl. Mater. Interfaces}, 13\penalty0 (10):\penalty0
  11940--11947, March 2021.
\newblock ISSN 1944-8252.
\newblock \doi{10.1021/acsami.0c22409}.

\bibitem[Krause et~al.(2023)Krause, Skibi{\'{n}}ska, Rox, Baumann, Marzec,
  Yang, Mutschke, {\.{Z}}abi{\'{n}}ski, Lasagni, and Eckert]{Krause2023}
L.~Krause, K.~Skibi{\'{n}}ska, H.~Rox, R.~Baumann, M.~M. Marzec, X.~Yang,
  G.~Mutschke, P.~{\.{Z}}abi{\'{n}}ski, A.~F. Lasagni, and K.~Eckert.
\newblock Hydrogen bubble size distribution on nanostructured ni surfaces:
  Electrochemically active surface area versus wettability.
\newblock \emph{ACS Appl. Mater. Interfaces}, 15\penalty0 (14):\penalty0
  18290--18299, apr 2023.
\newblock \doi{10.1021/acsami.2c22231}.

\bibitem[Skibińska et~al.(2023)Skibińska, Wojtaszek, Krause, Kula, Yang,
  Marzec, Wojnicki, and Żabiński]{Skibinska2023}
Katarzyna Skibińska, Konrad Wojtaszek, Lukas Krause, Anna Kula, Xuegeng Yang,
  Mateusz~M. Marzec, Marek Wojnicki, and Piotr Żabiński.
\newblock Tuning up catalytical properties of electrochemically prepared
  nanoconical co-ni deposit for her and oer.
\newblock \emph{Appl. Surf. Sci.}, 607:\penalty0 155004, January 2023.
\newblock ISSN 0169-4332.
\newblock \doi{10.1016/j.apsusc.2022.155004}.

\bibitem[Ren et~al.(2023)Ren, Feng, Ye, Xue, Lin, Eisenberg, Kou, Duoss, Zhu,
  and Li]{Ren2023}
Qiu Ren, Longsheng Feng, Congwang Ye, Xinzhe Xue, Dun Lin, Samuel Eisenberg,
  Tianyi Kou, Eric~B. Duoss, Cheng Zhu, and Yat Li.
\newblock Nanocone-modified surface facilitates gas bubble detachment for
  high-rate alkaline water splitting.
\newblock \emph{Adv. Energy Mater.}, 13\penalty0 (39):\penalty0 2302073, sep
  2023.
\newblock \doi{10.1002/aenm.202302073}.

\bibitem[Ömer Akay et~al.(2022)Ömer Akay, Poon, Robertson, Abdi, Cuenya,
  Giersig, and Brinkert]{Akay2022}
Ömer Akay, Jeffrey Poon, Craig Robertson, Fatwa~Firdaus Abdi, Beatriz~Roldan
  Cuenya, Michael Giersig, and Katharina Brinkert.
\newblock Releasing the bubbles: Nanotopographical electrocatalyst design for
  efficient photoelectrochemical hydrogen production in microgravity
  environment.
\newblock \emph{Adv. Sci.}, 9\penalty0 (8):\penalty0 2105380, 2022.
\newblock \doi{10.1002/advs.202105380}.

\bibitem[Brinkert et~al.(2018)Brinkert, Richter, Akay, Liedtke, Giersig,
  Fountaine, and Lewerenz]{Brinkert2018}
Katharina Brinkert, Matthias~H. Richter, {\"O}mer Akay, Janine Liedtke, Michael
  Giersig, Katherine~T. Fountaine, and Hans-Joachim Lewerenz.
\newblock Efficient solar hydrogen generation in microgravity environment.
\newblock \emph{Nat. Commun.}, 9\penalty0 (1):\penalty0 2527, July 2018.
\newblock ISSN 2041-1723.
\newblock \doi{10.1038/s41467-018-04844-y}.

\bibitem[R{\"a}nke et~al.(2024)R{\"a}nke, Baumann, Voisiat, Soldera, and
  Lasagni]{Raenke2024}
Fabian R{\"a}nke, Robert Baumann, Bogdan Voisiat, Marcos Soldera, and
  Andr{\'e}s~Fabi{\'a}n Lasagni.
\newblock Nano/microstructuring of nickel electrodes by combining direct laser
  interference patterning and polygon scanner processing for efficient hydrogen
  production.
\newblock \emph{Adv. Eng. Mater.}, 26\penalty0 (10):\penalty0 2301583, 2024.
\newblock \doi{10.1002/adem.202301583}.

\bibitem[Ränke et~al.(2022)Ränke, Baumann, Voisiat, and {Fabián
  Lasagni}]{Raenke2022}
Fabian Ränke, Robert Baumann, Bogdan Voisiat, and Andrés {Fabián Lasagni}.
\newblock High throughput laser surface micro-structuring of polystyrene by
  combining direct laser interference patterning with polygon scanner
  technology.
\newblock \emph{Mater. Lett. X}, 14:\penalty0 100144, 2022.
\newblock ISSN 2590-1508.
\newblock \doi{10.1016/j.mlblux.2022.100144}.

\bibitem[Baumann et~al.(2020)Baumann, Rauscher, Bernäcker, Zwahr,
  Weißgärber, Röntzsch, and Lasagni]{Baumann2020}
Robert Baumann, Thomas Rauscher, Christian~Immanuel Bernäcker, Christoph
  Zwahr, Thomas Weißgärber, Lars Röntzsch, and Andrés~Fabián Lasagni.
\newblock Laser structuring of open cell metal foams for micro scale surface
  enlargement.
\newblock \emph{J. Laser Micro/Nanoeng.}, pages 132--138, sep 2020.
\newblock \doi{10.2961/jlmn.2020.02.2010}.

\bibitem[Koj et~al.(2019)Koj, Gimpel, Schade, and Turek]{Koj2019}
Matthias Koj, Thomas Gimpel, Wolfgang Schade, and Thomas Turek.
\newblock Laser structured nickel-iron electrodes for oxygen evolution in
  alkaline water electrolysis.
\newblock \emph{Int. J. Hydrog. Energy}, 44\penalty0 (25):\penalty0
  12671--12684, may 2019.
\newblock \doi{10.1016/j.ijhydene.2019.01.030}.

\bibitem[Rauscher et~al.(2017)Rauscher, Müller, Gabler, Gimpel, Köhring,
  Kieback, Schade, and Röntzsch]{Rauscher2017}
Thomas Rauscher, Christian~Immanuel Müller, Andreas Gabler, Thomas Gimpel,
  Michael Köhring, Bernd Kieback, Wolfgang Schade, and Lars Röntzsch.
\newblock Femtosecond-laser structuring of ni electrodes for highly active
  hydrogen evolution.
\newblock \emph{Electrochim. Acta}, 247:\penalty0 1130--1139, sep 2017.
\newblock \doi{10.1016/j.electacta.2017.07.074}.

\bibitem[Neale et~al.(2014)Neale, Jin, Ouyang, Hughes, Hesp, Dhanak, Dearden,
  Edwardson, and Hardwick]{Neale2014}
Alex~R. Neale, Yang Jin, Jinglei Ouyang, Stephen Hughes, David Hesp, Vinod
  Dhanak, Geoff Dearden, Stuart Edwardson, and Laurence~J. Hardwick.
\newblock Electrochemical performance of laser micro-structured nickel
  oxyhydroxide cathodes.
\newblock \emph{J. Power Sources}, 271:\penalty0 42--47, December 2014.
\newblock ISSN 0378-7753.
\newblock \doi{10.1016/j.jpowsour.2014.07.167}.

\bibitem[Lasagni et~al.(2017)Lasagni, Gachot, Trinh, Hans, Rosenkranz, Roch,
  Eckhardt, Kunze, Bieda, Günther, Lang, and Mücklich]{Lasagni2017}
Andrés~F. Lasagni, Carsten Gachot, Kim~E. Trinh, Michael Hans, Andreas
  Rosenkranz, Teja Roch, Sebastian Eckhardt, Tim Kunze, Matthias Bieda, Denise
  Günther, Valentin Lang, and Frank Mücklich.
\newblock Direct laser interference patterning, 20 years of development: from
  the basics to industrial applications.
\newblock In Udo Klotzbach, Kunihiko Washio, and Rainer Kling, editors,
  \emph{Laser-based Micro- and Nanoprocessing XI}. SPIE, February 2017.
\newblock \doi{10.1117/12.2252595}.

\bibitem[Wahab et~al.(2016)Wahab, Ghazali, Yusoff, and Sajuri]{Wahab2016}
J.~A. Wahab, M.~J. Ghazali, W.~M.~W. Yusoff, and Z.~Sajuri.
\newblock Enhancing material performance through laser surface texturing: a
  review.
\newblock \emph{Trans. IMF}, 94\penalty0 (4):\penalty0 193--198, July 2016.
\newblock ISSN 1745-9192.
\newblock \doi{10.1080/00202967.2016.1191141}.

\bibitem[R{\"o}mer et~al.(2014)R{\"o}mer, Skolski, Oboňa, Ocel{\'i}k,
  de~Hosson, and in~'t Veld]{Roemer2014}
G.~R. B.~E. R{\"o}mer, J.~Z.~P. Skolski, J.~Vincenc Oboňa, V.~Ocel{\'i}k,
  J.~T.~M. de~Hosson, and A.~J.~Huis in~'t Veld.
\newblock {Laser-induced periodic surface structures, modeling, experiments,
  and applications}.
\newblock In Udo Klotzbach, Kunihiko Washio, and Craig~B. Arnold, editors,
  \emph{Laser-based Micro- and Nanoprocessing VIII}, volume 8968, page 89680D.
  International Society for Optics and Photonics, SPIE, 2014.
\newblock \doi{10.1117/12.2037541}.

\bibitem[Sipe et~al.(1983)Sipe, Young, Preston, and van Driel]{Sipe1983}
J.~E. Sipe, Jeff~F. Young, J.~S. Preston, and H.~M. van Driel.
\newblock Laser-induced periodic surface structure. i. theory.
\newblock \emph{Phys. Rev. B}, 27:\penalty0 1141--1154, 1983.
\newblock \doi{10.1103/PhysRevB.27.1141}.

\bibitem[Bonse et~al.(2017)Bonse, Kirner, H{\"o}hm, Epperlein, Spaltmann,
  Rosenfeld, and Kr{\"u}ger]{Bonse2017}
J{\"o}rn Bonse, Sabrina~V. Kirner, Sandra H{\"o}hm, Nadja Epperlein, Dirk
  Spaltmann, Arkadi Rosenfeld, and J{\"o}rg Kr{\"u}ger.
\newblock {Applications of laser-induced periodic surface structures (LIPSS)}.
\newblock In Udo Klotzbach, Kunihiko Washio, and Rainer Kling, editors,
  \emph{Laser-based Micro- and Nanoprocessing XI}, volume 10092, page 100920N.
  International Society for Optics and Photonics, SPIE, 2017.
\newblock \doi{10.1117/12.2250919}.

\bibitem[Hoffmann et~al.(2022)Hoffmann, Hoffmann, Schade, Turek, and
  Gimpel]{Hoffmann2022}
Viktor Hoffmann, Luise Hoffmann, Wolfgang Schade, Thomas Turek, and Thomas
  Gimpel.
\newblock Femtosecond laser molybdenum alloyed and enlarged nickel surfaces for
  the hydrogen evolution reaction in alkaline water electrolysis.
\newblock \emph{International Journal of Hydrogen Energy}, 47\penalty0
  (48):\penalty0 20729--20740, 2022.
\newblock ISSN 0360-3199.
\newblock \doi{https://doi.org/10.1016/j.ijhydene.2022.04.194}.

\bibitem[Cai et~al.(2023)Cai, Li, Zhang, Wei, Tian, Hao, Wei, and
  Yang]{Cai2023}
Zaiwei Cai, Zihao Li, Yingtao Zhang, Chiyi Wei, Hao Tian, Molei Hao, Xiaoming
  Wei, and Zhongmin Yang.
\newblock {High repetition rate ultrafast laser-structured nickel
  electrocatalyst for efficient hydrogen evolution reaction}.
\newblock \emph{Advanced Photonics Nexus}, 2\penalty0 (5):\penalty0 056009,
  2023.
\newblock \doi{10.1117/1.APN.2.5.056009}.

\bibitem[Ränke et~al.(2004)Ränke, Moghtaderifard, Zschach, Baumann, Voisiat,
  Soldera, Günther, Hilbert, Bernäcker, Weißgärber, and
  Lasagni]{Raenke2024b}
Fabian Ränke, Stephan Moghtaderifard, Lis Zschach, Robert Baumann, Bogdan
  Voisiat, Marcos Soldera, Leonard Günther, Sebastian Hilbert,
  Christian~Immanuel Bernäcker, Thomas Weißgärber, and Andrés~Fabián
  Lasagni.
\newblock Femtosecond direct laser interference patterning of nickel
  electrodesfor improving electrochemical properties.
\newblock \emph{J. Laser Micro Nanoeng.}, pages 102--110, September 2004.
\newblock ISSN 1880-0688.
\newblock \doi{10.2961/jlmn.2024.02.2002}.

\bibitem[Fabbri and Schmidt(2018)]{Fabbri2018}
Emiliana Fabbri and Thomas~J. Schmidt.
\newblock Oxygen evolution reaction—the enigma in water electrolysis.
\newblock \emph{ACS Catal.}, 8\penalty0 (10):\penalty0 9765--9774, September
  2018.
\newblock ISSN 2155-5435.
\newblock \doi{10.1021/acscatal.8b02712}.

\bibitem[Suen et~al.(2017)Suen, Hung, Quan, Zhang, Xu, and Chen]{Suen2017}
Nian-Tzu Suen, Sung-Fu Hung, Quan Quan, Nan Zhang, Yi-Jun Xu, and Hao~Ming
  Chen.
\newblock Electrocatalysis for the oxygen evolution reaction: recent
  development and future perspectives.
\newblock \emph{Chem. Soc. Rev.}, 46\penalty0 (2):\penalty0 337--365, 2017.
\newblock ISSN 1460-4744.
\newblock \doi{10.1039/c6cs00328a}.

\bibitem[Lee(2019)]{lee_statistical_2019}
Robert Lee.
\newblock Statistical {Design} of {Experiments} for {Screening} and
  {Optimization}.
\newblock \emph{Chem. Ing. Tech.}, 91\penalty0 (3):\penalty0 191--200, 2019.
\newblock ISSN 0009-286X, 1522-2640.
\newblock \doi{10.1002/cite.201800100}.

\bibitem[Lasagni and Voisiat(2021)]{Lasagni2021}
Andrés~Fabian Lasagni and Bogdan Voisiat.
\newblock Anordnung optischer elemente zur ausbildung von großflächigen
  strukturen mit linienförmigen strukturelementen, 2021.
\newblock No. DE102020204656A1.

\bibitem[Rox et~al.(2023)Rox, Bashkatov, Yang, Loos, Mutschke, Gerbeth, and
  Eckert]{Rox2023}
H.~Rox, A.~Bashkatov, X.~Yang, S.~Loos, G.~Mutschke, G.~Gerbeth, and K.~Eckert.
\newblock Bubble size distribution and electrode coverage at porous nickel
  electrodes in a novel 3-electrode flow-through cell.
\newblock \emph{Int. J. Hydrog. Energy}, 48\penalty0 (8):\penalty0 2892--2905,
  2023.
\newblock \doi{10.1016/j.ijhydene.2022.10.165}.

\bibitem[Heinrich et~al.(2024)Heinrich, Ränke, Schwarzenberger, Yang, Baumann,
  Marzec, Lasagni, and Eckert]{Heinrich2024}
Julian Heinrich, Fabian Ränke, Karin Schwarzenberger, Xuegeng Yang, Robert
  Baumann, Mateusz Marzec, Andrés~Fabián Lasagni, and Kerstin Eckert.
\newblock Functionalization of ti64 via direct laser interference patterning
  and its influence on wettability and oxygen bubble nucleation.
\newblock \emph{Langmuir}, 40\penalty0 (6):\penalty0 2918--2929, 2024.
\newblock \doi{10.1021/acs.langmuir.3c02863}.

\bibitem[Schmidt et~al.(2018)Schmidt, Weigert, Broaddus, and
  Myers]{Schmidt2018}
Uwe Schmidt, Martin Weigert, Coleman Broaddus, and Gene Myers.
\newblock Cell detection with star-convex polygons.
\newblock In \emph{Medical Image Computing and Computer Assisted Intervention -
  {MICCAI} 2018 - 21st International Conference, Granada, Spain, September
  16-20, 2018, Proceedings, Part {II}}, pages 265--273, 2018.
\newblock \doi{10.1007/978-3-030-00934-2_30}.

\bibitem[Weigert et~al.(2020)Weigert, Schmidt, Haase, Sugawara, and
  Myers]{Weigert2020}
Martin Weigert, Uwe Schmidt, Robert Haase, Ko~Sugawara, and Gene Myers.
\newblock Star-convex polyhedra for 3d object detection and segmentation in
  microscopy.
\newblock In \emph{The IEEE Winter Conference on Applications of Computer
  Vision (WACV)}, March 2020.
\newblock \doi{10.1109/WACV45572.2020.9093435}.

\bibitem[Allan et~al.(2023)Allan, Caswell, Keim, van~der Wel, and
  Verweij]{Allan2023}
Daniel~B. Allan, Thomas Caswell, Nathan~C. Keim, Casper~M. van~der Wel, and
  Ruben~W. Verweij.
\newblock soft-matter/trackpy: v0.6.1, February 2023.

\bibitem[Zhu et~al.(2019)Zhu, Zhang, Ni, Shen, and Lu]{Zhu2019}
HuaZhong Zhu, HongChao Zhang, XiaoWu Ni, ZhongHua Shen, and Jian Lu.
\newblock {Fabrication of superhydrophilic surface on metallic nickel by
  sub-nanosecond laser-induced ablation}.
\newblock \emph{AIP Advances}, 9\penalty0 (8):\penalty0 085308, 2019.
\newblock ISSN 2158-3226.
\newblock \doi{10.1063/1.5111069}.

\bibitem[Schäfer et~al.(2024)Schäfer, Delfino, Leonhard-Trautmann, Ott,
  Suarez, Stüber, Mücklich, and Pauly]{Schaefer2024}
Christian Schäfer, Pablo~Maria Delfino, Philipp Leonhard-Trautmann, Vincent
  Ott, Sebastian Suarez, Michael Stüber, Frank Mücklich, and Christoph Pauly.
\newblock Fabrication of smooth, periodic surface structures: Combining direct
  laser interference patterning and electropolishing.
\newblock \emph{Advanced Engineering Materials}, page 2400435, 2024.
\newblock \doi{10.1002/adem.202400435}.

\bibitem[Bonse et~al.(2012)Bonse, Krüger, Höhm, and Rosenfeld]{Bonse2012}
J.~Bonse, J.~Krüger, S.~Höhm, and A.~Rosenfeld.
\newblock {Femtosecond laser-induced periodic surface structures}.
\newblock \emph{Journal of Laser Applications}, 24\penalty0 (4):\penalty0
  042006, 2012.
\newblock ISSN 1042-346X.
\newblock \doi{10.2351/1.4712658}.

\bibitem[Bonse and Gräf(2020)]{Bonse2020}
Jörn Bonse and Stephan Gräf.
\newblock Maxwell meets marangoni—a review of theories on laser-induced
  periodic surface structures.
\newblock \emph{Laser \& Photonics Reviews}, 14\penalty0 (10):\penalty0
  2000215, 2020.
\newblock \doi{10.1002/lpor.202000215}.

\bibitem[Aguilar-Morales et~al.(2018)Aguilar-Morales, Alamri, and
  Lasagni]{Aguilar2018}
Alfredo~I. Aguilar-Morales, Sabri Alamri, and Andrés~Fabián Lasagni.
\newblock Micro-fabrication of high aspect ratio periodic structures on
  stainless steel by picosecond direct laser interference patterning.
\newblock \emph{Journal of Materials Processing Technology}, 252:\penalty0
  313--321, 2018.
\newblock ISSN 0924-0136.
\newblock \doi{10.1016/j.jmatprotec.2017.09.039}.

\bibitem[Liu et~al.(2019)Liu, Lin, Lin, Ji, and Hong]{Liu2019}
Huagang Liu, Wenxiong Lin, Zhenyuan Lin, Lingfei Ji, and Minghui Hong.
\newblock Self-organized periodic microholes array formation on aluminum
  surface via femtosecond laser ablation induced incubation effect.
\newblock \emph{Advanced Functional Materials}, 29\penalty0 (42):\penalty0
  1903576, 2019.
\newblock \doi{10.1002/adfm.201903576}.

\bibitem[Bernäcker et~al.(2022)Bernäcker, Gimpel, Bomm, Rauscher, Mauermann,
  Li, Hübner, Schade, and Röntzsch]{Bernaecker2022}
Christian~I. Bernäcker, Thomas Gimpel, Alexander Bomm, Thomas Rauscher,
  Sebastian Mauermann, Mingji Li, Eike~G. Hübner, Wolfgang Schade, and Lars
  Röntzsch.
\newblock Short pulse laser structuring as a scalable process to produce
  cathodes for large alkaline water electrolyzers.
\newblock \emph{J. Power Sources}, 538:\penalty0 231572, 2022.
\newblock \doi{10.1016/j.jpowsour.2022.231572}.

\bibitem[Pang et~al.(2020)Pang, Davis, Ill, and Esposito]{Pang2020}
Xueqi Pang, Jonathan~T. Davis, Albert D.~Harvey Ill, and Daniel~V. Esposito.
\newblock Framework for evaluating the performance limits of membraneless
  electrolyzers.
\newblock \emph{Energy Environ. Sci.}, 13:\penalty0 3663--3678, 2020.
\newblock \doi{10.1039/d0ee02268c}.

\bibitem[Craig et~al.(1993)Craig, Ninham, and Pashley]{Craig1993}
Vincent S.~J. Craig, Barry~W. Ninham, and Richard~M. Pashley.
\newblock The effect of electrolytes on bubble coalescence in water.
\newblock \emph{J. Phys. Chem.}, 97\penalty0 (39):\penalty0 10192--10197, sep
  1993.
\newblock \doi{10.1021/j100141a047}.

\bibitem[Beamson and Briggs(1993)]{beamson1993}
G.~Beamson and D.~Briggs.
\newblock High relution xps of organic polymers. the scienta esca 300 database.
\newblock \emph{J. Chem. Educ.}, 70:\penalty0 A25, 1993.
\newblock \doi{10.1021/ed070pA25.5}.

\bibitem[Rouxhet and Genet(2011)]{rouxhet2011}
Paul~G Rouxhet and Michel~J Genet.
\newblock Xps analysis of bio-organic systems.
\newblock \emph{Surf. Interface Anal.}, 43\penalty0 (12):\penalty0 1453--1470,
  2011.
\newblock \doi{10.1002/sia.3831}.

\bibitem[Liu et~al.(2020)Liu, Li, Zhao, Liu, Luo, Yuan, Luo, He, Yu, and
  Zhao]{liu2020}
Nan Liu, Tong Li, Ziqiong Zhao, Jing Liu, Xiaoguang Luo, Xiaohong Yuan, Kun
  Luo, Julong He, Dongli Yu, and Yuanchun Zhao.
\newblock From triazine to heptazine: origin of graphitic carbon nitride as a
  photocatalyst.
\newblock \emph{ACS Omega}, 5\penalty0 (21):\penalty0 12557--12567, 2020.
\newblock \doi{10.1021/acsomega.0c01607}.

\bibitem[Genet et~al.(2008)Genet, Dupont-Gillain, and Rouxhet]{genet2008}
Michel~J. Genet, Christine~C. Dupont-Gillain, and Paul~G. Rouxhet.
\newblock Xps analysis of biosystems and biomaterials.
\newblock In Egon Matijevic, editor, \emph{Medical applications of colloids},
  volume 177, pages 177--307. Springer, 2008.
\newblock \doi{10.1007/978-0-387-76921-9}.

\bibitem[Wagner et~al.(1982)Wagner, Passoja, Hillery, Kinisky, Six, Jansen, and
  Taylor]{wagner1982}
C.~D. Wagner, D.~E. Passoja, H.~F. Hillery, T.~G. Kinisky, H.~A. Six, W.~T.
  Jansen, and J.~A. Taylor.
\newblock Auger and photoelectron line energy relationships in aluminum--oxygen
  and silicon--oxygen compounds.
\newblock \emph{J. Vac. Sci. Technol.}, 21\penalty0 (4):\penalty0 933--944,
  1982.
\newblock \doi{10.1116/1.571870}.

\bibitem[Wagner et~al.(2003)Wagner, Naumkin, Kraut-Vass, Allison, Powell, and
  Rumble]{Wagner2003}
A.D. Wagner, A.V. Naumkin, A.~Kraut-Vass, J.W. Allison, C.J. Powell, and J.R.J.
  Rumble, 2003.
\newblock URL \url{http:/srdata.nist.gov/xps/}.

\bibitem[Fantauzzi et~al.(2015)Fantauzzi, Elsener, Atzei, Rigoldi, and
  Rossi]{fantauzzi2015}
Marzia Fantauzzi, Bernhard Elsener, Davide Atzei, Americo Rigoldi, and
  Antonella Rossi.
\newblock Exploiting xps for the identification of sulfides and polysulfides.
\newblock \emph{RSC Adv.}, 5\penalty0 (93):\penalty0 75953--75963, 2015.
\newblock \doi{10.1039/C5RA14915K}.

\bibitem[Biesinger et~al.(2009)Biesinger, Payne, Lau, Gerson, and
  Smart]{biesinger2009}
Mark~C. Biesinger, Brad~P. Payne, Leo W.~M. Lau, Andrea Gerson, and Roger
  St.~C. Smart.
\newblock X-ray photoelectron spectroscopic chemical state quantification of
  mixed nickel metal, oxide and hydroxide systems.
\newblock \emph{Surf. Interface Anal.}, 41\penalty0 (4):\penalty0 324--332,
  2009.
\newblock \doi{10.1002/sia.3026}.

\bibitem[Biesinger et~al.(2011)Biesinger, Payne, Grosvenor, Lau, Gerson, and
  Smart]{biesinger2011}
Mark~C. Biesinger, Brad~P. Payne, Andrew~P. Grosvenor, Leo W.~M. Lau, Andrea~R.
  Gerson, and Roger St.~C. Smart.
\newblock Resolving surface chemical states in xps analysis of first row
  transition metals, oxides and hydroxides: Cr, mn, fe, co and ni.
\newblock \emph{Appl. Surf. Sci.}, 257\penalty0 (7):\penalty0 2717--2730, 2011.
\newblock \doi{10.1016/j.apsusc.2010.10.051}.

\bibitem[Biesinger et~al.(2012)Biesinger, Lau, Gerson, and
  Smart]{biesinger2012}
Mark~C. Biesinger, Leo W.~M. Lau, Andrea~R. Gerson, and Roger St.~C. Smart.
\newblock The role of the auger parameter in xps studies of nickel metal,
  halides and oxides.
\newblock \emph{Phys. Chem. Chem. Phys.}, 14\penalty0 (7):\penalty0 2434--2442,
  2012.
\newblock \doi{10.1039/C2CP22419D}.

\end{thebibliography}

\clearpage

\renewcommand{\thefigure}{S\arabic{figure}}
\renewcommand{\thetable}{S\arabic{table}}
\renewcommand{\theequation}{S\arabic{equation}}
\renewcommand{\thesection}{S\arabic{section}}
\setcounter{section}{0}
\setcounter{equation}{0}
\setcounter{figure}{0}
\setcounter{table}{0}
\section*{Electronic supplementary information:\\ Boosting electrode performance and bubble management via Direct Laser Interference Patterning}

\section{Design of Experiments}

\subsection*{Full-factorial design}

\begin{table}[h]
	\small
	\centering
	\caption{\ Randomized experiment sequence according to full-factorial design for galvanostatic measurements with dropped experiments for $\Lambda = \SI{30}{\micro\metre}$ and $AR = 1.0$ as this structure could not be manufactured reproducibly}
	\begin{tabular*}{0.5\textwidth}{@{\extracolsep{\fill}}llll}
		\hline
		Exp. No. & $\Lambda$ (\si{\micro\metre}) & $AR$ (-) & $j$ (\si{\milli\ampere\per\centi\metre\squared})\\
		\hline
		1 & 6.00 & 0.33 & 10.00\\
		2 & 30.00 & 0.33 & 100.00 \\
		3 & 30.00 & 0.33 & 31.62 \\
		4 & 6.00 & 1.00 & 10.00 \\
		5 & 6.00 & 1.00 & 100.00 \\
		6 & 15.00 & 1.00 & 31.62 \\
		7 & 15.00 & 0.67 & 31.62 \\
		8 & 6.00 & 0.33 & 31.62 \\
		9 & 15.00 & 0.67 & 100.00 \\
		10 & 30.00 & 0.67 & 31.62 \\
		11 & 6.00 & 0.67 & 10.00 \\
		12 & 15.00 & 1.00 & 10.00 \\
		13 & 15.00 & 0.33 & 10.00 \\
		14 & 30.00 & 0.67 & 100.00 \\
		15 & 6.00 & 0.33 & 100.00 \\
		16 & 15.00 & 0.33 & 100.00 \\
		17 & 6.00 & 1.00 & 31.62 \\
		18 & 15.00 & 0.33 & 31.62 \\
		19 & 15.00 & 0.67 & 31.62 \\
		20 & 6.00 & 0.67 & 31.62 \\
		21 & 15.00 & 0.67 & 10.00 \\
		22 & 6.00 & 0.67 & 100.00 \\
		23 & 30.00 & 0.33 & 10.00 \\
		24 & 15.00 & 1.00 & 100.00 \\
		25 & 15.00 & 0.67 & 31.62 \\
		26 & 30.00 & 0.67 & 10.00 \\
		27 & \multicolumn{2}{c}{NSE} & 10.00\\
		28 & \multicolumn{2}{c}{NSE} & 31.62\\
		29 & \multicolumn{2}{c}{NSE} & 100.00 \\
		\hline
	\end{tabular*}
	\label{tbl:appendix_doe_galvanostatic}
\end{table}

\section{Preliminary experiments on DLIP}
In the first set of experiments, nickel substrates were irradiated using the two-beam DLIP configuration to generate line shaped surface features, exhibiting spatial periods of 6.0, 15.0 and \SI{30.0}{\micro\metre}. As process parameter the number of consecutive passes $N$ were varied from 1 to 45 to evaluate their influence on the resulting structure morphology and aspect ratio $AR$. For this purpose, the total amount of energy that is used to irradiate a certain area (cumulated laser fluence $\Phi_\text{cum}$) has been calculated using Eq. \ref{eq:appendix_nbr_pulses} and \ref{eq:appendix_fluence}:
\begin{align}
	N_\text{pulses} &= \frac{d_\text{y} \cdot f_\text{rep}}{v_\text{scan}} \label{eq:appendix_nbr_pulses}\\
	\Phi_\text{cum} &= \frac{E_\text{p}}{A_\text{spot}} \cdot N_\text{pulses} \cdot N \label{eq:appendix_fluence},
\end{align}
where $N_\text{pulses}$ denotes the number of laser pulses irradiating the same effective area. In this context, the pulse-to-pulse distance of \SI{5}{\micro\metre} corresponds to 16 accumulated laser pulses per area. $A_\text{spot} = \pi \cdot d_\text{x} \cdot d_\text{y}$ describes the area of the interfering laser beams and $N$ the number of consecutive scans. In all experiments, the repetition rate remained constant, resulting in pulse energies $E_\text{p}$ of \SI{612}{\micro\joule}, with calculated cumulated laser fluence values $\Phi_\text{cum}$ spanning from 4.1 to \SI{122.3}{\joule\per\centi\metre\squared}. 

\begin{figure*}[h]
	\centering
	\includegraphics[width=0.5\textwidth]{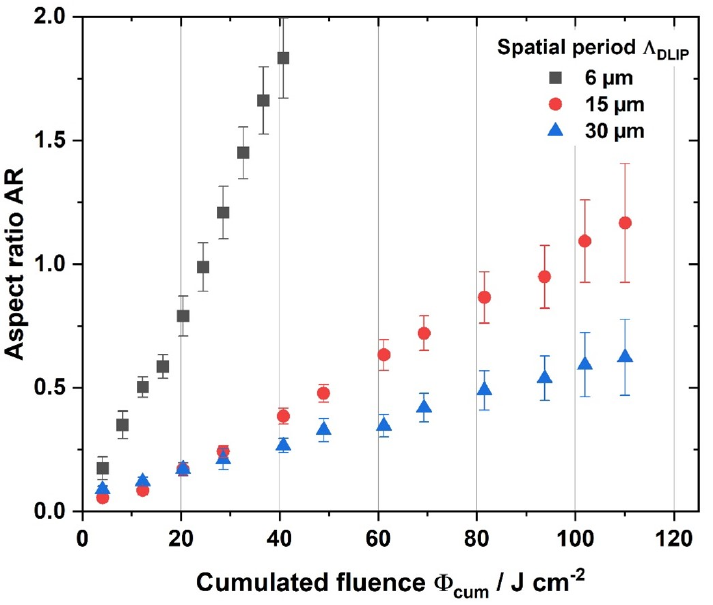}
	\caption{\ Aspect ratio $AR$ of resulting line-like DLIP structures in dependency of the cumulated laser fluence $\Phi_\text{cum}$ for spatial periods $\Lambda$ of 6, 15 and \SI{30}{\micro\metre} fabricated with a single pulse fluence $\Phi_\text{sp}$ of \SI{0.25}{\joule\per\centi\metre\squared} and a pulse-to-pulse distance $PtP$ of \SI{5}{\micro\metre} for different number of scans $N$.}
	\label{fgr:appendix_laser_fluence}
\end{figure*}

The generated $AR$ for the single-line experiments (without laser beam hatching) in relation to the applied cumulated fluences are presented in Fig. \ref{fgr:appendix_laser_fluence}. For this analysis, only the central area of the ablated zone, corresponding to the peak fluence of the Gaussian laser spot, was considered. Across all spatial periods $\Lambda$, an increase in cumulative laser fluence resulted in a linear rise of $AR$ for the DLIP line-like features.

The steepest increase in the $AR$ curve was observed for the smallest spatial period of \SI{6}{\micro\metre}, while larger structure periods led to a continuous flattening of the aspect ratio curves. This flattening could be attributed to the enlargement of the maxima regions within the interference profile, resulting in lower ablation rates. Therefore, it can be concluded that higher energy densities (cumulative fluences) are necessary to achieve higher aspect ratios for larger structure periods. For the \SI{6}{\micro\metre} period, the highest $AR$ of 1.8 was achieved, with a greater volume of material vaporized due to the laser-material interaction compared to other structure periods. Maximum aspect ratios of 1.2 and 0.6 were reached for structure periods of \SI{15}{\micro\metre} and \SI{30}{\micro\metre}, respectively.

Based on these results, nickel electrode areas of $\SI{25}{\milli\metre} \times \SI{10}{\milli\metre}$ were equipped with line-like DLIP features aiming to generate aspect ratios of 0.33, 0.67 and 1.0 for all mentioned spatial periods. It should be noted that an aspect ratio of 1.0 was not achieved for the structure period of \SI{30}{\micro\metre}. The reason was the significant bending of the nickel foils caused by high thermal stresses during processing with elevated cumulative energy densities. As a result of this bending, the alignment of the individual laser lines could no longer be maintained, leading to partial destruction of the microstructure. 

\section{Working electrode holder}
\begin{figure*}[h]
	\centering
	\includegraphics[width=0.9\textwidth]{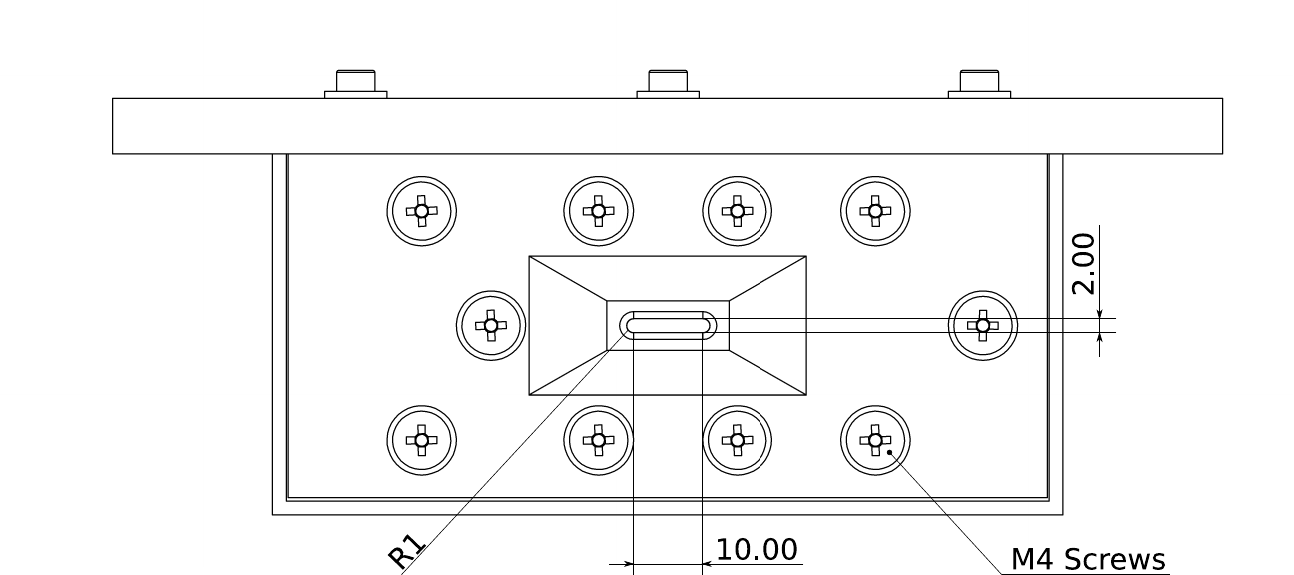}
	\caption{\ Drawing of the working electrode holder with dimensions of the open area.}
	\label{fgr:appendix_we_holder}
\end{figure*}
\newpage
\section{Image processing}
\begin{figure*}[h]
	\centering
	\includegraphics[width=0.9\textwidth]{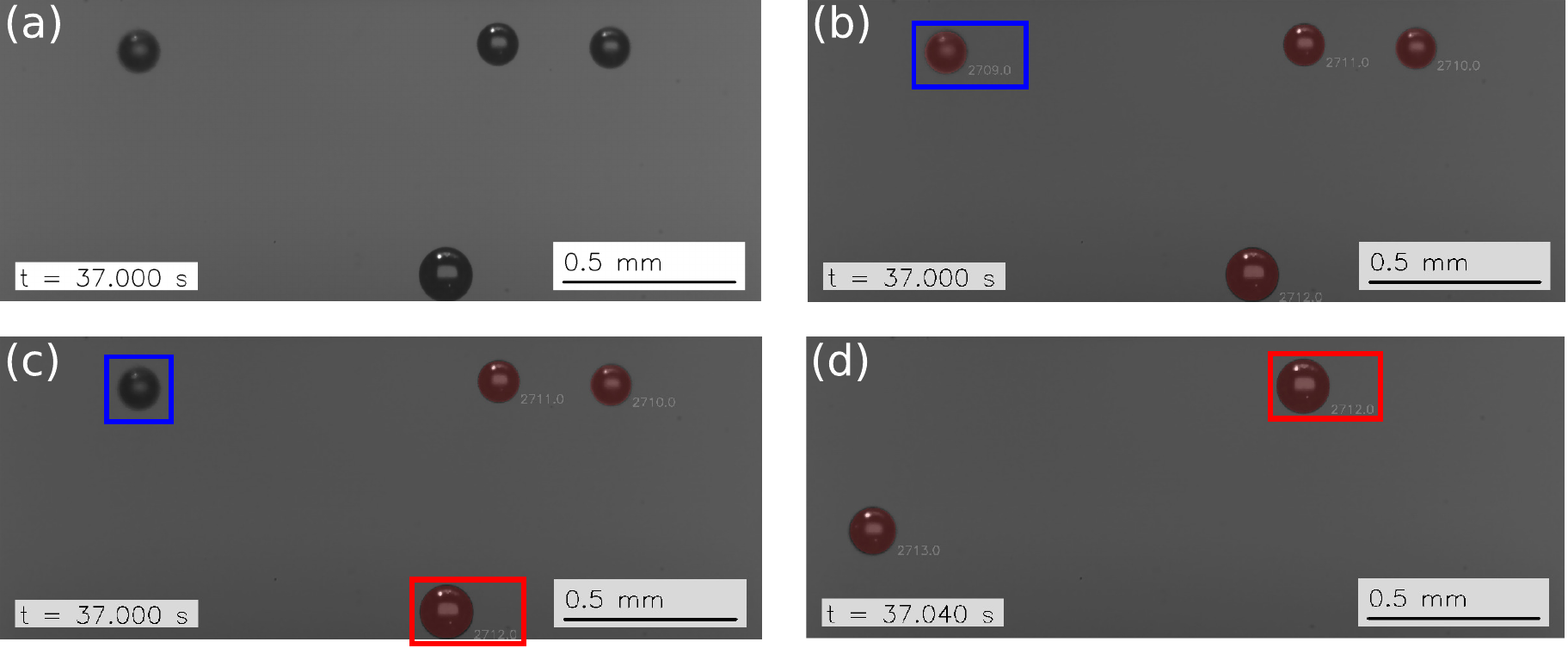}
	\caption{\ Procedure of image processing of sideview images: (a) Raw image, where (b) all bubble-like objects are segmented and linked (e.g. tracked bubble highlighted in red rectangle in (c-d)) using stardist and trackpy, respectively. Afterwards, by calculating the size-normalized variance of the bubble image Laplacian (Var$(\Delta) \cdot d_\text{B}$) blurred bubbles, like the highlighted bubble in the blue rectangle can be excluded.}
	\label{fgr:appendix_image_processing_sideview}
\end{figure*}

\begin{figure*}[h]
	\centering
	\includegraphics[width=0.9\textwidth]{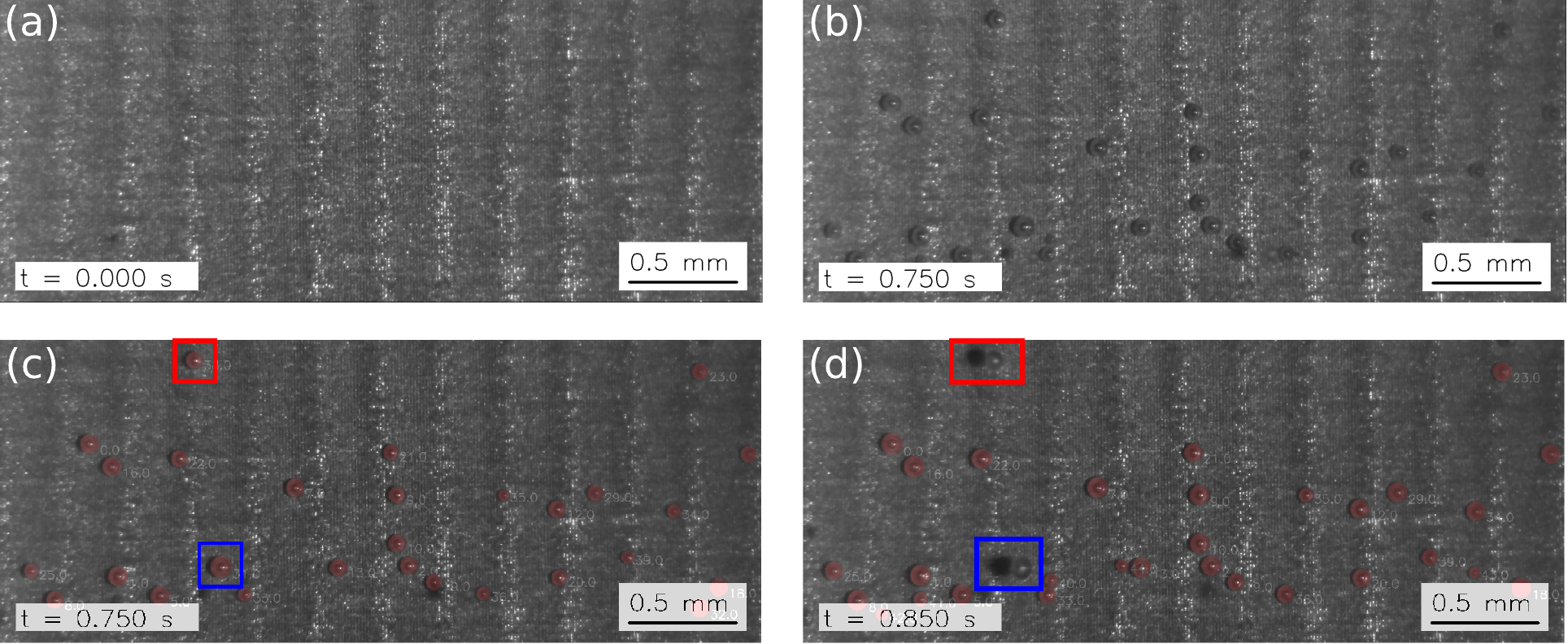}
	\caption{\ Procedure of image processing of topview images: (a) Clean electrode surface at beginning of experiment and (b) evolving O$_2$-bubbles after applying $j$. (c-d) Segmented and linked bubbles sitting on electrode using stardist and trackpy, respectively. The highlighted bubbles show the distinction between bubble sitting on the electrode and detached, rising bubble with a shadow cast on the electrode surface.}
	\label{fgr:appendix_image_processing_topview}
\end{figure*}
\newpage
\section{Electrode surface}
\begin{figure*}[ht]
	\centering
	\includegraphics[width=\textwidth]{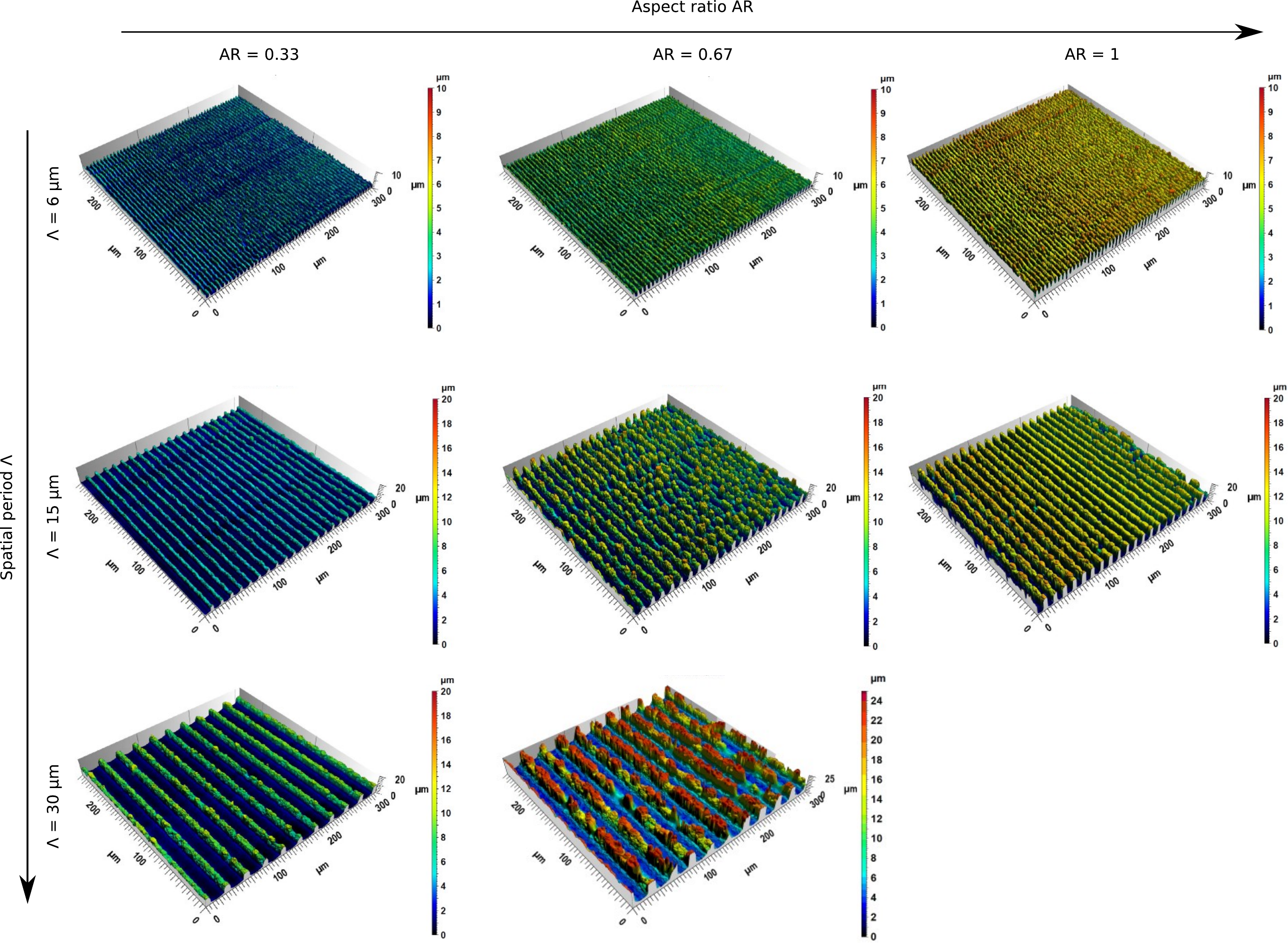}
	\caption{\ 3D confocal images of all DLIP line-like structures.}
	\label{fgr:appendix_electrode_structure}
\end{figure*}
\begin{figure*}[ht!]
	\centering\includegraphics[width=\textwidth]{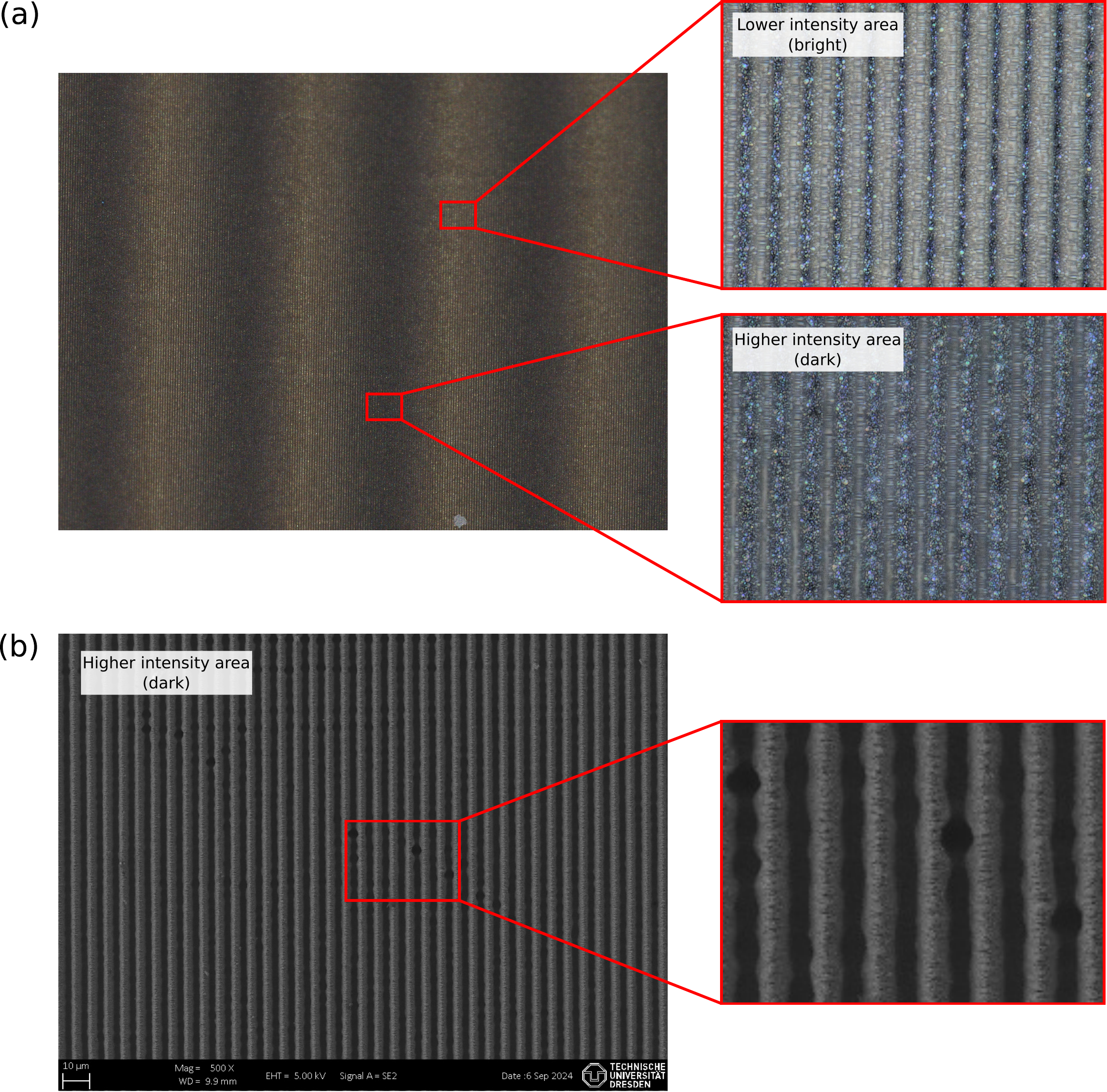}
	\caption{(a) Images of electrode \#4 ($\Lambda = \SI{6}{\micro\metre}$ and $AR = 1$) taken with a Keyence VHX Digital Microscope showing a more shallow profile for the brighter area. (b) SEM image of the higher intensity (dark) area of electrode \#4 showing microholes in the maxima region of the interference pattern, which are not present in the lower intensity area.}
	\label{fgr:appendix_electrode_structure_p6}
\end{figure*}

\clearpage

\section{Wetting of electrodes}
\begin{figure*}[h]
	\centering
	\includegraphics[width=0.9\textwidth]{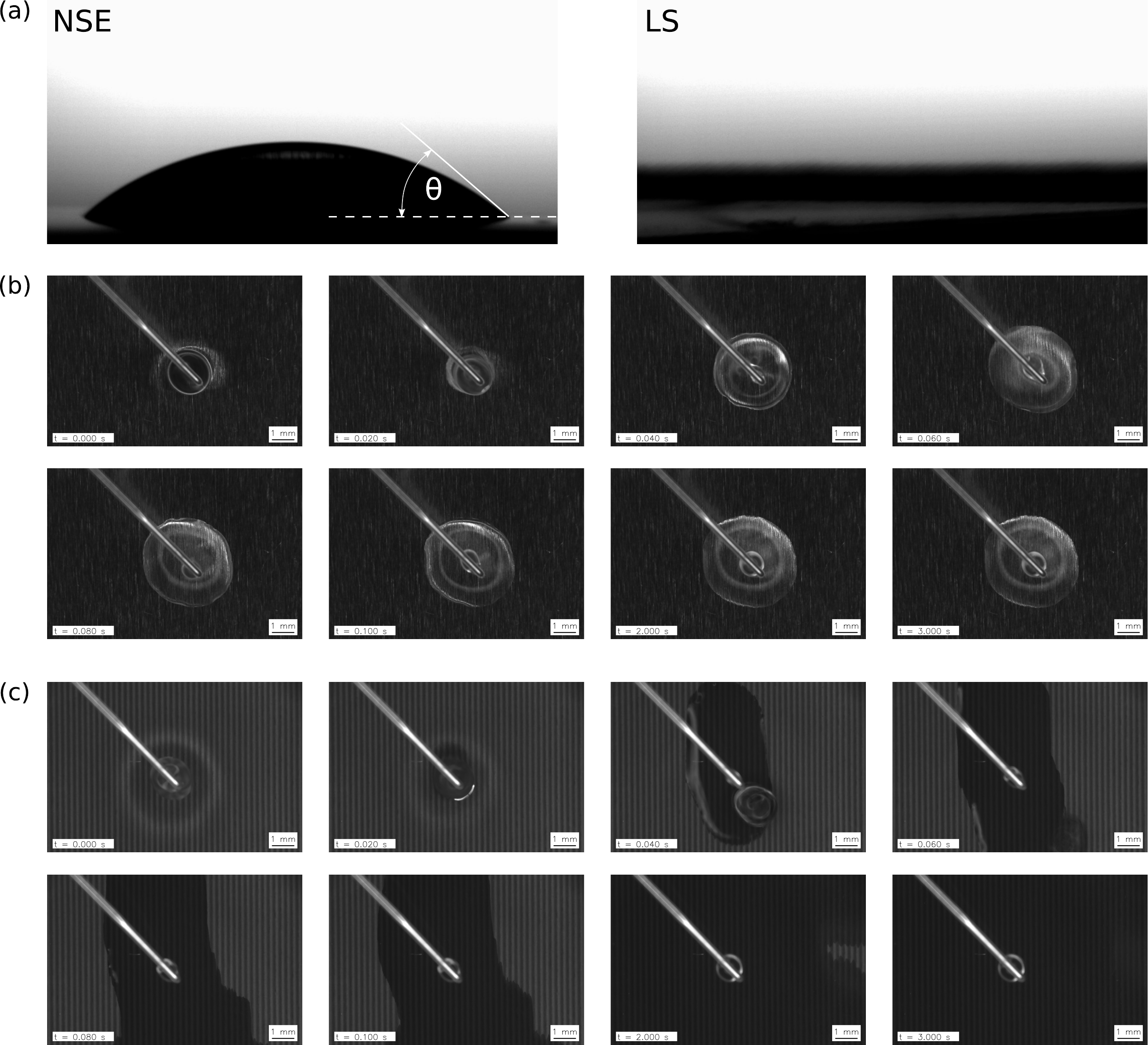}
	\caption{\ (a) Contact angle measurement of electrodes with highlighted $\theta_\text{NSE} \approx  \SI{38.5}{\degree}$ and the non-visible droplet on the DLIP-structured electrode due to superhydrophilic surface. Wetting behaviour of (b) non-structured and (c) laser-structured Ni-foil showing superhydrophilic wetting of laser-structured surface by applying a droplet of $\approx$ \SI{0.2}{ml} of DI water on the surface with a \SI{0.4}{mm} needle and the droplet spreads within less than \SI{3}{\second} across entire surface.}
	\label{fgr:appendix_wetting}
\end{figure*}
\newpage

\section{Electrochemical characterization of electrodes}
\subsection{Measurement of double-layer capacitance}
\begin{figure*}[h]
	\centering
	\includegraphics[width=\textwidth]{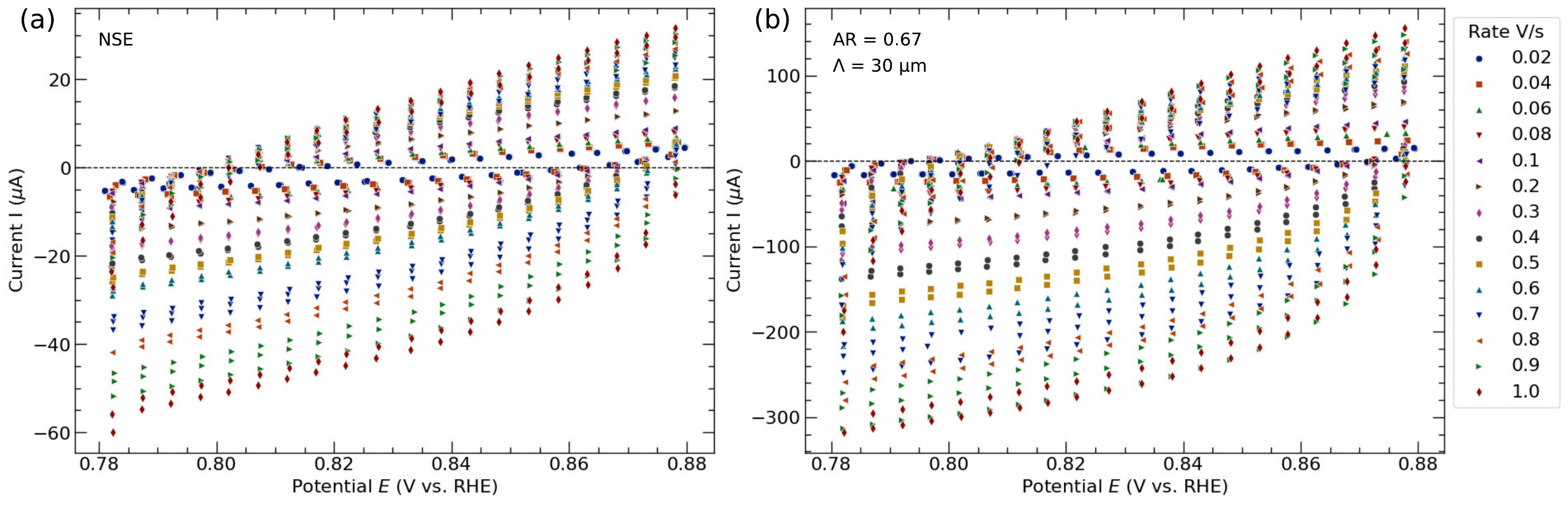}
	\caption{\ CVs at different scan rates $\nu$ for (a) NSE and (b) DLIP-structured electrode.}
	\label{fgr:appendix_ecsa}
\end{figure*}
\subsection{Measurement of onset potential}
\begin{figure*}[h]
	\centering
	\includegraphics[width=\textwidth]{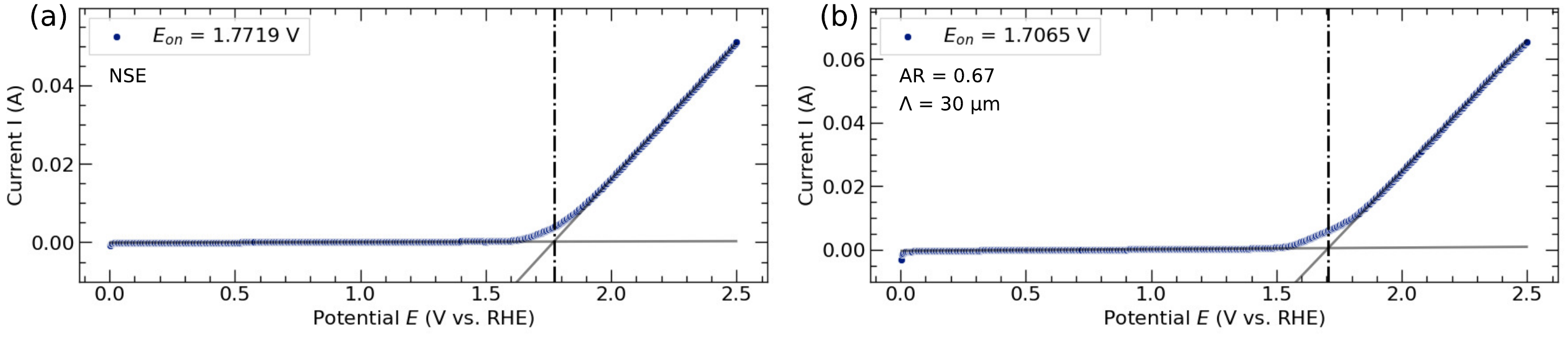}
	\caption{\ LSVs for (a) NSE and (b) DLIP-structured electrode with fitted tangents to calculate $E_\text{on}$.}
	\label{fgr:appendix_onset_potential}
\end{figure*}
\newpage
\section{XPS Spectra}
\subsection{Deconvolution and fitting of XPS Spectra}
For the fitting of the detailed study on the influence of the LIPSS structures (dark/bright pattern) on the electrode surfaces, slightly different values of the binding energy were used for individual spectra. In theses cases, the binding energy  used is given in brackets.

The C 1s spectra for all samples were fitted with four components. First line centered at \SI{285.0}{eV} arise from aliphatic carbon C-C, second line lies at \SI{286.5}{eV} and indicate presence of C-O and/or C-N bonds, third line centered at \SI{288.2}{eV} indicate presence of C=O and/or N-C=O bonds\cite{beamson1993, rouxhet2011}, and fourth line at \SI{289.3}{eV} indicate presence of O-C=O and or CO$_3$$^{2-}$ type compounds\cite{beamson1993}.

The N 1s spectra were fitted with up to three lines: first centered at \SI{398.3}{eV} indicate presence of N=C type bonds, second line at \SI{400.1}{eV} originates from central three-coordinated nitrogen N-C$_3$ and/or amine type groups and third line positioned at \SI{402.6}{eV} which comes from NH$_4^+$ type ions presence\cite{beamson1993, liu2020}. For the LIPSS study only a single line centered at \SI{400.0}{eV} was used for fitting the  N 1s spectra, indicating the presence of N-C=O and/or C-NH type groups\cite{beamson1993, liu2020}.

The O 1s spectra are similar for all samples and were fitted using three lines, with first line centered at  \SI{529.8}{eV} (\SI{529.5}{eV}) which indicates presence of metal oxide (O-Ni), second line at \SI{531.5}{eV} (\SI{531.2}{eV}) indicates presence of defective oxygen in metal oxides and/or O=C and/or O-Si type bonds and/or CO$_3$$^{2-}$ groups, and the last line found at \SI{532.2}{eV} (\SI{532.6}{eV}) which can originate either from O-H and/or C-O type bonds and/or adsorbed H$_2$O\cite{beamson1993, genet2008, wagner1982}.

The P 2p spectra were fitted with doublet structure (p$_{3/2}$ – p$_{1/2}$ doublet separation equals \SI{0.84}{eV}) with main 2p$_{3/2}$ line centered at \SI{133.2}{eV} which indicates presence of P$^{5+}$ oxidation state like in PO$_4^{3-}$\cite{Wagner2003}.

For the LIPSS study instead of the P 2p spectra, the Si 2p spectra show two doublet structures (doublet separation p$_{3/2}$ – p$_{1/2}$ equals \SI{0.6}{eV}) with first 2p$_{3/2}$ line centered at \SI{102.0}{eV} which indicate presence of C-Si-O type bonds like in silicones/siloxanes\cite{Wagner2003} and second 2p$_{3/2}$ line centered at \SI{103.7}{eV} which indicate presence of silica type compounds like in e.g. SiO$_2$\cite{wagner1982, Wagner2003}.

The S 2p spectra were fitted with doublet structure (p$_{3/2}$ – p$_{1/2}$ doublet separation equals \SI{1.16}{eV}) with main 2p$_{3/2}$ line centered at \SI{168.3}{eV} which indicate presence of SO$_3$$^{2-}$ ions\cite{Wagner2003, fantauzzi2015}.

The spectra collected at Ni 2p$_{3/2}$ region are similar for all samples where nickel was detected. Each spectrum was fitted with up to six lines. First asymmetric line centered at \SI{852.3}{eV} indicate presence of metallic nickel whereas second line found at \SI{853.8}{eV} indicate the Ni$^{2+}$ in nickel oxide NiO and/or hydroxide\cite{biesinger2009, biesinger2011, biesinger2012}. The four lines within energy range of 855 – \SI{866}{eV} are due to the multiplet splitting phenomena.
\newpage
\subsection{XPS Spectra and surface composition}
\begin{figure*}[ht]
	\centering
	\includegraphics[width=\textwidth]{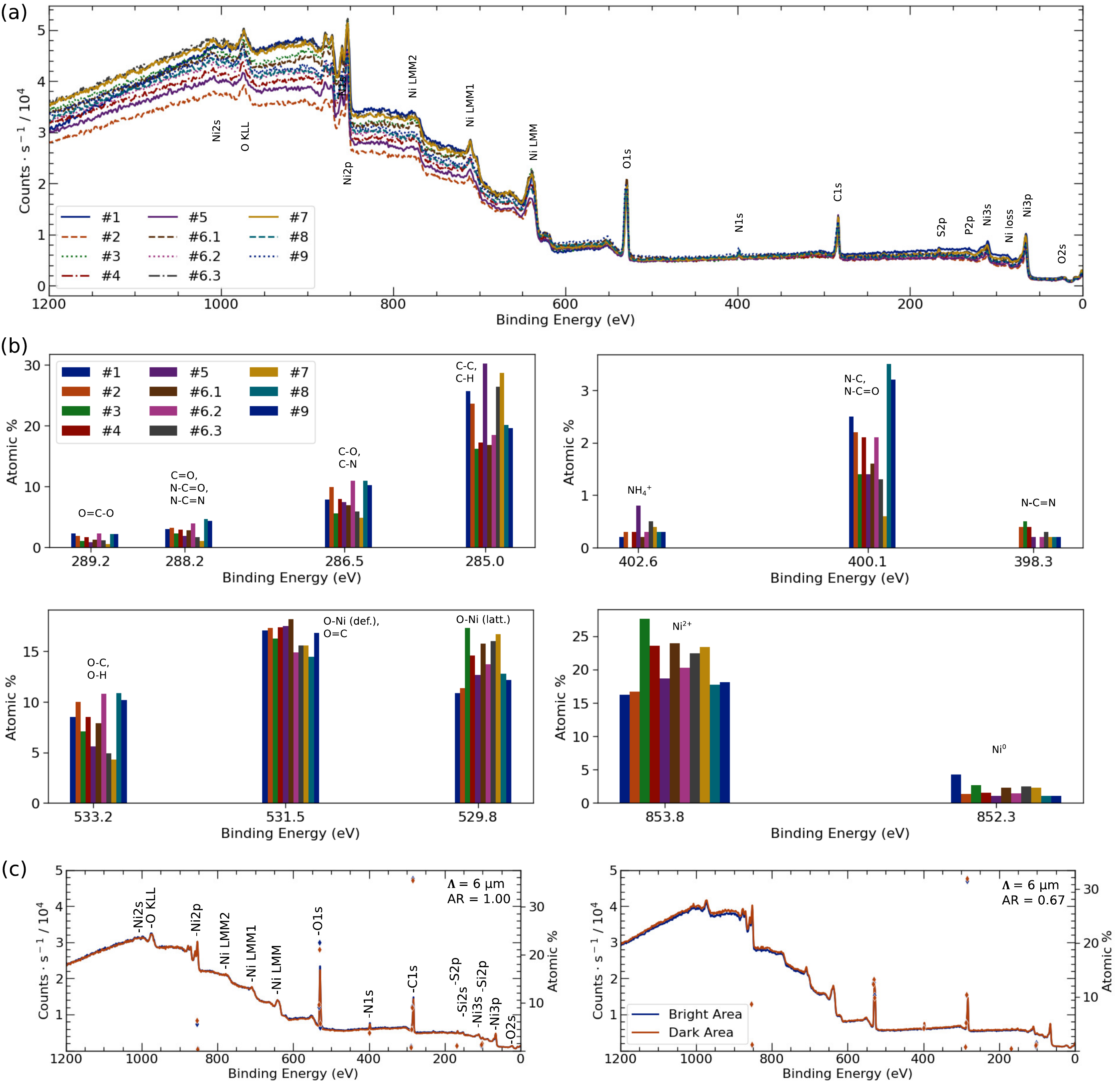}
	\caption{\ (a) Survey spectra of XPS measurements and (b) surface composition for the Elements C, N, O and Ni for all studied electrodes. (c) Survey spectra of XPS measurements and surface composition inside and outside of $HD$ (dark/bright pattern, see Fig. \ref{fgr:appendix_electrode_structure_p6}) of the two electrodes for which linear patterned bubble nucleation could be observed.}
	\label{fgr:appendix_xps}
\end{figure*}
\newpage
\begin{landscape}
	
	\topskip0pt
	\begin{table}
		\centering
		\small
		\caption{\ Surface composition (in atomic \%) determined by fitting XPS spectra for all studied electrodes. The position of the \SI{100}{\micro\metre} X-ray focus spot was chosen randomly on the electrode surface. Thus, no distinction is made between HIA and LIA.}
		\label{tbl:appendix_xps}
		\begin{tabular*}{24cm}{@{\extracolsep{\fill}}lcccccccccccccc}
			\hline
			\textbf{Element} & \multicolumn{4}{c}{\textbf{C}} & \multicolumn{3}{c}{\textbf{N}} & \multicolumn{3}{c}{\textbf{O}} & \textbf{P} & \textbf{S} & \multicolumn{2}{c}{\textbf{Ni}} \\
			BE (eV) & 285.0 & 286.5 & 288.2 & 289.2 & 398.2 & 400.1 & 402.6 & 529.8 & 531.5 & 533.2 & 133.2 & 168.3 & 852.3 & 853.8\\
			Groups, & C-C, & C-O, & C=O, & \multirow{3}{*}{O=C-O} & \multirow{3}{*}{N-C=N} & N-C & \multirow{3}{*}{NH$_4^+$} & \multirow{3}{*}{O-Ni (latt.)} & O-Ni(def.), & O-C, & \multirow{3}{*}{PO$_4^{3-}$} & \multirow{3}{*}{SO$_3^{2-}$} & \multirow{3}{*}{Ni$^0$} & \multirow{3}{*}{Ni$^{2+}$}\\
			Ox. state & C-H & C-N & N-C=O & & & N-C=O & & & O=C & O-H & & & & \\
			& & & N-C=N & & & & & & & & & &  \\
			\hline
			\#1 & 25.7 & 7.8 & 3.0 & 2.3 & 0.0 & 2.5 & 0.2 & 10.9 & 17.1 & 8.5 & 0.6 & 0.8 & 4.3 & 16.2 \\
			\#2 & 23.6 & 9.9 & 3.2 & 1.9 & 0.4 & 2.2 & 0.3 & 11.4 & 17.3 & 10.0 & 0.7 & 1.1 & 1.3 & 16.7 \\
			\#3 & 16.2 & 5.6 & 2.3 & 1.1 & 0.5 & 1.4 & 0.0 & 17.3 & 16.3 & 7.1 & 1.5 & 0.4 & 2.7 & 27.6 \\
			\#4 & 17.2 & 8.0 & 2.9 & 1.7 & 0.4 & 2.1 & 0.3 & 14.6 & 17.4 & 8.5 & 1.2 & 0.6 & 1.5 & 23.5 \\
			\#5 & 30.2 & 7.4 & 1.9 & 0.8 & 0.2 & 1.4 & 0.8 & 12.7 & 17.5 & 5.6 & 0.0 & 1.7 & 1.1 & 18.7 \\
			\#6.1 & 16.8 & 6.9 & 2.8 & 1.3 & 0.0 & 1.6 & 0.2 & 15.8 & 18.2 & 7.9 & 1.9 & 0.5 & 2.3 & 23.9 \\
			\#6.2 & 18.5 & 10.9 & 3.9 & 2.3 & 0.2 & 2.1 & 0.3 & 13.7 & 14.9 & 10.8 & 0.0 & 0.9 & 1.4 & 20.3 \\
			\#6.3 & 26.4 & 5.9 & 1.7 & 1.2 & 0.3 & 1.3 & 0.5 & 16.0 & 15.6 & 4.9 & 0.0 & 1.5 & 2.5 & 22.4 \\
			\#7 & 28.6 & 4.9 & 1.1 & 0.5 & 0.2 & 0.6 & 0.4 & 16.7 & 15.6 & 4.3 & 0.0 & 1.4 & 2.3 & 23.4 \\
			\#8 & 20.1 & 10.9 & 4.7 & 2.2 & 0.2 & 3.5 & 0.3 & 12.8 & 14.5 & 10.9 & 0.7 & 0.7 & 1.1 & 17.7 \\
			\#9 & 19.6 & 10.2 & 4.3 & 2.2 & 0.2 & 3.2 & 0.3 & 12.2 & 16.8 & 10.2 & 0.3 & 0.7 & 1.1 & 18.1 \\
			\hline 
		\end{tabular*}
	\end{table}
\end{landscape}
\section{Models of the multiple regression analysis}
\begin{figure*}[h]
	\centering
	\includegraphics[width=0.9\textwidth]{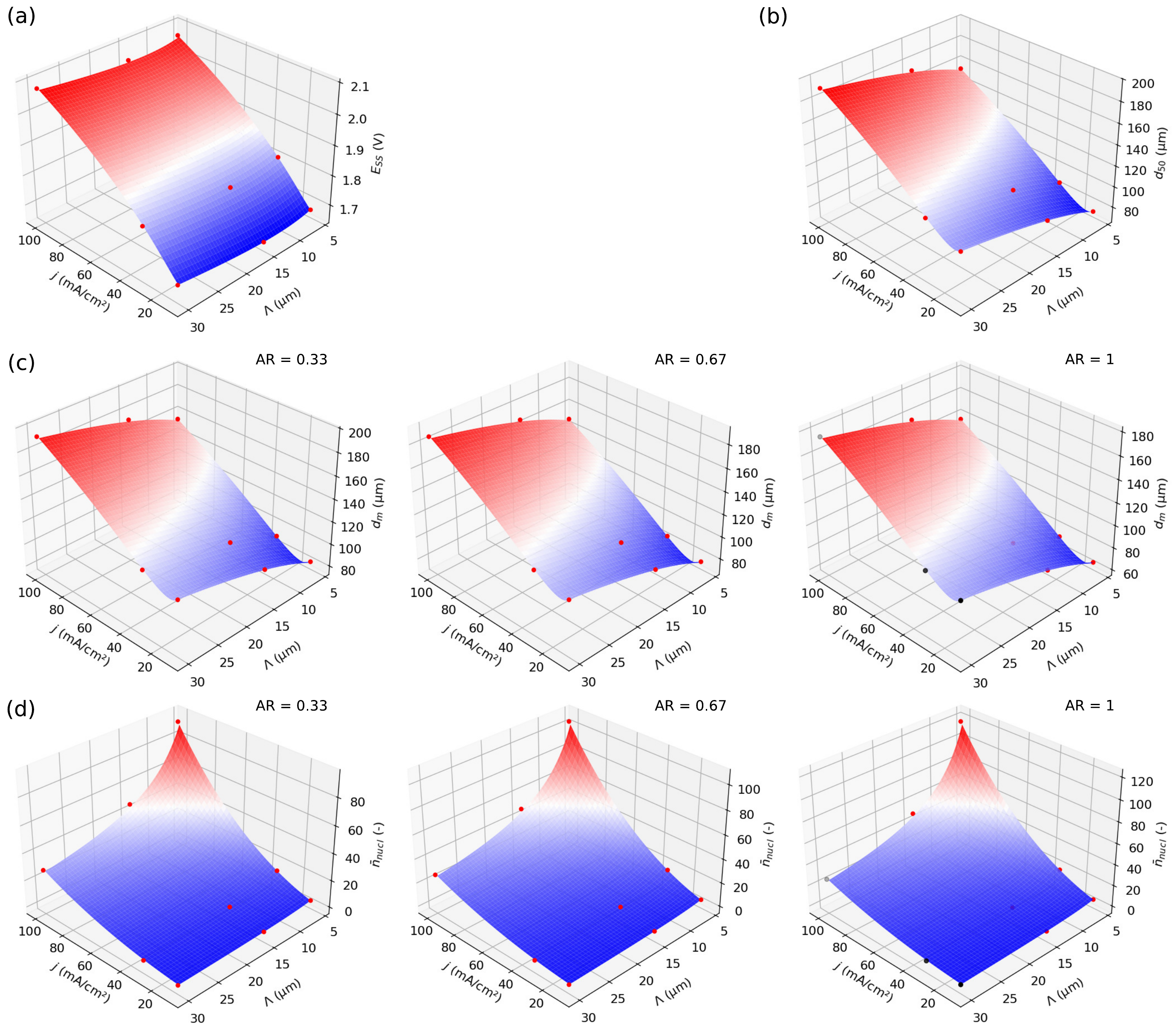}
	\caption{\ Surface plots of the determined models with highlighted measuring points of (a) $E_\text{SS}$, (b) $d_{50}$, (c) $d_\text{m}$ and (d) $\bar{n}_\text{nucl}$. Measurement points at $AR = 1$ in (c) and (d) plotted in black mark the extrapolated data points of the non-existent electrode with $\Lambda = \SI{30}{\micro\metre}$ and $AR = 1$.}
	\label{fgr:appendix_models}
\end{figure*}

For all studied responses
\begin{itemize}
	\itemsep-5pt 
	\item Double-layer capacitance ($C_\text{dl}$),
	\item Onset potential ($E_\text{on}$),
	\item Quasi-steady state potential ($E_\text{SS}$),
	\item Number of nucleation sites ($n_\text{nucl}$),
	\item Mode ($d_\text{m}$) and median ($d_{50}$) value of the bubble size distributions,
\end{itemize}
response surface models were derived using multiple regression analysis for a better understanding of the influence of these factors:
\begin{itemize}
	\itemsep-5pt 
	\item Spatial period ($\Lambda$)
	\item Aspect ratio ($AR$)
	\item Current density ($j$)
\end{itemize}
The resulting surface plots of these models are shown in Fig. \ref{fgr:appendix_models}.
\newpage
\section{Error analysis of experiments}

\begin{table}[h]
	\small
	\centering
	\caption{\ Measurement noise ($\sigma$) of experiments calculated with the central and reference point of DoE for double-layer capacitance ($C_\text{dl}$), onset potential ($E_\text{on}$), quasi-steady state potential ($E_\text{SS}$), number of nucleation sites ($n_\text{nucl}$) and the mode ($d_\text{m}$) and median value ($d_{50}$) of the bubble size distributions}
	\label{tbl:appendixs_measurement_noise}
	\begin{tabular*}{0.48\textwidth}{@{\extracolsep{\fill}}lcc}
		\hline
		\multirow{2}{*}{Parameter} & \multicolumn{2}{c}{Measurement point} \\ & Central point & Reference point \\
		\hline
		$C_\text{dl}$ & \SI{4.6991}{\micro\farad} & \SI{1.7194}{\micro\farad} \\
		$E_\text{on}$ & \SI{0.0050}{\volt} & \SI{0.0215}{\volt} \\
		$E_\text{SS}$ & \SI{0.0244}{\volt} & - \\
		$n_\text{nucl}$ & 3.5447 & - \\
		$d_\text{m}$ & \SI{17.0762}{\micro\metre} & - \\
		$d_{50}$ & \SI{19.3189}{\micro\metre} & - \\
		\hline
	\end{tabular*}
\end{table}
\section{Data and videos}
Sample data sets with raw images, electrochemical measurement data, and results can be found at \href{https://doi.org/10.14278/rodare.3064}{10.14278/rodare.3064}. Due to the size of the complete data, the remaining image data can be made available upon request.

The provided characteristic videos are named after following scheme:
\begin{center}
	\textit{Perspective\_Electrode\_CurrentDensity} $\rightarrow$ \textit{E.g.: Sideview\_\#1\_NSE\_100mAcm-2}
\end{center}

\end{document}